\documentclass[review,number,sort&compress]{elsarticle}

\usepackage[utf8]{inputenc}
\usepackage{amssymb}
\usepackage{graphicx}
\usepackage{amsmath}
\usepackage{amsthm}
\usepackage{wasysym}
\usepackage{url}
\usepackage{multirow}
\usepackage{lineno}
\usepackage{color}

\def\units#1{\hbox{$\,{\rm #1}$}}

\journal{Nuclear Instruments \& Methods A}

\begin{document}

\begin{frontmatter}

\title{Identification of particles with Lorentz factor up to $10^{4}$ with
Transition Radiation Detectors based on micro-strip silicon detectors}

\author[add1]{J.~Alozy}
\author[add2]{N.~Belyaev}
\author[add1]{M.~Campbell}
\author[add3]{M.~Cherry}
\author[add1,add4]{F.~Dachs}
\author[add2]{S.~Doronin}
\author[add2]{K.~Filippov}
\author[add5,add6]{P.~Fusco}
\author[add6]{F.~Gargano}
\author[add1]{E.~Heijne}
\author[add7]{S.~Konovalov}
\author[add2]{D.~Krasnopevtsev}
\author[add1]{X.~Llopart}
\author[add5,add6]{F.Loparco\corref{cor}}
\ead{francesco.loparco@ba.infn.it}
\author[add8,add9]{V.~Mascagna}
\author[add6]{M.~N.~Mazziotta\corref{cor}}
\ead{marionicola.mazziotta@ba.infn.it}
\author[add1]{H.~Pernegger}
\author[add2]{D.~Ponomarenko}
\author[add8,add9]{M.~Prest}
\author[add2]{D.~Pyatiizbyantseva}
\author[add2]{R.~Radomskii}
\author[add1]{C.~Rembser}
\author[add2]{A.~Romaniouk\corref{cor}}
\ead{anatoli.romaniouk@cern.ch}
\author[add2,add10]{A.~A.~Savchenko}
\author[add11]{D.~Schaefer}
\author[add1]{E.~J.~Schioppa}
\author[add7]{D.~Shchukin}
\author[add2,add10]{D.~Yu~Sergeeva}
\author[add2]{E.~Shulga}
\author[add2]{S.~Smirnov}
\author[add2]{Y.~Smirnov}
\author[add8,add9]{M.~Soldani}
\author[add5,add6]{P.~Spinelli}
\author[add2]{M.~Strikhanov}
\author[add2]{P.~Teterin}
\author[add7]{V.~Tikhomirov}
\author[add2,add10]{A.~A.~Tishchenko}
\author[add12]{E.~Vallazza}
\author[add13]{M.~van~Beuzekom}
\author[add13]{B.~van~der~Heijden}
\author[add2]{K.~Vorobev}
\author[add7]{K.~Zhukov}

\cortext[cor]{Corresponding authors}

\address[add1]{CERN, the European Organization for Nuclear Research, Esplanade des Particules 1, 1211 Geneva, Switzerland }
\address[add2]{National Research Nuclear University MEPhI (Moscow Engineering Physics Institute), Kashirskoe highway 31, Moscow, 115409, Russia}
\address[add3]{Dept.\ of Physics \& Astronomy, Louisiana State University, Baton Rouge, LA 70803 USA}
\address[add4]{Technical University of Vienna, Karlsplatz 13, 1040 Vienna, Austria}
\address[add5]{ Dipartimento di Fisica “M. Merlin” dell'Universit\`a e del Politecnico di Bari, Via G. Amendola 173, 70126 Bari, Italy }
\address[add6]{ Istituto Nazionale di Fisica Nucleare, Sezione di Bari,  Via E. Orabona 4, 70126 Bari, Italy }
\address[add7]{P.\ N.\ Lebedev Physical Institute of the~Russian Academy of Sciences, Leninsky prospect 53, Moscow, 119991, Russia}
\address[add8]{INFN Milano Bicocca, Piazza della Scienza 3, 20126 Milano}
\address[add9]{Universit\'a degli Studi dell'Insubria, Via Valleggio 11, 22100, Como}
\address[add10]{National Research Center “Kurchatov Institute”, Akademika Kurchatova pl. 1, Moscow, Russia}
\address[add11]{University of Chicago, 5801 S Ellis Ave, Chicago IL 60637, USA}
\address[add12]{INFN Trieste, Padriciano 99, 34149 Trieste, Italy}
\address[add13]{Nikhef, Science Park 105, 1098 XG Amsterdam, The Netherlands}

\begin{abstract}

This work is dedicated to the study of a technique for hadron identification
in the $\units{TeV}$ momentum range, based on the simultaneous measurement of
the energies and of the emission angles of the Transition Radiation (TR) X-rays
with respect to the radiating particles. A detector setup has been built
and tested with particles in a wide range of Lorentz factors (from
about $10^{3}$ to about $4 \times 10^{4}$ crossing different types
of radiators. The measured double-differential (in energy and angle) 
spectra of the TR photons are in a reasonably good agreement 
with TR simulation predictions.

\end{abstract}

\begin{keyword}
Transition Radiation Detectors \sep Particle identification

\end{keyword}

\end{frontmatter}

\section{Introduction}

Particle identification (PID) is a crucial aspect in all high-energy physics experiments. 
There exist only 3 techniques which can identify particles without a significant effect of interaction with 
the detector material: Time-of-flight (TOF), Cherenkov radiation and Transition Radiation (TR). 
Transition radiation detectors (TRDs) are used for PID for many decades and can be the
only option for PID in certain ranges of the Lorentz gamma factor.

Predicted by Ginzburg and Frank~\cite{Ginzburg:1945zz}, TR is emitted whenever a charged particle crosses the interface between two media 
with different dielectric constants. In an ultrarelativistic case, the TR appears in a X-ray energy range and has a sort of 
threshold effect as a function of the gamma factor~\cite{Garibian:1958aa,Garibian:1960aa}.
Since the average number of TR photons per interface is small (usually one per $30$ foils) radiators consisting of  
multiple foils (typical thickness of a few tens of $\units{\mu m}$) are used. 
Regularly spaced foils (the spacing from fraction to a few $\units{mm}$) allow to obtain a 
coherent sum of amplitudes from different foils and enhance the TR yield (see for instance~\cite{Artru:1975xp}).  
In conventional detectors the TR emission usually starts from a threshold Lorentz factor $\gamma_{th}$ of 
about $5\cdot10^{2}$ (this value depends on the radiator material and foil thickness) and saturates at a 
Lorentz factor of about $10$ times larger than the threshold (defined by the ratio between the spacing and the foil thickness). 
TRDs are widely used for electron-hadron separation in  both accelerator~\cite{Aad:2008zzm,Aamodt:2008zz} 
and cosmic-ray experiments~\cite{Kirn:2013bca} (see also the review in Ref.~\cite{Tanabashi:2018oca}). 
Such detectors efficiently separate  electrons (positrons) from pions with momenta up to more than $100\units{GeV/c}$ and 
from antiprotons (protons) in a momentum range extending  up to about $1 \units{TeV/c}$.  
The major factor which defines the performance of any TRD is its overall length. 
A typical TRD consists of a radiator followed by a X-ray detector (gaseous detectors are usually chosen); configurations with 
multiple TRD modules are often implemented to reach required particle separation. In order to achieve a  
hadron contamination at the level of  $0.1\%$ while keeping a $90\%$ electron identification efficiency, a TRD length 
of at least $1 \units{m}$ is needed (see fig.~34.16 in ref.~\cite{Tanabashi:2018oca}).
Alternative configurations have been also proposed, in which silicon strip detectors are used in conjunction with 
a magnetic field to separate the radiating particles from the TR photons in relatively modest 
distances~\cite{Brigida:2005yt,Brigida:2006ma,Brigida:2007zza}.

Hadron identification with TRDs is usually more challenging, since the relative differences between 
the masses of the particles involved in the selection are smaller. In the past, several authors reported the 
performances of various TRD prototypes equipped with gaseous detectors, used to separate pions from protons 
and kaons at momenta up to $200\units{GeV/c}$~\cite{Fabjan:1980hv,Oganesian:1977sh,Commichau:1980tb,Deutschmann:1980xe,Ludlam:1980tm}.
They showed that pion contamination values of a few percent can be achieved while keeping a kaon selection 
efficiency of $90\%$ at $140 \units{GeV/c}$ momentum with 15 sets of TRD modules.
The results also show that, in order to achieve similar performance as in the electron-hadron case, longer TRDs are required. 

With growing energies of particles at modern or planned accelerator experiments as well as in various 
cosmic-ray experiments there is a need to identify particles including hadrons with gamma factors up to $\sim 10^5$.
An example of such an experiment can be the measurement of the inclusive cross sections in the forward region 
for the production of charged particles in proton-proton, proton-nucleus and nucleus-nucleus interactions  
at center-of-mass energies of $14 \units{TeV}$~\cite{Albrow:2018kxz}. In these collisions, secondary hadrons 
(mainly pions, kaons, protons) are produced with momenta in the $\units{TeV}$ range. Hadron identification 
in this range is highly challenging, since all particles have $1-\beta<10^{-5}$, and Cherenkov detectors become ineffective. 
The only detector which can perform hadron identification in the $\units{TeV}$ region is therefore the TRD.

In conventional TRDs, PID is performed exploiting the information about the energy loss in the detectors, 
which includes ionization losses and the energies of absorbed TR photons. 
Particles are separated by measuring the total energy deposited in the detector or by counting  
candidate TR photons. This approach is proven to be very effective for separation of particles which 
produce TR from those which do not produce it. Various species of hadrons with momenta in the $\units{TeV}$ 
region have close masses and emit some amount of TR. In this case a set-up which includes sets of detectors having 
different TR production thresholds must be used (see for instance~\cite{Belyaev:2017iyt}). 
This unavoidably ends up with a quite large material budget (in terms of radiation lengths and nuclear interaction lengths). 

In order to enhance the PID performance of TRDs it is important to use all available information about TR production,  
e.g. to measure simultaneously the number of X-rays, their energies and the angles at which they are produced. This kind of 
measurement requires a high-granularity detector with high X-ray detection efficiency at a distance of about $1 \units{m}$
from the radiator in order to ensure particle-photon separations at the level of a hundred $\units{\mu m}$.
Real detector performance strongly depends on the radiator 
parameters and particle gamma factors. Parameters of the detector can be optimized with simulations; 
however this study requires basic measurements and detailed comparisons with simulations. 
In the past, a few measurements of the TR angular distribution were made~\cite{Deutschmann:1980xe,Belyaev:2017fyq}.
In this  paper for the first time the results of a high-precision measurement of the TR photon energies and production 
angles are presented.

The measurements were performed using a $300 \units{\mu m}$ thick double-face silicon strip detector (DSSD) 
with a readout strip pitch of $50 \units{\mu m}$. They were carried out at the CERN SPS facility with $20\units{GeV/c}$ 
electrons and with muons from $120$ to $290 \units{GeV/c}$ crossing different types of radiators.  
Data are compared with the predictions of a dedicated simulation. 
Similar measurements were also performed using a silicon pixel detector. 
Preliminary results are published in ~\cite{Pixel} and detailed results will be published elsewhere.

\section{Detector description and test beam setup}
\label{sec:detector}

\subsection{The silicon detector and the readout ASICs}

A double-side high-resistivity $300 \units{\mu m}$ thick silicon micro-strip detector with a sensitive area 
of $1.92 \times 1.92 \units{cm^{2}}$ developed by the INSULAB 
group~\cite{Lietti:2013dfa} was used in the test beam studies. The main features of the detector are summarized 
in Table~\ref{tab:detector}. On the $p$-side (``junction side'') a readout scheme with one floating strip is implemented;
it has $768$ strips, with a pitch of $25\units{\mu m}$ and with a readout pitch of $50\units{\mu m}$.  
The $n$-side (``ohmic side'') is equipped with $384$ 
strips, orthogonal to those on the junction side, with a pitch 
of $50 \units{\mu m}$. All the strips on the ohmic side are readout. Hence the
detector has $384$ readout strips with a $50\units{\mu m}$ pitch on both sides.    
The detector is fully depleted with a bias voltage of $50\units{V}$. The average
leakage current is of the order of $1.5 \units{nA}$ per strip.

\begin{table}[!t]
\begin{center}
\begin{tabular}{ll}
\hline
\hline
Detector manufacturer & CSEM~\cite{CSEM} \\
ASIC & VA2 \\
\hline
Sensitive area ($\units{cm^{2}}$) & $1.92 \times 1.92$ \\
Si sensor thickness ($\units{\mu m}$) & $300$ \\
Number of readout channels & $768$ \\
Leakage current ($\units{nA/strip}$) & $1.5$ \\
Full depletion bias voltage ($\units{V}$) & $50$ \\
\hline
$p$-side (junction) & \\
Strip pitch ($\units{\mu m}$) & $25$ \\
Readout pitch ($\units{\mu m}$) & $50$ \\
Floating scheme & yes \\
\hline
$n$-side (ohmic side) & \\
Strip pitch ($\units{\mu m}$) & $50$ \\
Readout pitch ($\units{\mu m}$) & $50$ \\
Floating scheme & no \\
\hline 
\hline
\end{tabular}
\end{center}
\caption{General features of the silicon micro-strip detector installed at the test beam line.}
\label{tab:detector}
\end{table}

\begin{figure}
\begin{center}
\begin{tabular}{c}
\includegraphics[width=0.75\columnwidth]{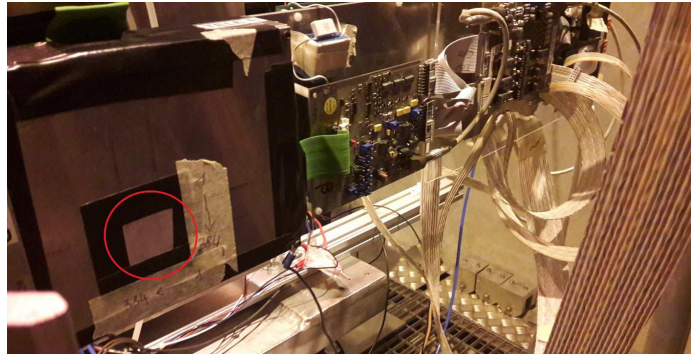}
\end{tabular}
\end{center}
\caption{ The silicon micro-strip detector installed on the beam line.
The red circle indicates the kapton foil placed in front of the silicon detector.}
\label{fig:detector}
\end{figure}

Each detector side is read-out by three $128$ channels VA2 ASICs~\cite{VA2,Toker:1993dd,Barbiellini:2002vf},  
with main parameters shown in table~\ref{tab:va2}. The strips are capacitively coupled to the input of the 
ASICs using custom $128$ channel capacitor chips built on a quartz substrate. 
Each ASIC channel consists of a low-noise charge sensitive amplifier, a CR-RC shaper and a sample-and-hold (S\&H) circuit. 
The signal from each strip is amplified and shaped, and its peak amplitude is sampled by a hold signal, which is generated by the readout electronics
after it receives a beam trigger signal. The $128$ S\&H outputs from each ASIC are then  multiplexed into one output line.
The peaking times of the ASICs connected to the junction and  to the ohmic sides are set respectively to $1.5\units{\mu s}$ 
and $2 \units{\mu s}$, while the gains are adjusted to properly match the dynamic range of ADCs used for the digitization of the signals. 
A view of the detector installed in the test beam set-up is shown in Figure~\ref{fig:detector}. 
The red circle indicates a $70 \units{\mu m}$ thick kapton foil, which is placed in front of the silicon detector to protect 
it from the light and to minimize the amount of passive materials along the beam.

\begin{table}[!t]
\begin{center}
\begin{tabular}{ll}
\hline
\hline
Number of channels & $128$ \\
ENC at $1 \units{\mu s}$ of peaking time ($\units{e^{-} RMS}$) & $80 + 15C_{d} ($\units{pF}$)$ \\
Power consumption ($\units{mW}$) & $170$ \\
Shaper peaking time ($\units{\mu s}$) & $1-3$ \\
Dynamic range ($\units{MIPs}$) & $\pm 4$ \\
Current gain ($\units{\mu A/fC}$) & $25$ \\
\hline
\hline
\end{tabular}
\end{center}
\caption{Main technical parameters of the VA2 readout ASICs.}
\label{tab:va2}
\end{table}

The data acquisition system (DAQ) is controlled by a VME SBS Bit3 620 board~\cite{SBS}  
optically linked to a PC running Linux and includes a custom ADC board, a dedicated
VME readout/memory board and a custom trigger board~\cite{Bonfanti:2012bda}.
The ADC board is equipped with a 12-bit AD9220 ADC~\cite{ADC} for the 
signal digitization. 
 
The pedestals of each strip are  defined in dedicated pedestal runs. Noise level is defined on the basis of 
rms of the  pedestal distributions and it is about $1.39 \units{keV}$ for the strips on the junction side  
and of about $1.78\units{keV}$ for those on the ohmic side. Here the noise level is expressed in energy scale using 
a calibration procedure described in Section~\ref{sec:partclus}. Noisy strips are also identified in pedestal runs 
and are masked in the  data analysis. A total of $18$ noisy strips ($5$ on the junction side and $13$ on the ohmic side) 
have been masked. Nearly all masked strips are located at the edges of the detector.

\subsection{Beam test setup}
\label{sec:setup}

\begin{figure}[!t]
\begin{center}
\includegraphics[width=0.95\columnwidth]{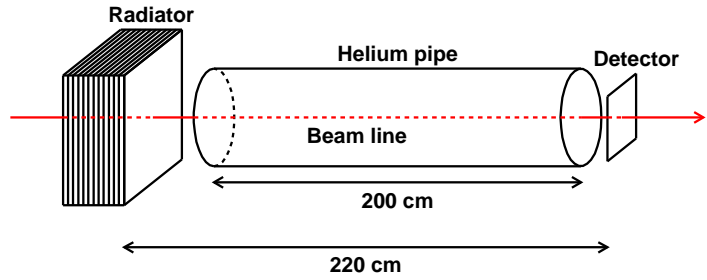}
\end{center}
\caption{Schematic view of the experimental setup.}
\label{fig:setup}
\end{figure}

Measurements were performed at the CERN-SPS H8 facility, using a mixed electron-pion beam with 
a momentum of $20 \units{GeV/c}$ and muon beams with momenta of $290 \units{GeV/c}$,
$180 \units{GeV/c}$ and $120 \units{GeV/c}$. Figure~\ref{fig:setup} shows a schematic view 
of the experimental setup. In order to separate the TR X-rays from the beam particles, 
the silicon detector was installed $220 \units{cm}$ downstream the radiator,
with the junction side facing to the radiator. A $200 \units{cm}$ long pipe filled with helium 
was placed between the radiator  and the detector to minimize the absorption of the X-rays along the path.
A trigger system, which separates particles of different types, included scintillator counters, 
an upstream Cherenkov counter, a preshower detector and a lead-glass calorimeter. A multiplicity counter 
was used as veto to remove upstream showers and multi-particle events. 

Periodic and irregular radiators made of different materials and with different properties were used in the tests. 
The main properties of the radiators are summarized in Table~\ref{tab:radiators}. 
Radiators were made as sets. Each set of radiator was assembled inside its own support frame.  
The radiators consisting of multiple sets are stacked in such a way that the distance between foils 
of adjacent sets is the same as within one set.

\begin{table}[!t]
\begin{center}
\begin{tabular}{lcccc}
\hline
\hline
 Radiator & Foil/gap material & $d_{1}$ & $d_{2}$ & $N_{f}$ \\
\hline
\hline
Mylar (1 set) & mylar/air & $50 \units{\mu m}$ & $2.97 \units{mm}$ & 30 \\
\hline
Mylar (3 sets) & mylar/air & $50 \units{\mu m}$ & $2.97 \units{mm}$ & 90 \\
\hline
Polyethylene (1 set) & polyethylene/air & $270 \units{\mu m}$ & $3.3 \units{mm}$ & 30 \\
\hline
Polyethylene (3 sets) & polyethylene/air & $270 \units{\mu m}$ & $3.3 \units{mm}$ & 90 \\
\hline
 Polypropylene (10 sets) & polypropylene/air & $15 \units{\mu m}$ & $210 \units{\mu m}$ & $360$ \\
\hline
 Fibers (1 set) & polypropylene/air & $12 \units{\mu m}$ & $180 \units{\mu m}$ & $520$ \\
\hline
\hline
\end{tabular}
\end{center}
\caption{Parameters of radiators used in the beam test: $d_{1}$ and $d_{2}$ are the thickness of the foils and of the gaps respectively; 
$N_{f}$ is the number of foils. For the fiber radiator average values calculated from the density and from individual fiber parameters are presented.}
\label{tab:radiators}
\end{table}

\section{Data analysis and results}

\subsection{Selection of particle clusters and energy calibration}
\label{sec:partclus}

A charged particle crossing a silicon micro-strip detector induces electrical  signals on several adjacent strips. 
The ADC charge associated to a cluster is the sum of the charges of individual strips in the cluster. 
A particle cluster is defined starting from a ``seed'', i.e. a strip with Signal-to-Noise level  $S/N > 20$, 
and associating to it all the adjacent strips with $S/N > 3$. In addition, $S/N > 30$ is required for the whole cluster. 
The cut on the seed  allows to efficiently select the strips hit by the beam particles without contaminations from noisy strips.

In our analysis we selected samples of events with only one particle cluster on both sides of the detector. 
This choice allows the selection of events with only one particle crossing the detector. It makes also possible to unambiguously 
determine the position of the particle on each side of the detector, as the center of gravity of  the corresponding particle cluster. 
To perform the energy calibration, the ADC count distributions of particle clusters are fitted with a Landau distribution 
folded with a Gaussian function. These distributions are compared with those  obtained using the simulation code 
described in Ref.~\cite{Brigida:2004ff}, following the prescriptions of Ref.~\cite{Bichsel:1988if}. The calculated most 
probable value of the ionization energy loss is $83.5\units{keV}$ for  $290 \units{GeV/c}$ muons in $300\units{\mu m}$ silicon.
The values of the ionization energy losses calculated for the other configurations (i.e. with $20\units{GeV/c}$ electrons 
and with $180\units{GeV/c}$ and $120\units{GeV/c}$ muons) are not significantly different.

\begin{figure}[!t]
\begin{center}
\includegraphics[width=0.47\columnwidth]{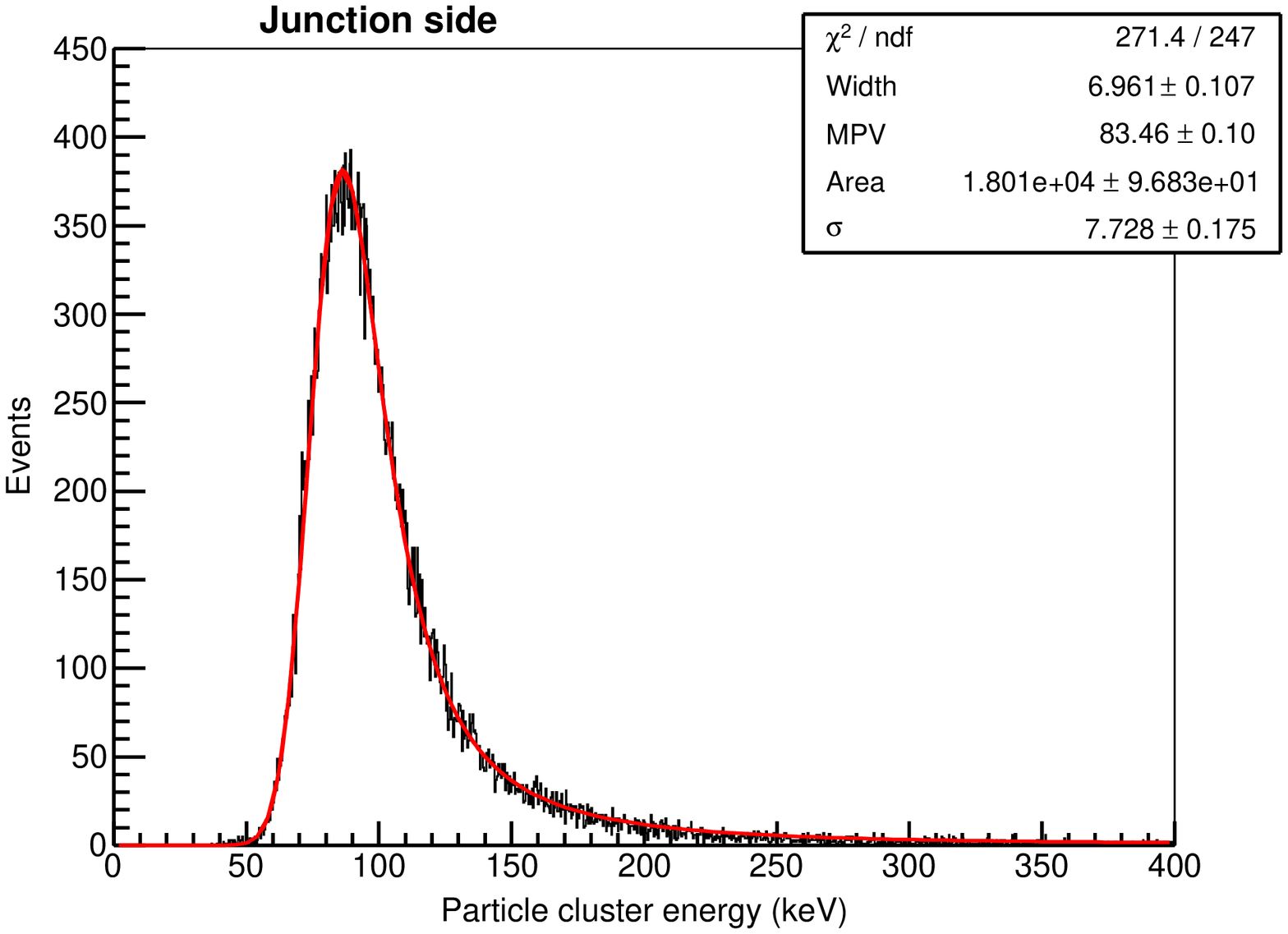}
\includegraphics[width=0.47\columnwidth]{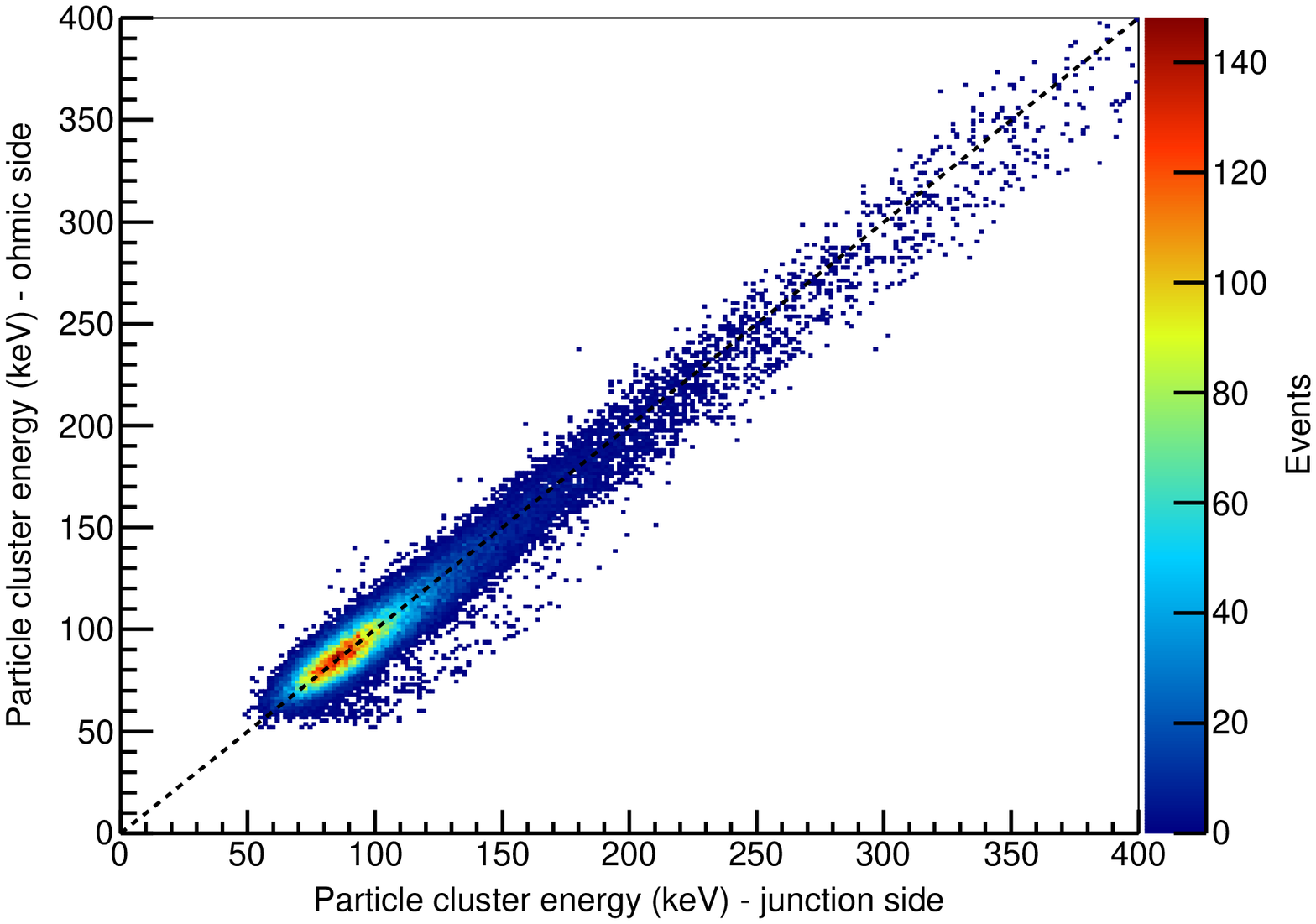}
\end{center}
\caption{Energy depositions associated to particle clusters in a run with $180\units{GeV/c}$ muons without radiators. The left plot shows
the energy distribution associated to particle clusters on the junction side. The distribution is fitted by a Landau distribution folded with a 
Gaussian function. The right plot shows the energy of the particle clusters on the ohmic side versus the energy of those on the junction side. 
The dashed black line indicates the points in which the two energies are equal. 
}
\label{fig:epart}
\end{figure}

In the left panel of Figure~\ref{fig:epart}, the energy distribution of particle clusters 
on the junction side is shown 
for  $180\units{GeV/c}$ muons without radiators. It is well fitted by a
Landau distribution folded with a Gaussian function, with a most probable value of
about $83.5\units{keV}$. The scatter plot on the right panel shows that 
the energies associated to the particle clusters on the ohmic side and on the junction 
side are clearly correlated. Since the noise level of the strips on the junction side is less than for those on the ohmic one 
($1.39 \units{keV}$ against $1.78 \units{keV}$), the particle energy deposition associated to the cluster 
on the junction side will be considered in the analysis.

\subsection{Selection of X-ray clusters}
\label{sec:xrayclus}

An X-ray cluster is defined in a way similar to a particle cluster. As TR energy is less than the ionization loss of particle, 
the search for an X-ray cluster starts from a seed strip with $S/N > 4$. Adjacent strips with $S/N > 1$ are associated to the seed strip. 
This  approach allows to obtain average energy thresholds for X-ray clusters of $5.6 \units{keV}$ on the junction side and 
of $7.1 \units{keV}$ on the ohmic side. For each X-ray cluster on the two sides, a separation of at least one strip from  
the particle cluster and from other X-ray clusters is also required. To study the TR properties, events with only one X-ray cluster 
on each of the detector sides were selected. As in the case of the particle clusters, this selection allows an unambiguous matching of 
the X-ray clusters on the two sides of the detector. The horizontal and vertical coordinates of each X-ray are defined as 
the centers of gravity of the strips in the X-ray clusters on the two detector sides, while the energy is taken from the junction side. 

\begin{figure}[!t]
\begin{center}
\includegraphics[width=0.85\columnwidth]{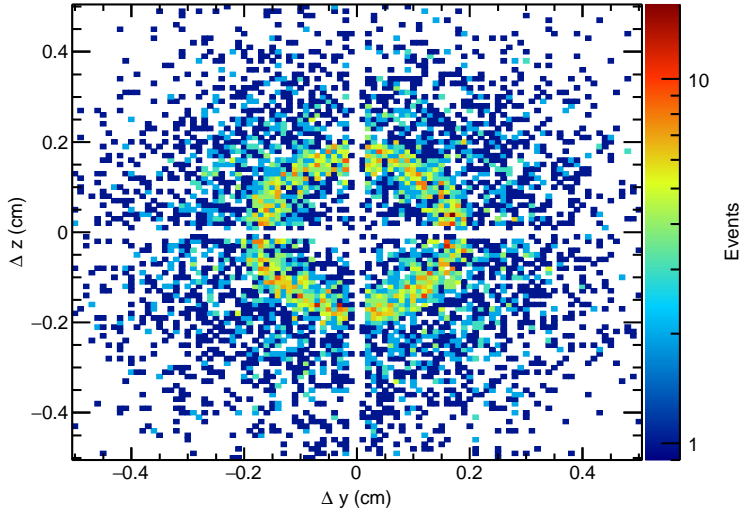}
\end{center}
\caption{Relative positions of the X-ray clusters with respect to the particle
clusters for $290\units{GeV/c}$ muons crossing the 360 foil regular polypropylene radiator (see Table~\ref{tab:radiators}). 
The $y$ and $z$ coordinates refer respectively to strip positions on the junction side 
and on the ohmic side of the detector.}
\label{fig:rings}
\end{figure}

Figure~\ref{fig:rings} shows the relative positions of X-rays with respect to the radiating particles in a run 
with $20\units{GeV/c}$ electrons crossing the polypropylene radiator. 
As expected, the figure shows that the TR X-rays are distributed in rings centered on the radiating particle. 
Due to the event selection implemented in the present analysis, 
the regions $|\Delta y | < 50 \units{\mu m}$ and $|\Delta z | < 50\units{\mu m}$ 
in the plot are not populated (here  $y$ and $z$ are the coordinates in the junction side and in the ohmic side respectively). 

\subsection{Results}

As discussed in Section~\ref{sec:detector}, the measurements were performed using beams of 
$20\units{GeV/c}$ electrons and of muons with momenta of $120$, $180$ and
$290 \units{GeV/c}$.  These beams cover a wide Lorentz factor range, from $\gamma \approx 1.1 \times 10^{3}$ 
to $\gamma \approx 3.9 \times 10^{4}$. 

\begin{figure}[!tbp]
\begin{center}
\includegraphics[width=0.4\columnwidth]{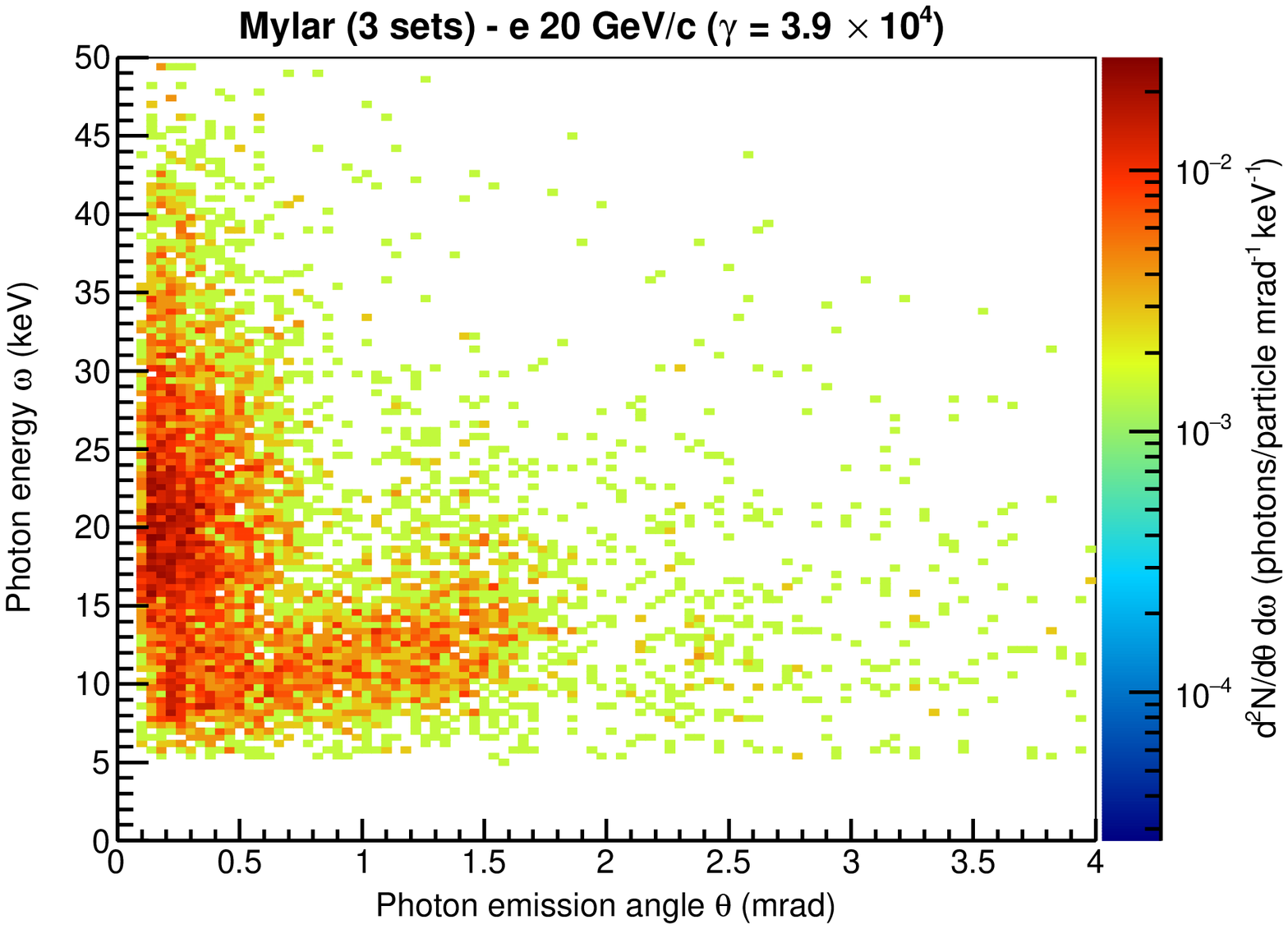}
\includegraphics[width=0.4\columnwidth]{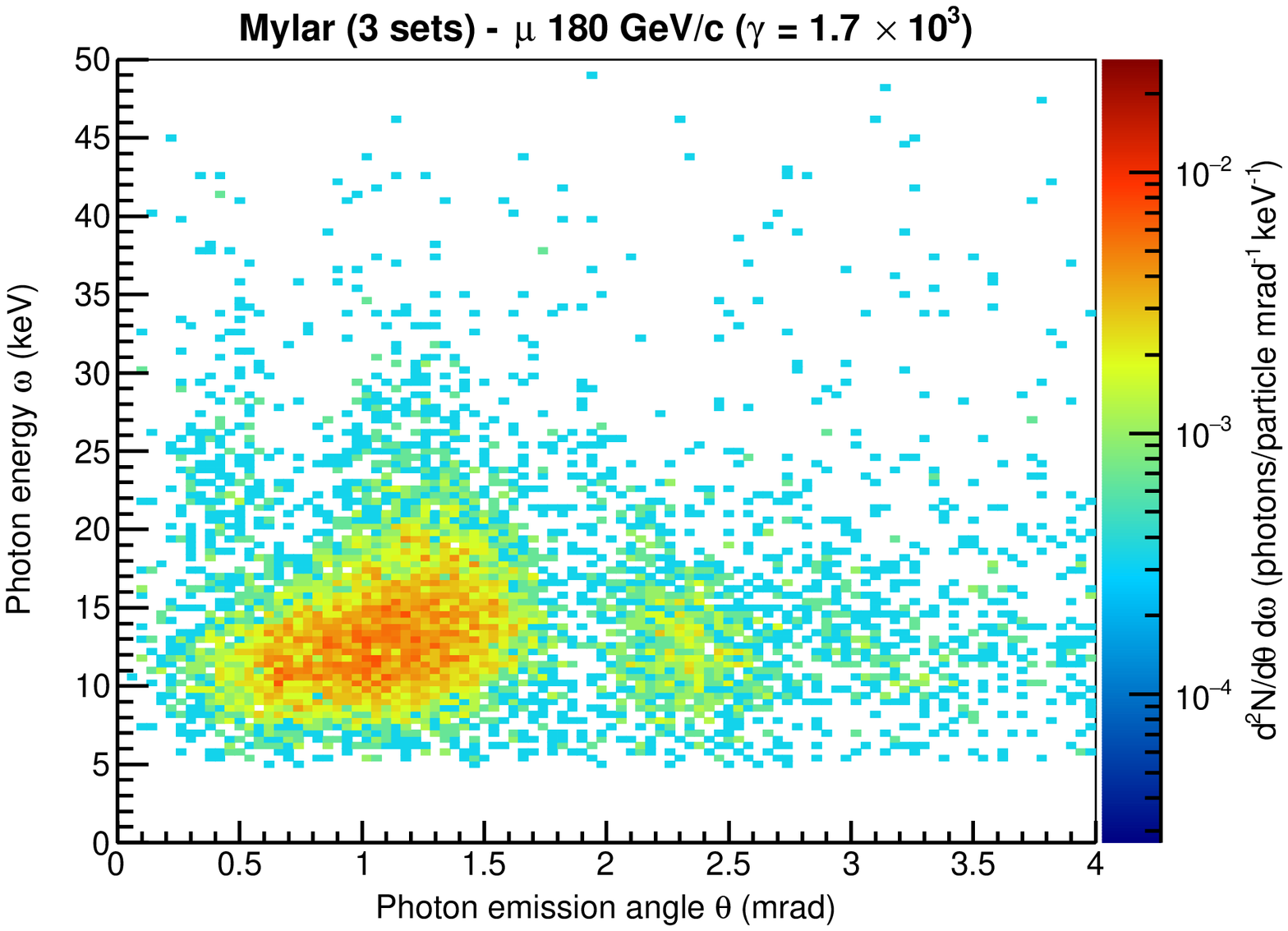}
\includegraphics[width=0.4\columnwidth]{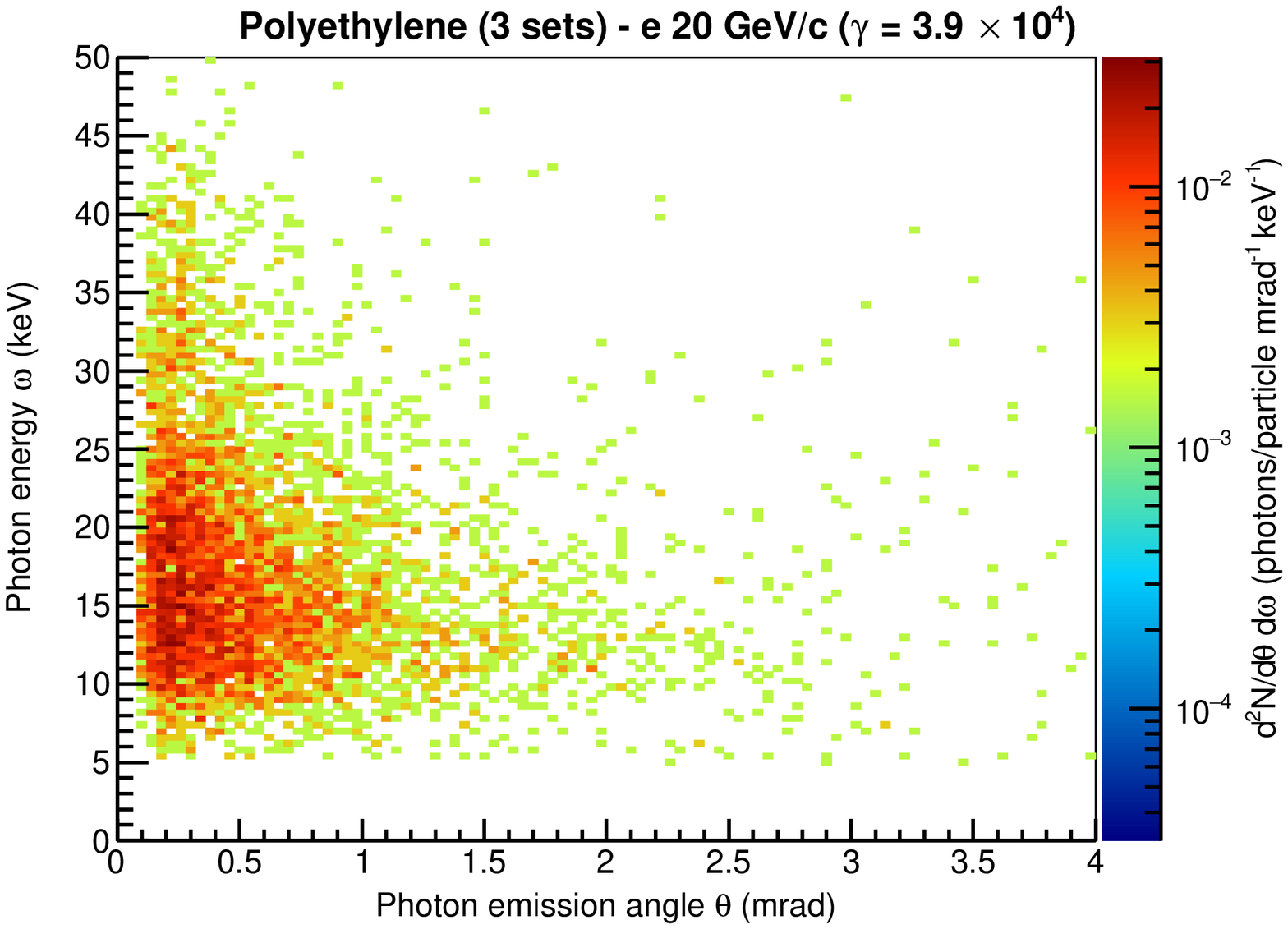}
\includegraphics[width=0.4\columnwidth]{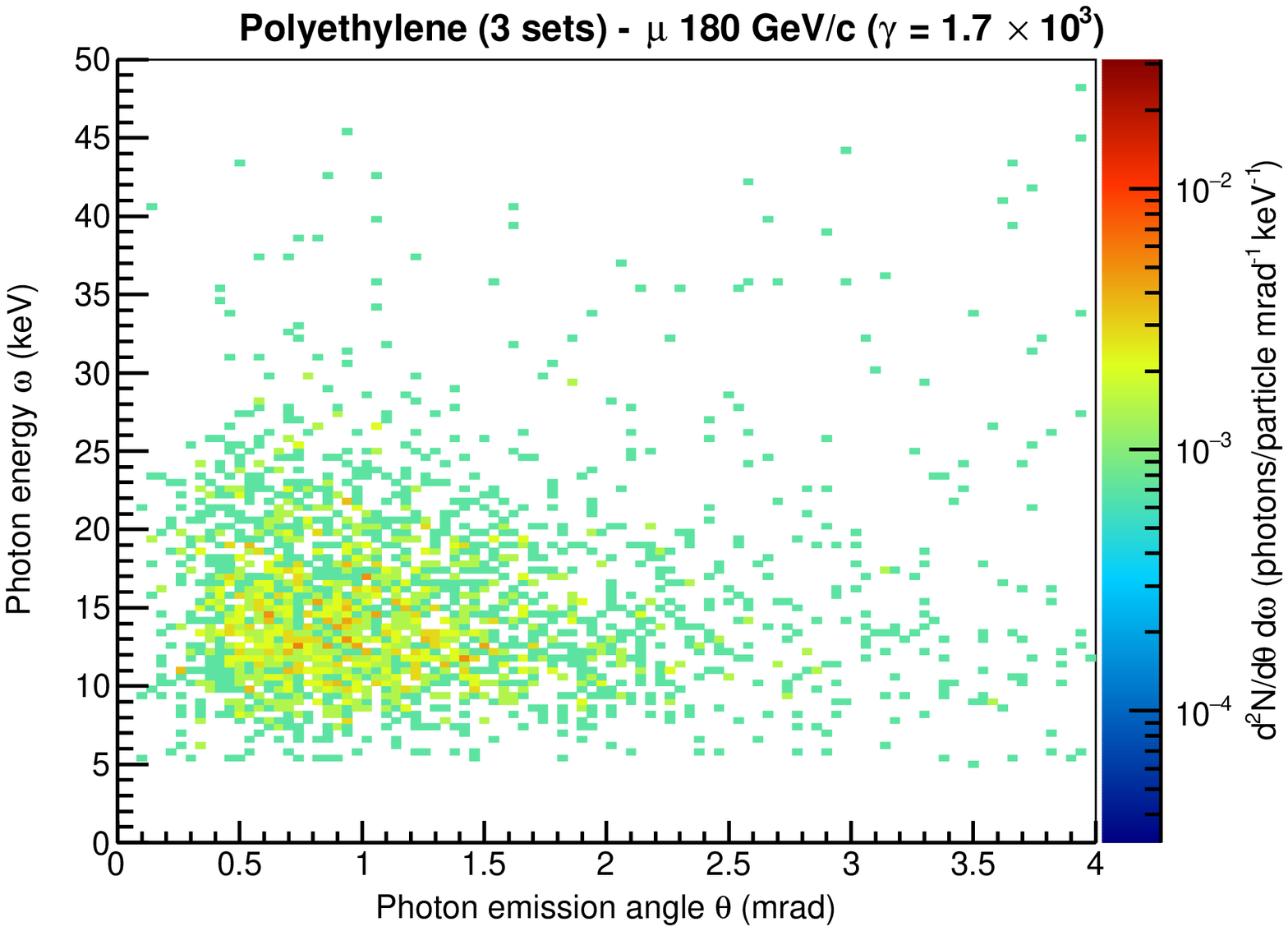}
\includegraphics[width=0.4\columnwidth]{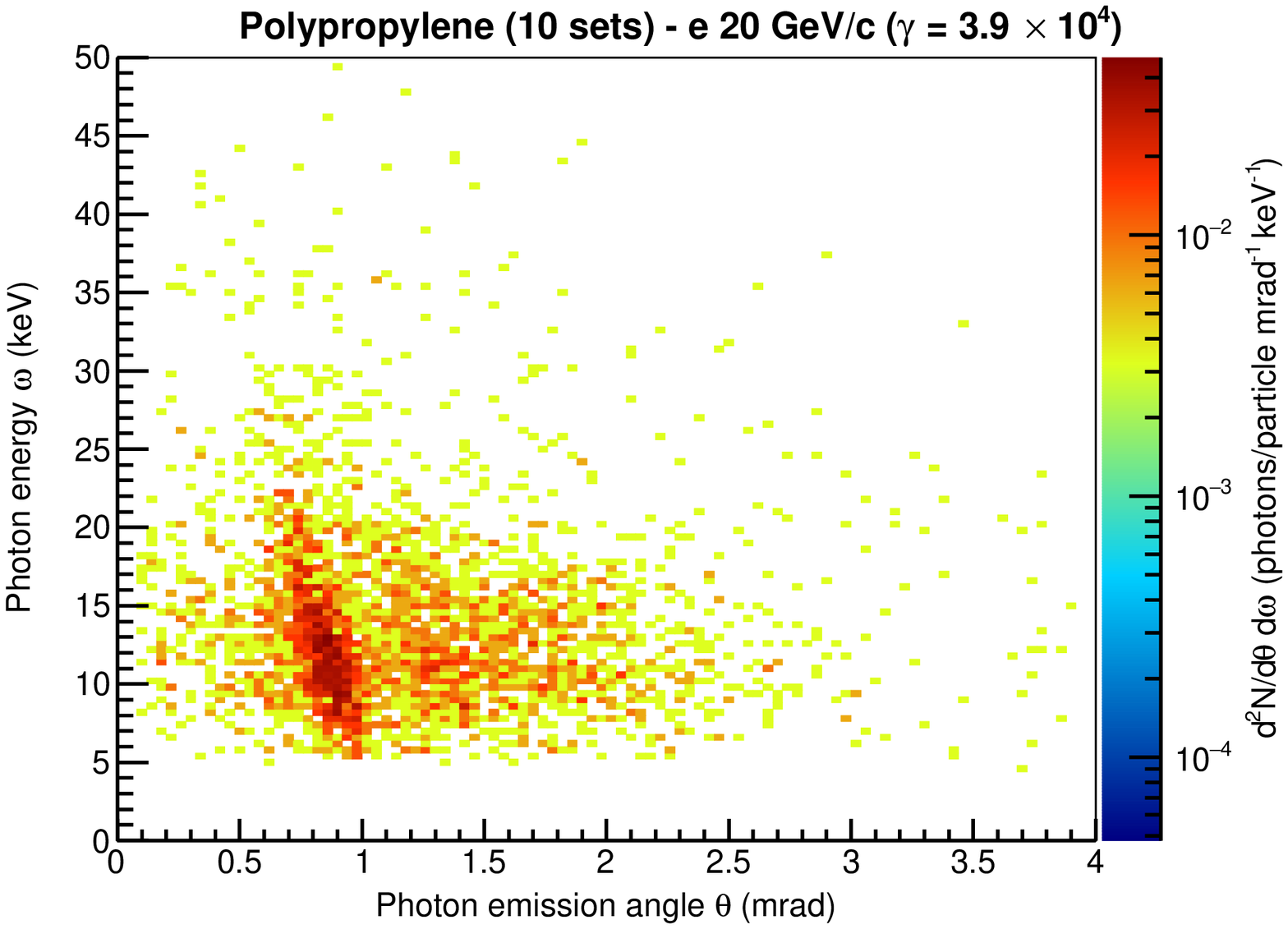}
\includegraphics[width=0.4\columnwidth]{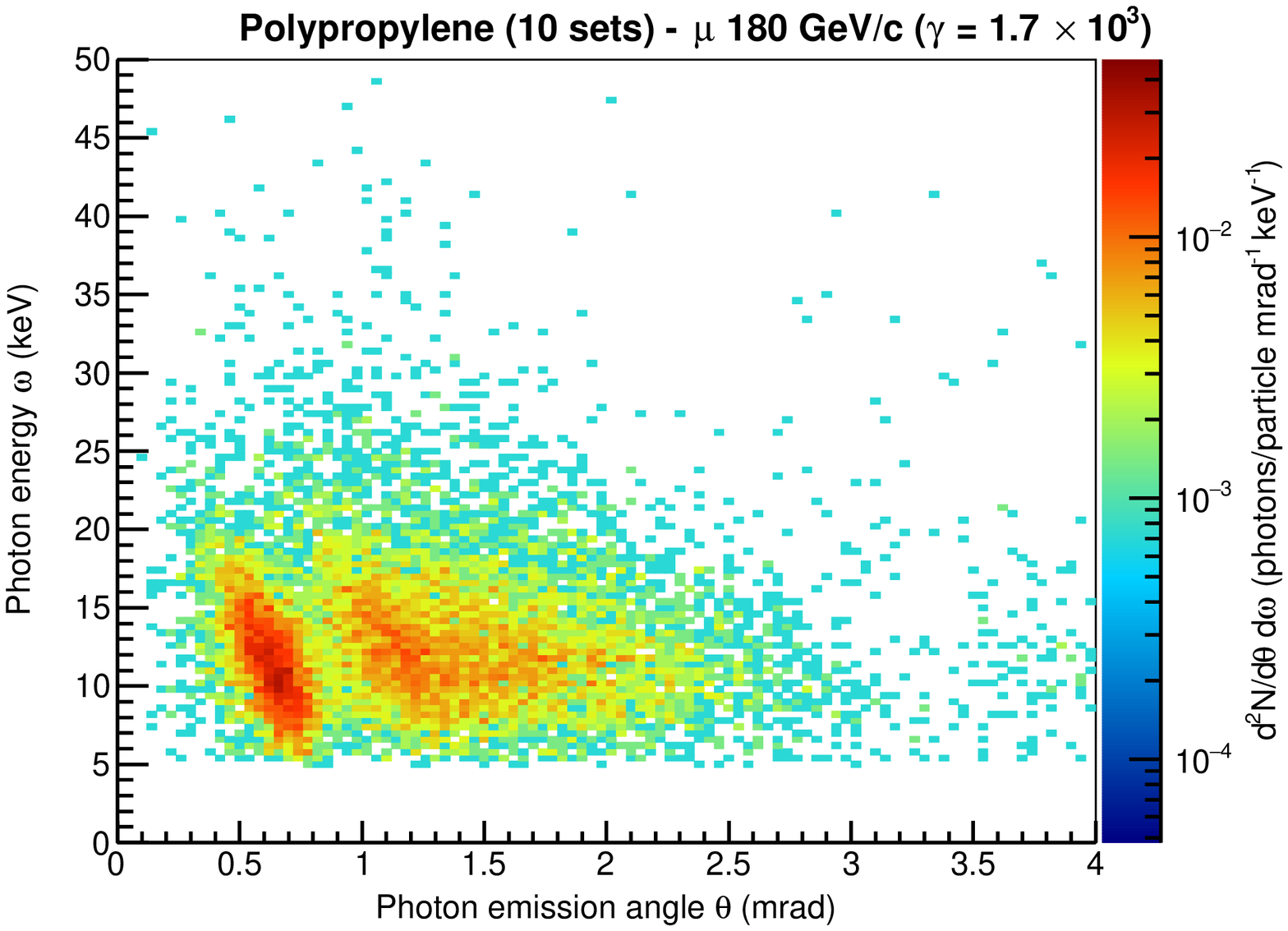}
\includegraphics[width=0.4\columnwidth]{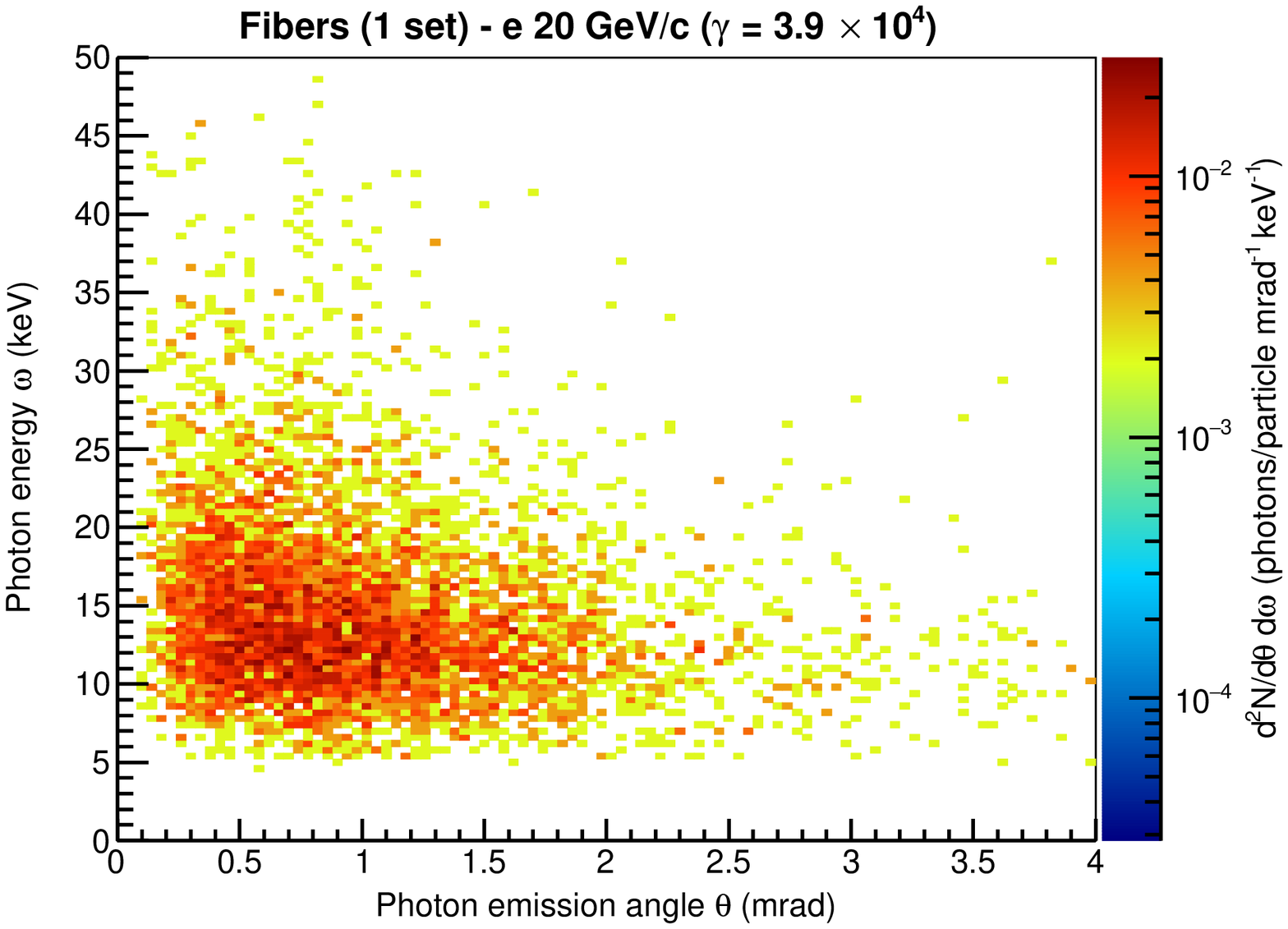}
\includegraphics[width=0.4\columnwidth]{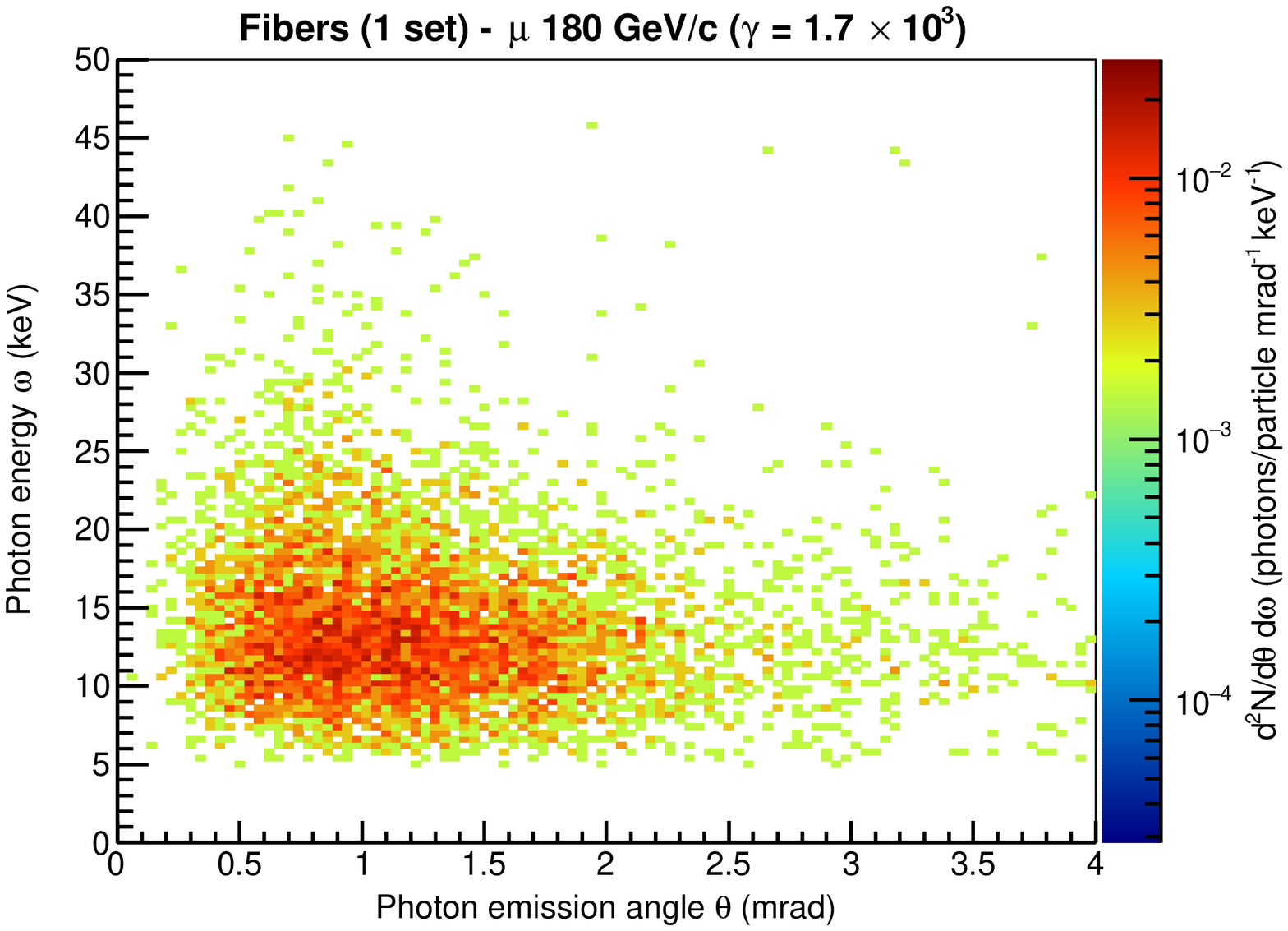}
\end{center}
\caption{Measured two-dimensional distributions of  energy (Y-axis) vs angle (X-axis) of the TR X-rays produced 
by $20 \units{GeV/c}$ electrons (left column) and $180 \units{GeV/c}$ muons (right column) crossing four different radiators.}
\label{fig:evsang}
\end{figure}

Two-dimensional distributions of  energy (Y-axis) vs  angle (X-axis) of the TR X-rays produced 
by $20 \units{GeV/c}$  electrons (left column) and $180 \units{GeV/c}$ muons (right column) 
crossing mylar, polyethylene, polypropylene and fiber radiators (from top to bottom) are shown 
in Figure~\ref{fig:evsang}.  Here and later, the distributions are normalized to the total number 
of incident particles, i.e. to the number of events with only one particle 
cluster on both detector sides. The TR emission angle is evaluated from the distance of the absorbed 
X-ray from the particle and from the known distance between the center of radiator and the detector plane.

These distributions clearly show the main features of the TR production. 
First of all, comparing  the TR spectra from the mylar radiator for $20 \units{GeV}$ electrons with those 
for $180 \units{GeV}$ muons (top row, left and right plots) one can distinguish two TR energy ranges with different 
angular and Lorentz factor dependencies. TR photons with energies $\omega <15 \units{keV}$ exhibit wide angular distributions, 
which do not follow the commonly quoted law $\theta \approx 1/\gamma$. The width of these distributions is practically 
the same for electrons with $\gamma \approx 3.9 \times 10^{4}$ and for muons with $\gamma \approx 1.7 \times 10^{3}$. 
On the other hand, high-energy TR photons ($\omega > 15 \units{keV}$) exhibit a much narrower angular distribution for 
electrons, while their yield is almost suppressed in the case of muons.  
Basic equations for TR generation (see for instance ~\cite{Artru:1975xp,Cherry:1974de,Cherry:1978xk}) 
predict significant interference effects between the photons  emitted at the various detector interfaces. 
The global structure of distributions shown in these plots is defined by the properties of the single foil. 
Multi-foil interference does not change the energy distribution of the TR photons, but may significantly change their 
angular distribution. This is clearly seen from the distributions of TR from the polypropylene radiator, 
which has $15 \units{\mu m}$ foils spaced by $210 \units{\mu m}$ gaps (third row from the top). 
Instead of an expected concentration of TR at angles $<0.1 \units{mrad}$ for electrons, the most intensive TR peak 
is found at an angle of $\sim 0.8\units{mrad}$. Contrary to simple considerations, 
this peak moves to smaller angles $\sim 0.6 \units{mrad}$ for muons with a lower gamma factor. 
For irregular radiators, such as the fiber radiator, which has no periodic structure, these patterns are smeared 
out due to the fluctuations of the material and gap thicknesses (see for instance ref.~\cite{Garibian:1975xx}).

\begin{figure}[!t]
\begin{center}
\includegraphics[width=0.39\columnwidth]{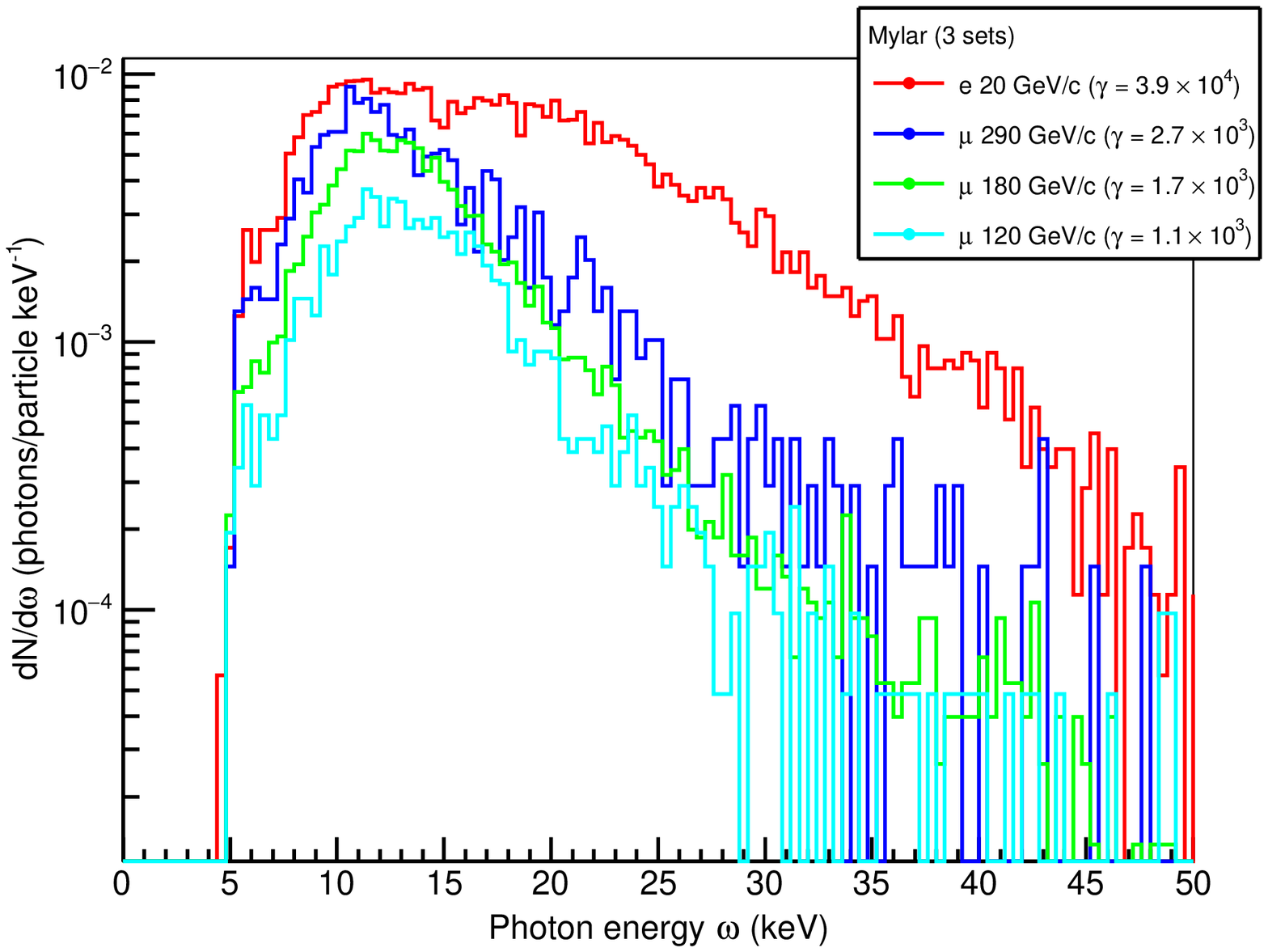}
\includegraphics[width=0.39\columnwidth]{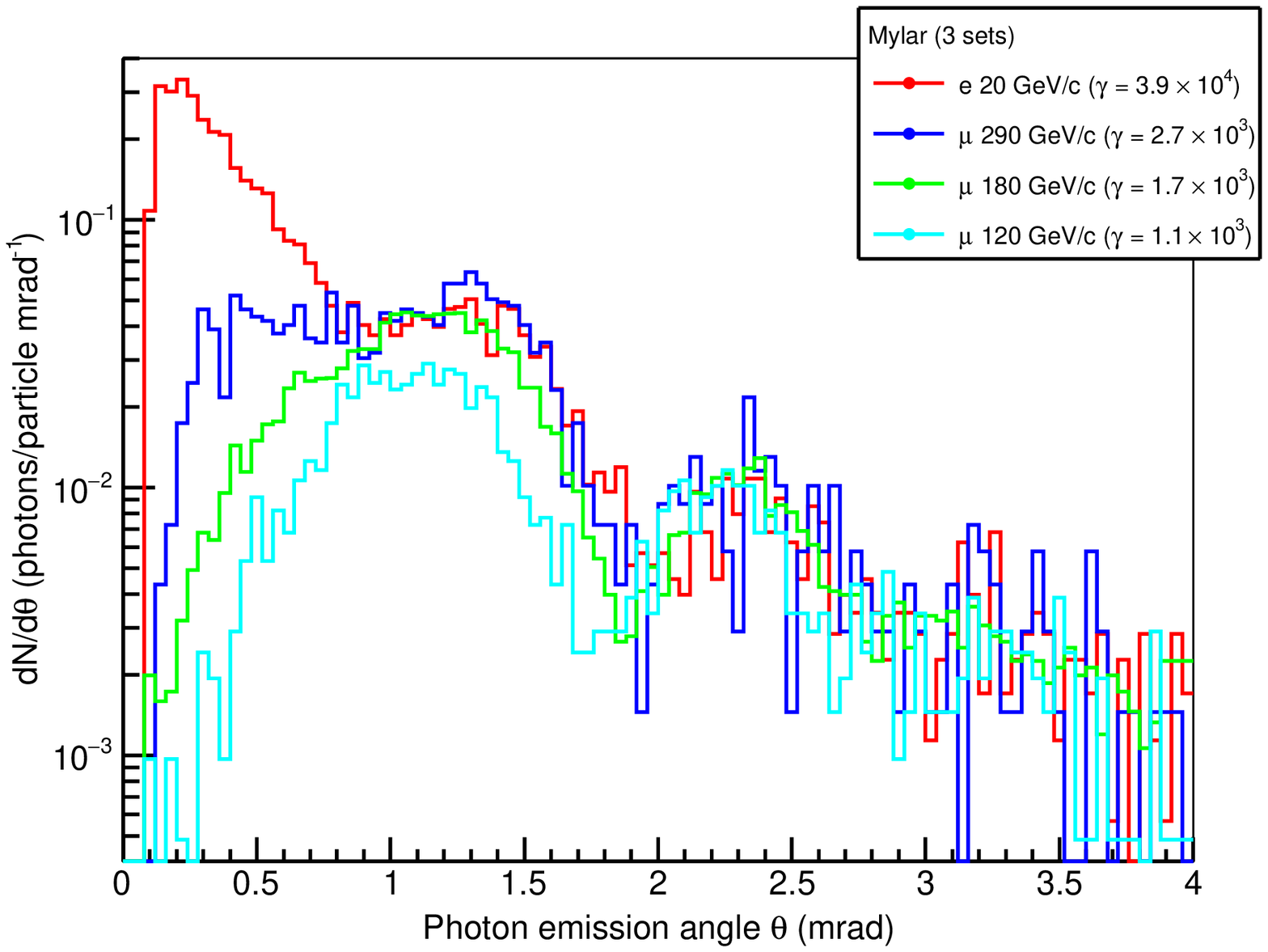}
\includegraphics[width=0.39\columnwidth]{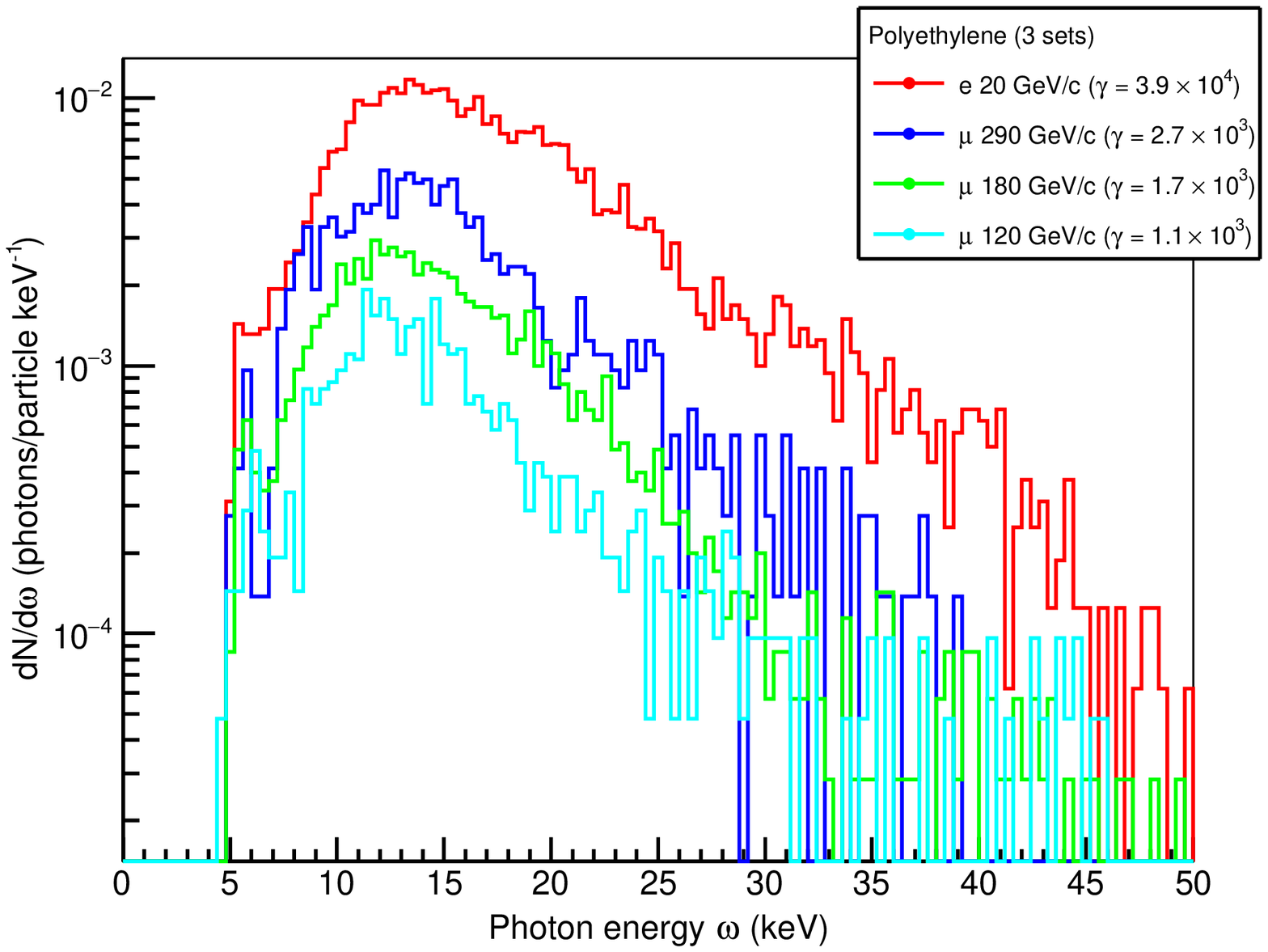}
\includegraphics[width=0.39\columnwidth]{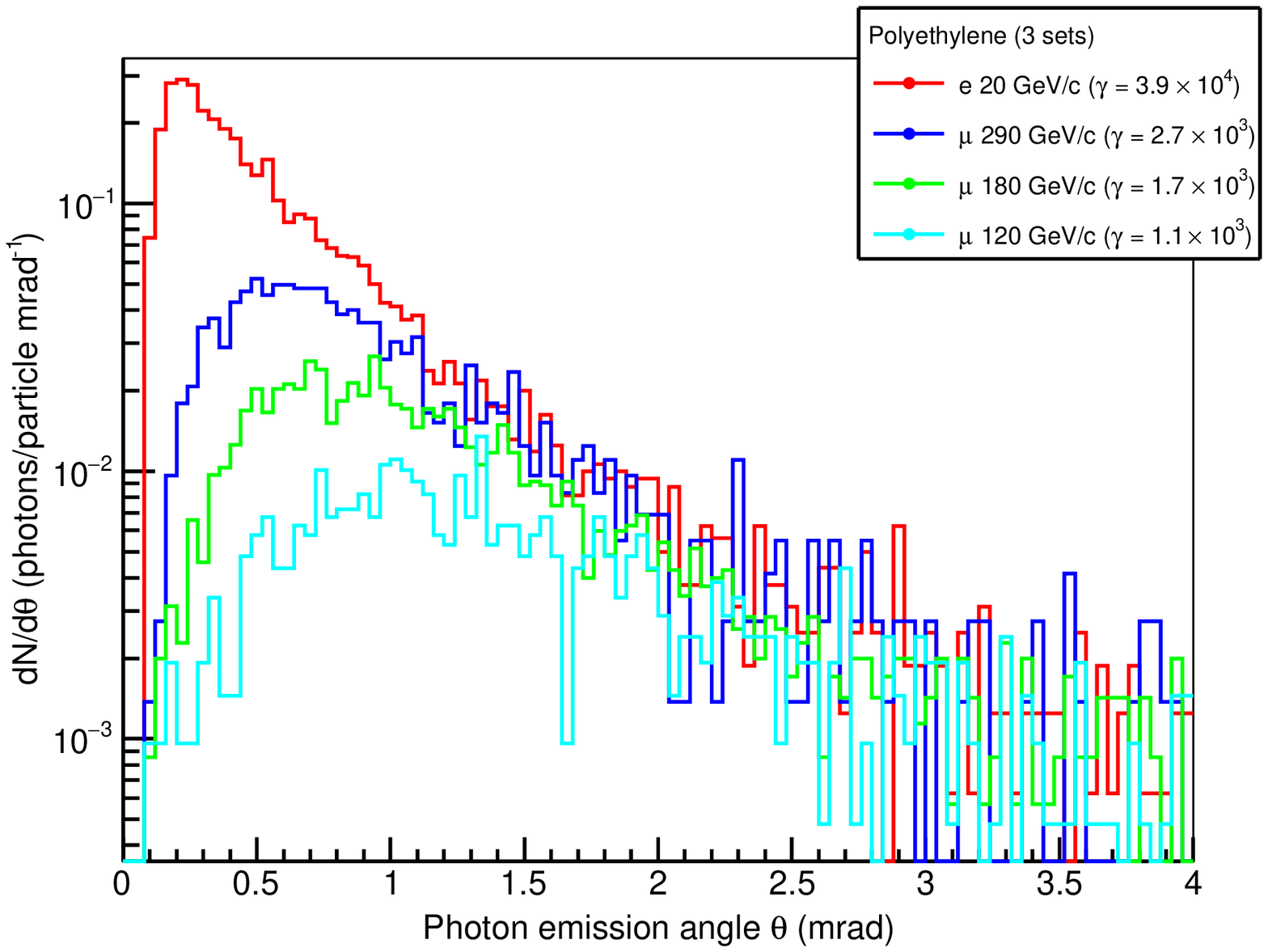}
\includegraphics[width=0.39\columnwidth]{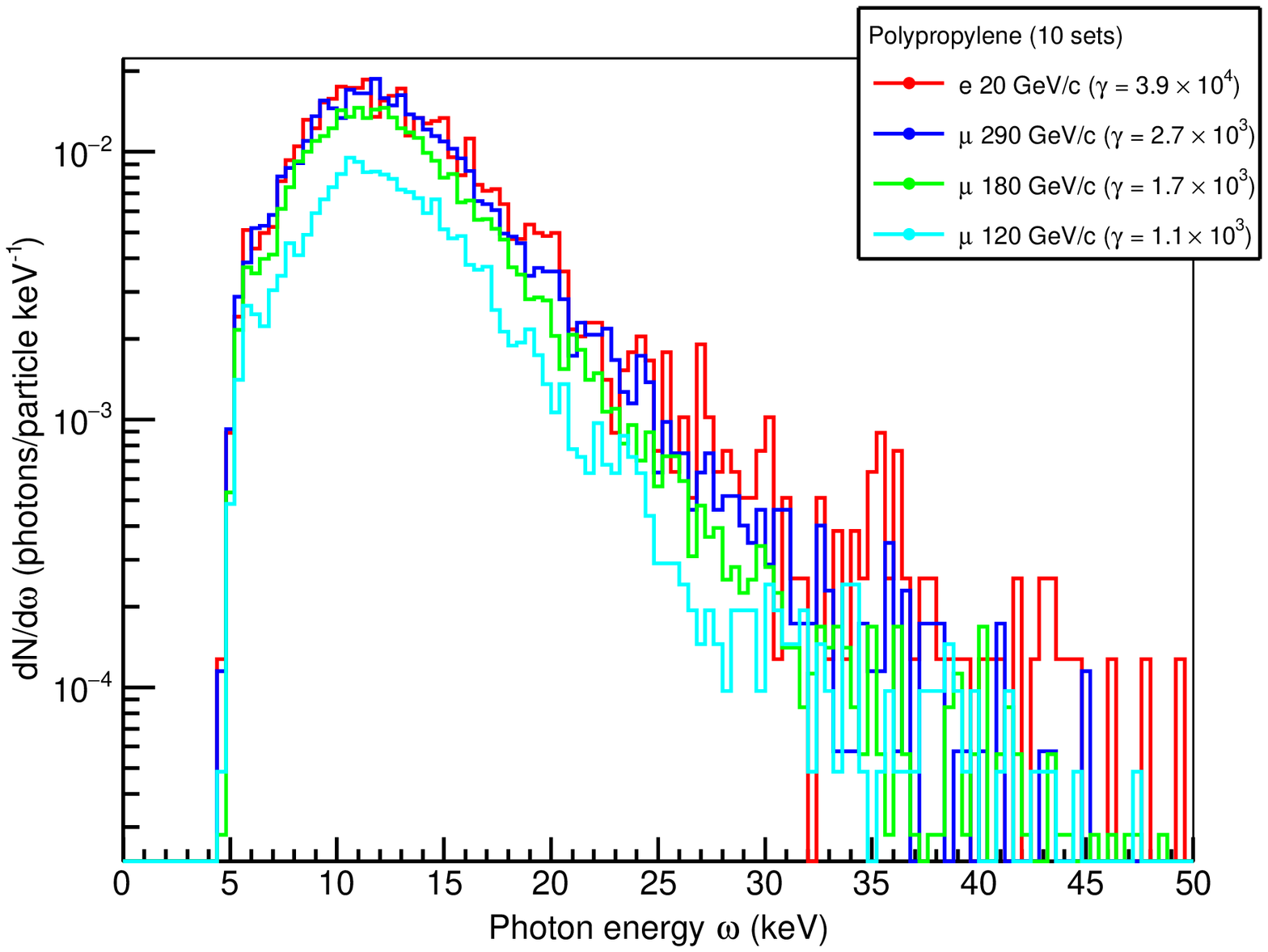}
\includegraphics[width=0.39\columnwidth]{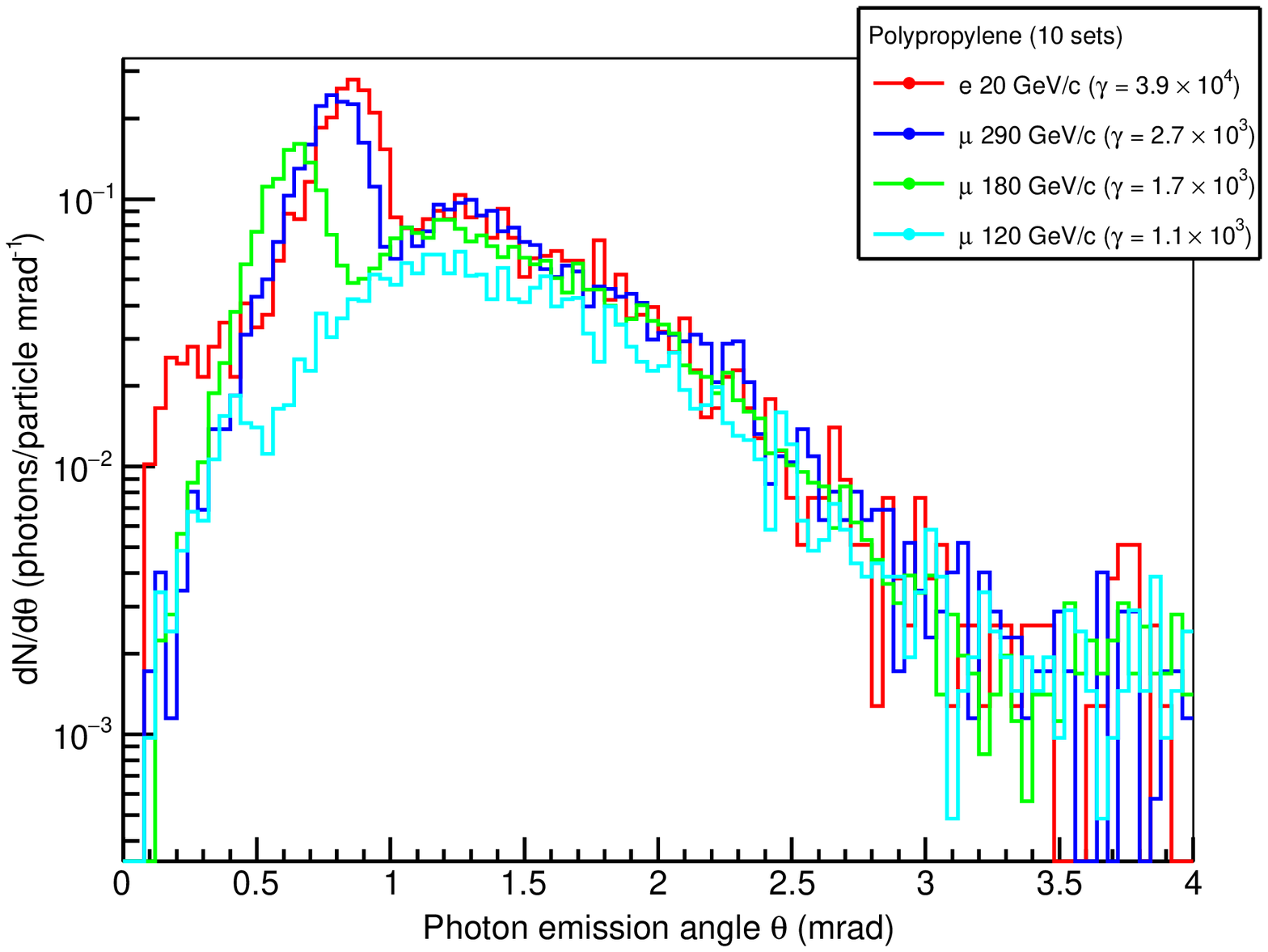}
\includegraphics[width=0.39\columnwidth]{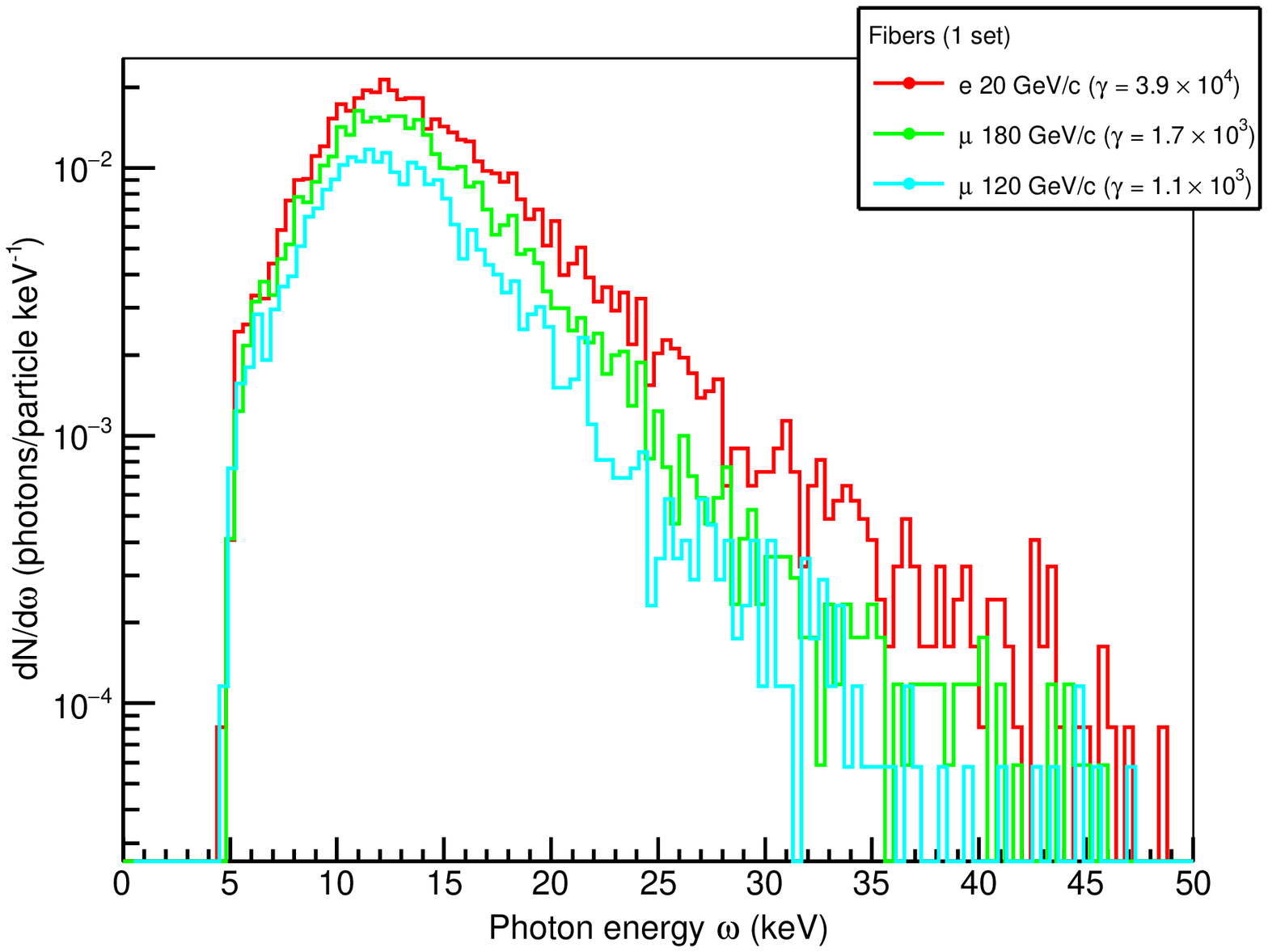}
\includegraphics[width=0.39\columnwidth]{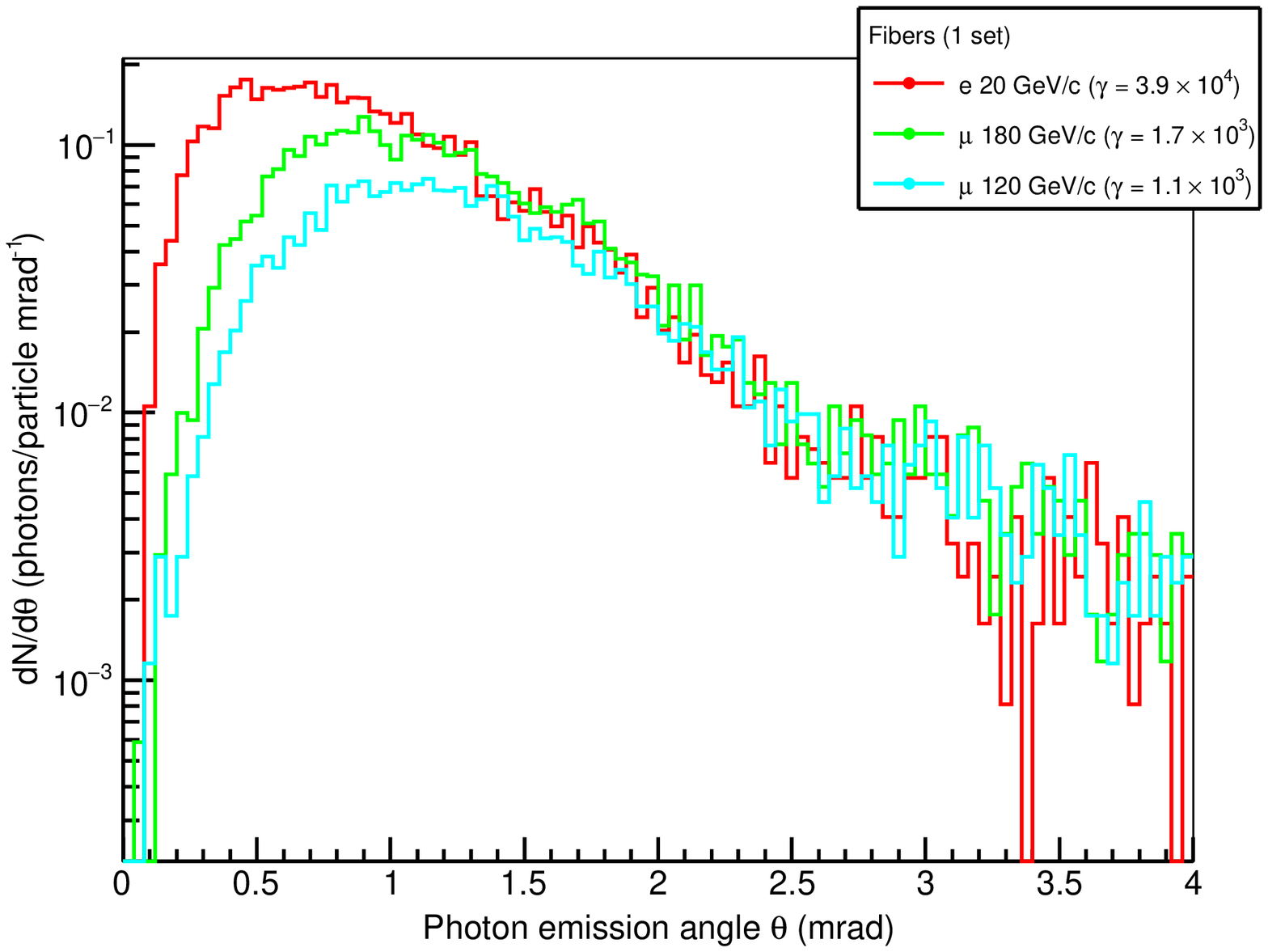}
\end{center}
\caption{Measured differential energy (left column) and  angular (right column) spectra of the TR X-rays 
produced by $20 \units{GeV/c}$ electrons and $290$, $180$ and $120 \units{GeV/c}$ muons crossing four different 
radiators (mylar, polyethylene, polypropylene and fibers).}
\label{fig:diffspectra}
\end{figure}

Projections of two-dimensional distributions on the energy axis (Y) and on the angle axis (X) 
are shown respectively in the left and in the right column of Figure~\ref{fig:diffspectra}. 
The spectra for different beam configurations are overlapped on the same plots. Results for different 
radiators are presented in the same order as in Figure~\ref{fig:evsang}. Data for the mylar as well as for 
the polyethylene radiators clearly show that high-energy photons irradiated at small angles have a much higher 
TR production threshold than those of lower energies ($< 15 \units{keV}$). The angular distribution of 
low-energy photons is practically the same for particles with Lorentz factors from $\gamma \approx 1.1 \times 10^{3}$ 
to $\gamma \approx 3.9 \times 10^{4}$. 

\subsection{Background evaluation}
\label{sec:dummy}

\begin{figure}[!t]
\begin{center}
\includegraphics[width=0.39\columnwidth]{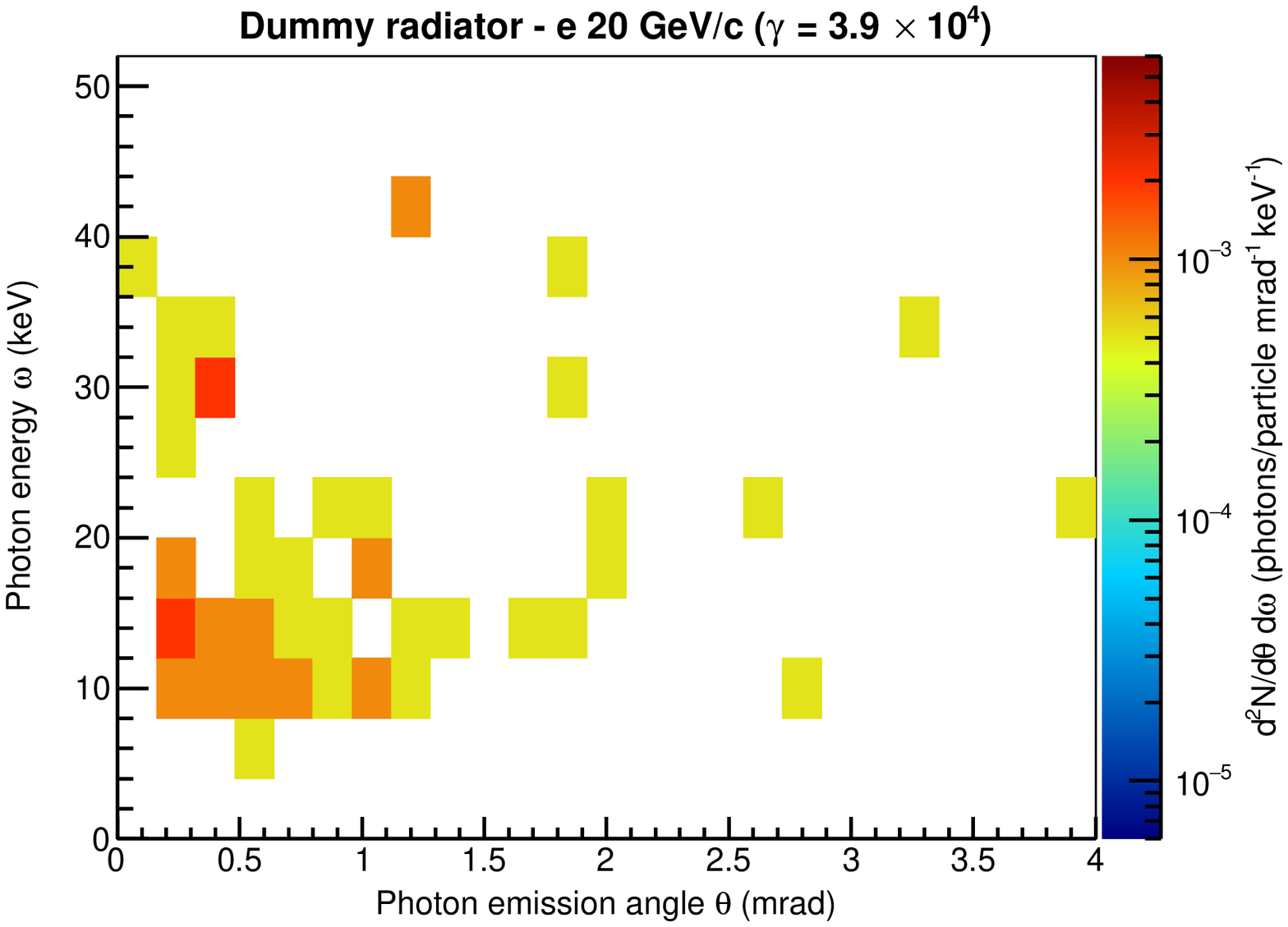}
\includegraphics[width=0.39\columnwidth]{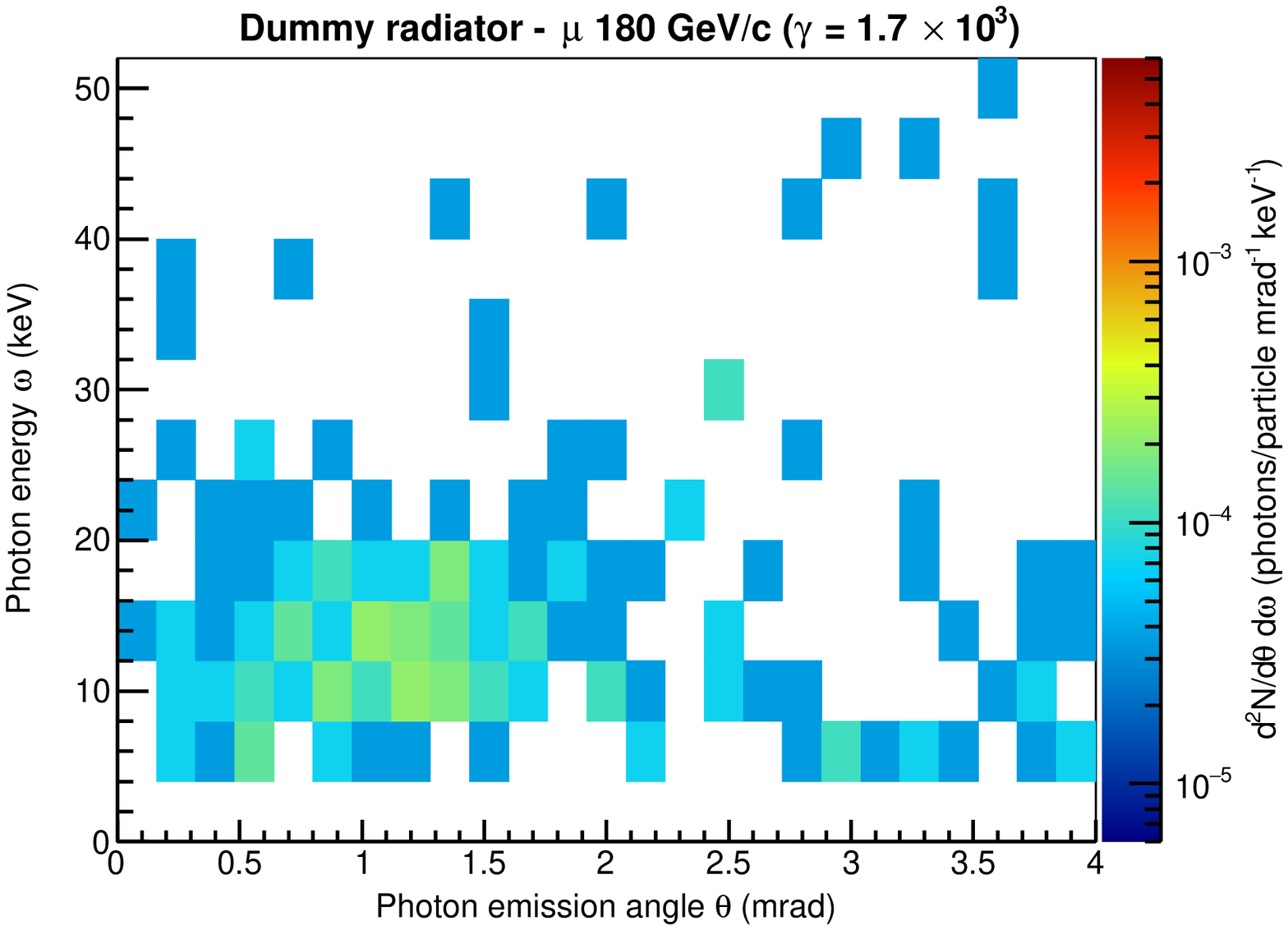}
\includegraphics[width=0.39\columnwidth]{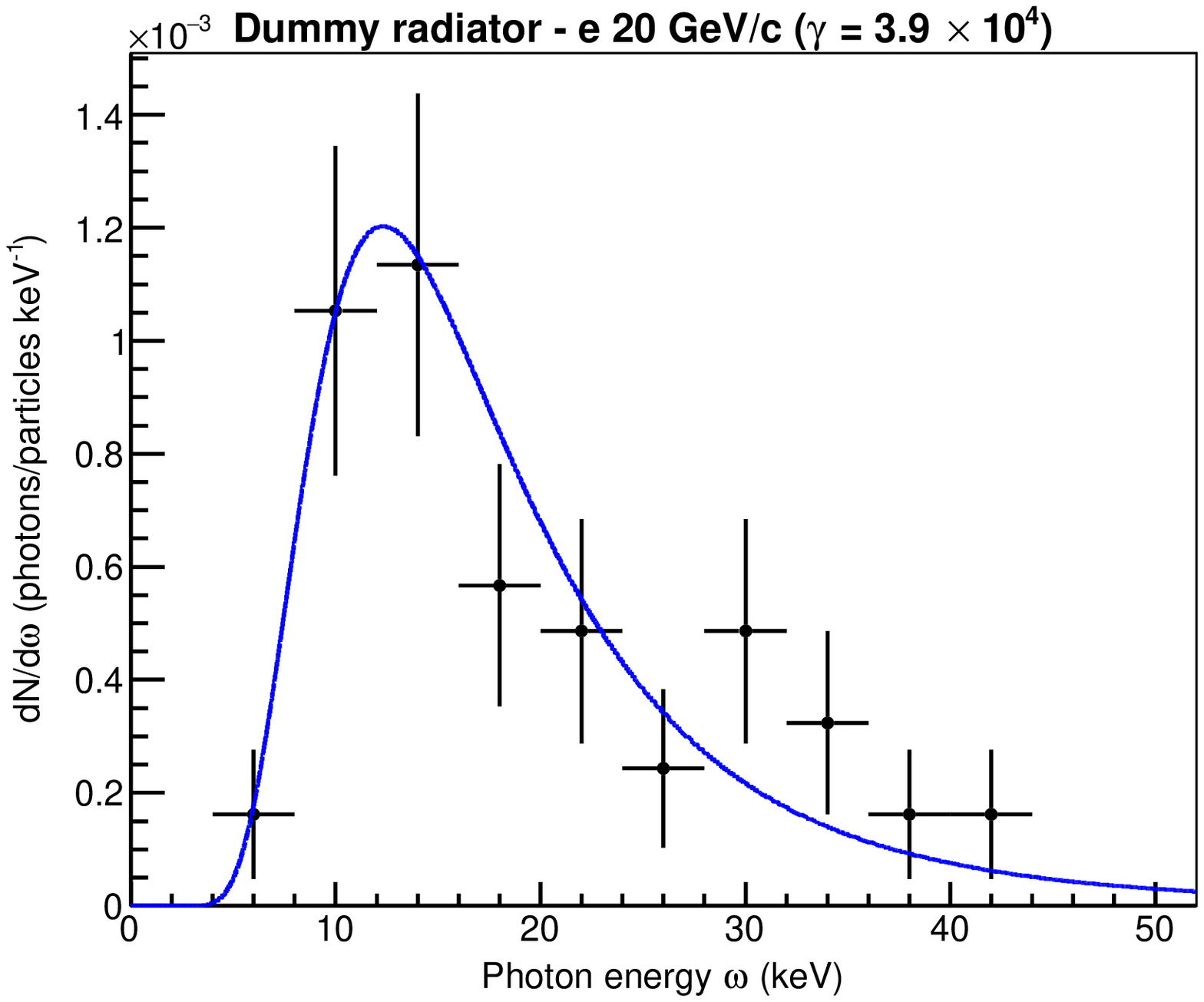}
\includegraphics[width=0.39\columnwidth]{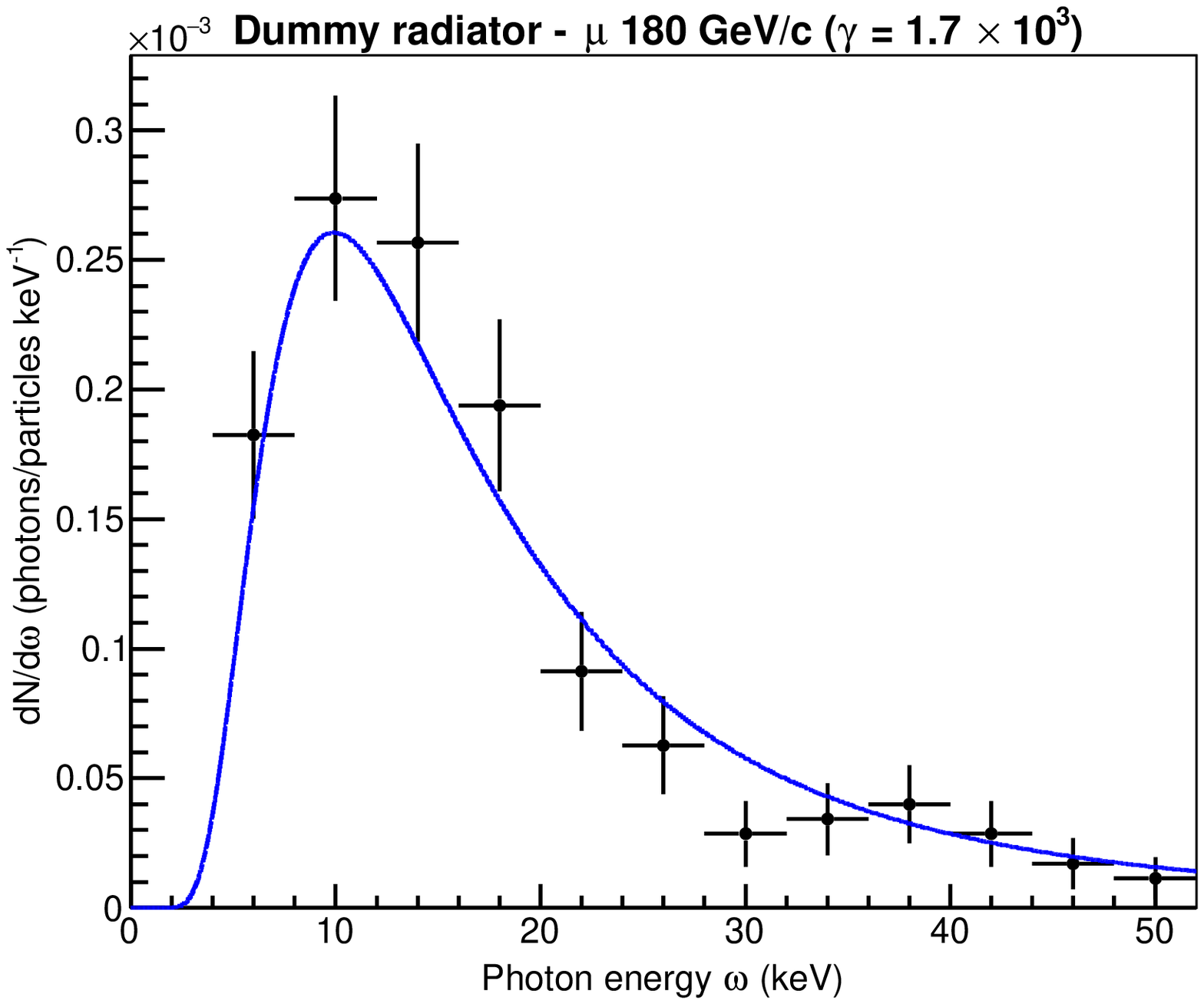}
\includegraphics[width=0.39\columnwidth]{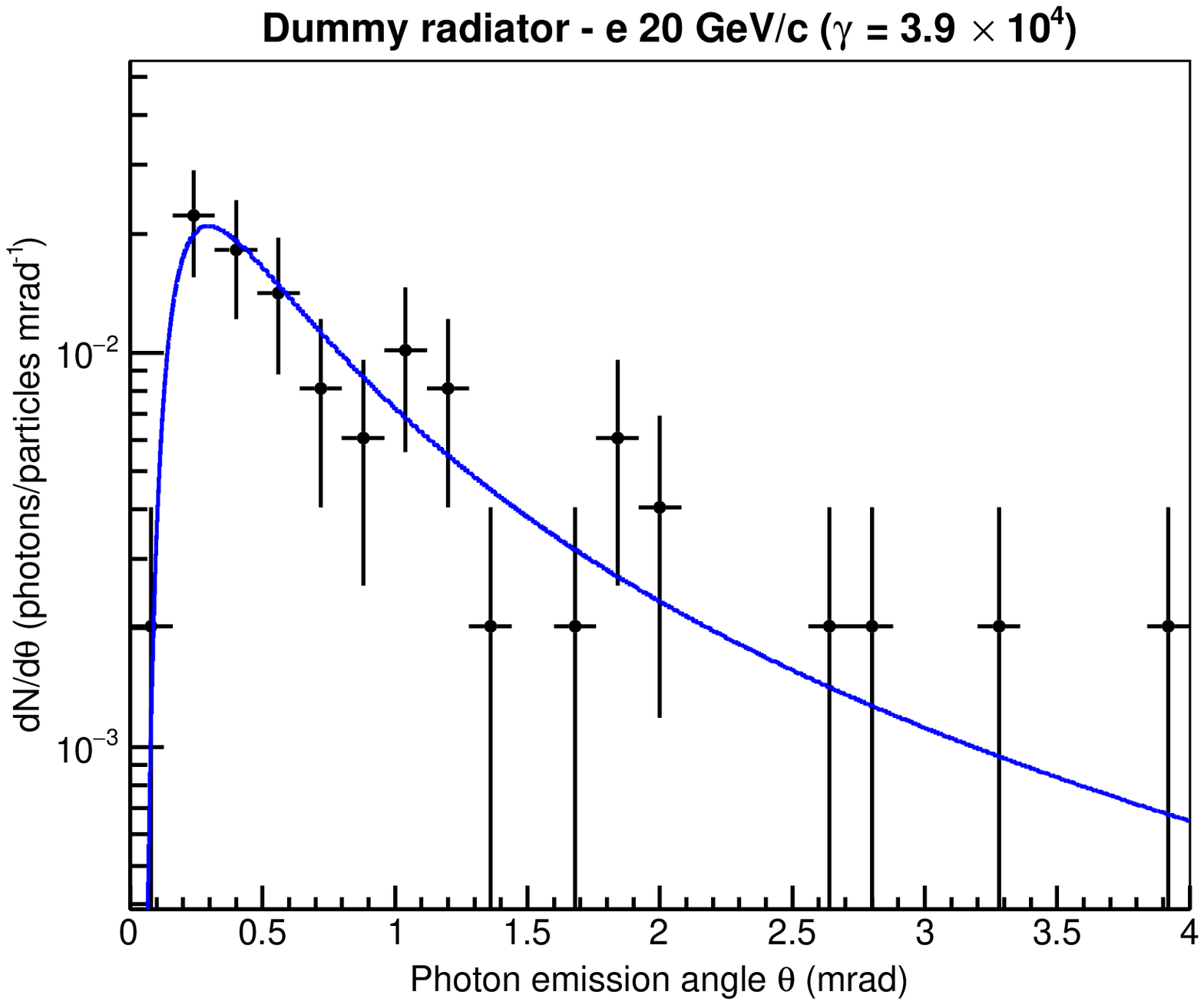}
\includegraphics[width=0.39\columnwidth]{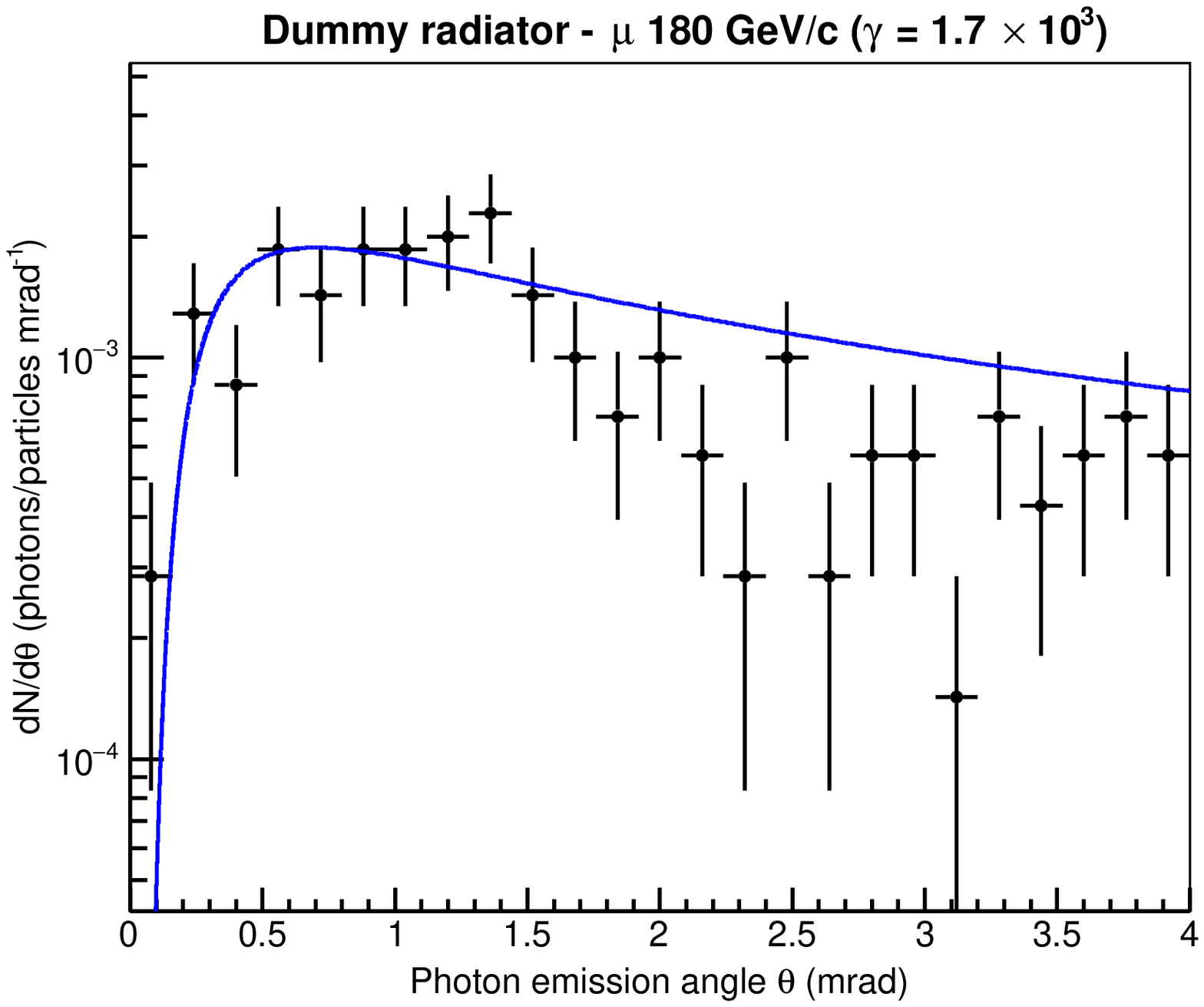}
\end{center}
\caption{Measured spectra of the X-rays produced by $20 \units{GeV/c}$ electrons (left column) and $180 \units{GeV/c}$ muons 
(right column) crossing the dummy radiator. The plots in the top panels show the two-dimentional distributions (energy vs angle); 
the plots in the middle panels show the differential energy spectra; the plots in the bottom panel show the differential angular spectra. 
The blue curves superimposed to the data points are the results of the fit with the template model.}
\label{fig:dummy}
\end{figure}

In order to make accurate comparison of the experimental results with simulations, 
one has to exclude possible background effects caused by non-TR X-rays or delta-rays produced by the 
beam particles interacting with the materials upstream the detector or in the radiator itself. 
Special runs were taken with a ``dummy'' radiator, e.g. a radiator consisting of a polyethylene slab with 
the same thickness (in radiation length units) as the 3-sets mylar radiator. 
Figure~\ref{fig:dummy} shows the X-ray spectra obtained in a run with $20\units{GeV/c}$ electrons and 
in a run with $180\units{GeV/c}$ muons crossing the dummy radiator. 
In the electron run the average number of hits which can be associated to photons is about $0.02$ per particle, 
while in the muon runs it is less than $0.005$. This difference is expected, since the probability to produce 
bremsstrahlung X-rays is higher for electrons than for muons.

The statistics of X-rays obtained in the data samples with dummy radiator is quite poor with respect to the statistics 
in the runs with radiators. This  makes impossible a direct bin-by-bin subtraction of the background. In order to subtract properly 
the background  from data, the X-ray spectra measured with dummy radiator were fitted with a phenomenological template model:
\begin{equation}
 \cfrac{d^{2}N_{bkg}}{d\theta d\omega} = k \theta^{-\alpha} \omega^{-\beta} 
\exp \left[ {-\cfrac{\theta_{0}}{\theta} - \cfrac{\omega_{0}}{\omega}} \right]
\end{equation}
where $\alpha$, $\beta$, $\theta_{0}$ and $\omega_{0}$ are free parameters and the constant factor $k$ is fixed by the overall normalization
(i.e. the average number of photons per particle). In the plots in the middle and in the bottom panels of Figure~\ref{fig:dummy} 
the fitted functions (integrated either on the angles or on the energies) are superimposed to the measured energy and angular spectra. 
In both cases the template model fit reproduces reasonably the measured spectra.

\section{Monte Carlo simulations and discussion}

\subsection{Monte Carlo simulation model}

The Monte Carlo model used for an interpretation of the experimental data  
includes a full simulation of the physics processes and of the detector response. 
The spectrum of the TR X-rays absorbed in the detector is described as:
\begin{equation}
 \cfrac{d^{2}N_{det}}{d\theta d\omega} = \cfrac{d^{2}N_{prod}}{d\theta d\omega} 
 \times e^{-\sum_{i}{x_{i}}/\lambda_{i}}
 \times \left( 1 - e^{-x_{det}/\lambda_{det}} \right).
 \label{eq:specdet}	
\end{equation}
Where  $\omega$ is the X-ray energy and $\theta$ is the emission angle of X-rays with respect to the radiating particle. 
The first term on the right-hand side of the equation is the TR production spectrum, the second one takes into account 
the absorption of X-rays along their path from the radiator to the detector and the last term takes into account X-ray absorption 
in the detector itself. Here $x_{i}$ and $\lambda_{i}$ are the thicknesses and the absorption lengths of the various materials 
between the radiator and the detector (the helium in the  pipe, the polyethylene windows of the pipe, the air gaps and 
a kapton window before the detector), while $x_{det}$ and $\lambda_{det}$ are the detector thickness ($300 \units{\mu m}$) 
and the silicon absorption length. The values of the absorption lengths have been taken from the NIST reference database~\cite{NIST}.     

For the production spectra the approach described in Refs.~\cite{Cherry:1974de} and~\cite{Mazziotta:1999ji} was used. It also includes 
X-ray attenuation in the radiator material and in the air gaps between the foils. According to this approach, 
the TR production spectrum in a periodic radiator consisting of $N$ foils is given by:
\begin{equation}
 \cfrac{d^{2}N_{prod}}{d\theta d\omega} = \cfrac{d^{2}N_{0}}{d\theta d\omega}
 \times \left( 1 + e^{-\sigma_{1}} -2 e^{-\sigma_{1}/2} \cos \varphi_{1} \right) 
 \times I_{N}(\theta, \varphi).
 \label{eq:specreg}
\end{equation}
The first term on the right-hand side of Eq.~\ref{eq:specreg} represents the single
interface spectrum, the second factor takes into account the single foil interference
term, and the third term describes the $N$-foils interference. The single interface spectrum is given by:
\begin{equation}
 \cfrac{d^{2}N_{0}}{d\theta d\omega} = \cfrac{2 \alpha \hslash \theta^{3}}{\pi \omega}
 \left( \cfrac{1}{\theta^{2} + \gamma^{-2} + \omega_{p1}^{2}/\omega^{2}} 
 - \cfrac{1}{\theta^{2} + \gamma^{-2} + \omega_{p2}^{2}/\omega^{2}} \right)^{2}
 \label{eq:specint}
\end{equation}
where $\gamma$ is the Lorentz factor of the radiating particle, $\alpha$ is the fine
structure constant and $\omega_{p1}$ and $\omega_{p2}$ are the plasma frequencies of the foil and of the gap (air) 
materials respectively. The $N$-foil interference term which takes into account absorption within radiator is given by:
\begin{equation}
 I_{N}(\theta, \varphi) = 
 \cfrac{1 + e^{-N \sigma} - 2e^{-N\sigma/2} cos N\varphi}{1 + e^{-\sigma} - 2e^{-\sigma/2}cos \varphi}
 \label{eq:nfoils}
\end{equation}
with $\sigma = \sigma_{1} + \sigma_{2}$ and $\varphi = \varphi_{1} + \varphi_{2}$, where
$\varphi_{i}$ and $\sigma_{i}$ ($i=1,2$)  are the phase shifts and the attenuations in the foils and in the air gaps defined as:

\begin{equation}
 \varphi_{i} = \cfrac{\omega d_{i}}{2} \left( \theta^{2} + \gamma^{-2} + \omega_{pi}^{2}/\omega^{2} \right) 
\end{equation}
\begin{equation}
 \sigma_{i} = d_{i}/\lambda_{i}
\end{equation}
In this expressions  $d_{i}$ and $\lambda_{i}$ are the foil (gap) thicknesses and absorption lengths.

For simulations of irregular radiators, the Garibian formulas were used~\cite{Garibian:1975xx}. 
In this approach it is assumed that the foil and gap thicknesses $d_{1}$ and $d_{2}$ follow gamma distributions. 
Calculations show that fluctuations of $d_{1}$ and $d_{2}$ do not change the total X-ray yield, but result in a smearing of the TR spectra.

Multiple scattering of particles passing different materials was taken into account as described in~\cite{Tanabashi:2018oca}. 
For each particle, two angles $\theta_{xy}$ and $\theta_{xz}$ are extracted from a Gaussian distribution  with null mean and with sigma given by:
\begin{equation}
 \theta_{0} = \cfrac{13.6\units{MeV}}{\beta c p} \sqrt{\cfrac{x}{X_{0}}}
 \left[ 1 + 0.038 \ln \left( \cfrac{x}{X_{0}} \right) \right]
\end{equation}
where $x/X_{0}$ is the thickness of the radiator and of the absorbers in radiation length units. 
The impact position of the particle on the detector is then obtained using
the distance $L$ between the radiator and the detector.\footnote{The multiple scattering angles 
are usually small for muons, but can be not negligible for electrons. In fact, for $20 \units{GeV/c}$ 
electrons crossing the $90$ foils of polyethylene radiator (total thickness $x/X_{0} \sim 0.03$) 
the scattering angle is $\theta_{0} \sim 0.14 \units{mrad}$.}

The energy deposition of the particle in the detector is extracted from a Landau distribution folded with a Gaussian 
function with a most probable value of $83.5\units{keV}$ and with $\sigma=7.7\units{keV}$, as the one shown in Figure~\ref{fig:epart}. 
The number $n_{i}$ of hit strips on two detector sides are extracted from the strip multiplicity distributions of 
particle clusters in real data. The energy deposited by the particle is then shared among the $n_{i}$ strips nearest 
to the particle impact point according to a Gaussian profile: the strips closer to the impact point have larger fractions of energy than the farther ones. 

The number of TR X-rays associated to each particle is extracted from a Poisson distribution with an average 
value corresponding to the integral of the simulated TR spectrum absorbed in the detector. 
The energy and the polar angle $\theta$ of each photon with respect to the direction of the radiating particle 
are randomly extracted from the distribution in Eq.~\ref{eq:specdet}, while the azimuth angle $\phi$ is extracted 
from a uniform distribution in $[0, 2\pi]$. The generation point of each TR photon is randomly chosen along the 
particle track within the radiator, and the photon is then propagated to the detector.

Since the X-ray energy is usually significantly smaller than the particle energy loss
and since the photoelectric absorption is a point-like event, the photon energy is shared between 
two strips on each detector side which are closest to the X-ray absorption point. 
This corresponds to the hit multiplicity  observed in data.
The detector noise is taken into account in the simulations according to experimentally defined values.
Finally, the charge  associated to each strip corresponds to the sum of the noise and of the energies released 
by the particle and by the TR photons.
The simulated data are then analyzed applying the same event selection algorithm
as in real data. 

\subsection{Mylar and polyethylene radiators}

\begin{figure}[!t]
\begin{center}
\includegraphics[width=0.475\columnwidth]{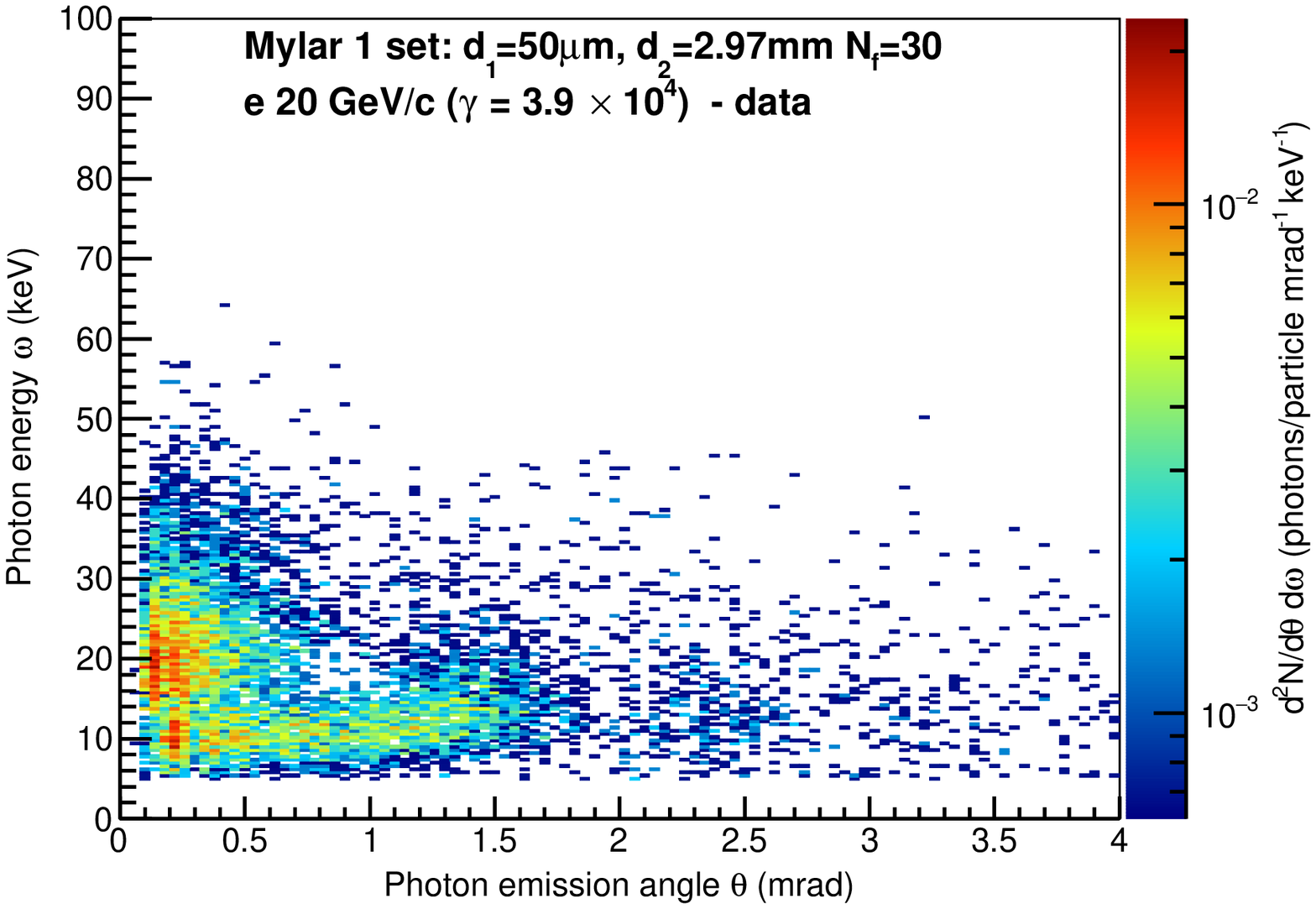}
\includegraphics[width=0.475\columnwidth]{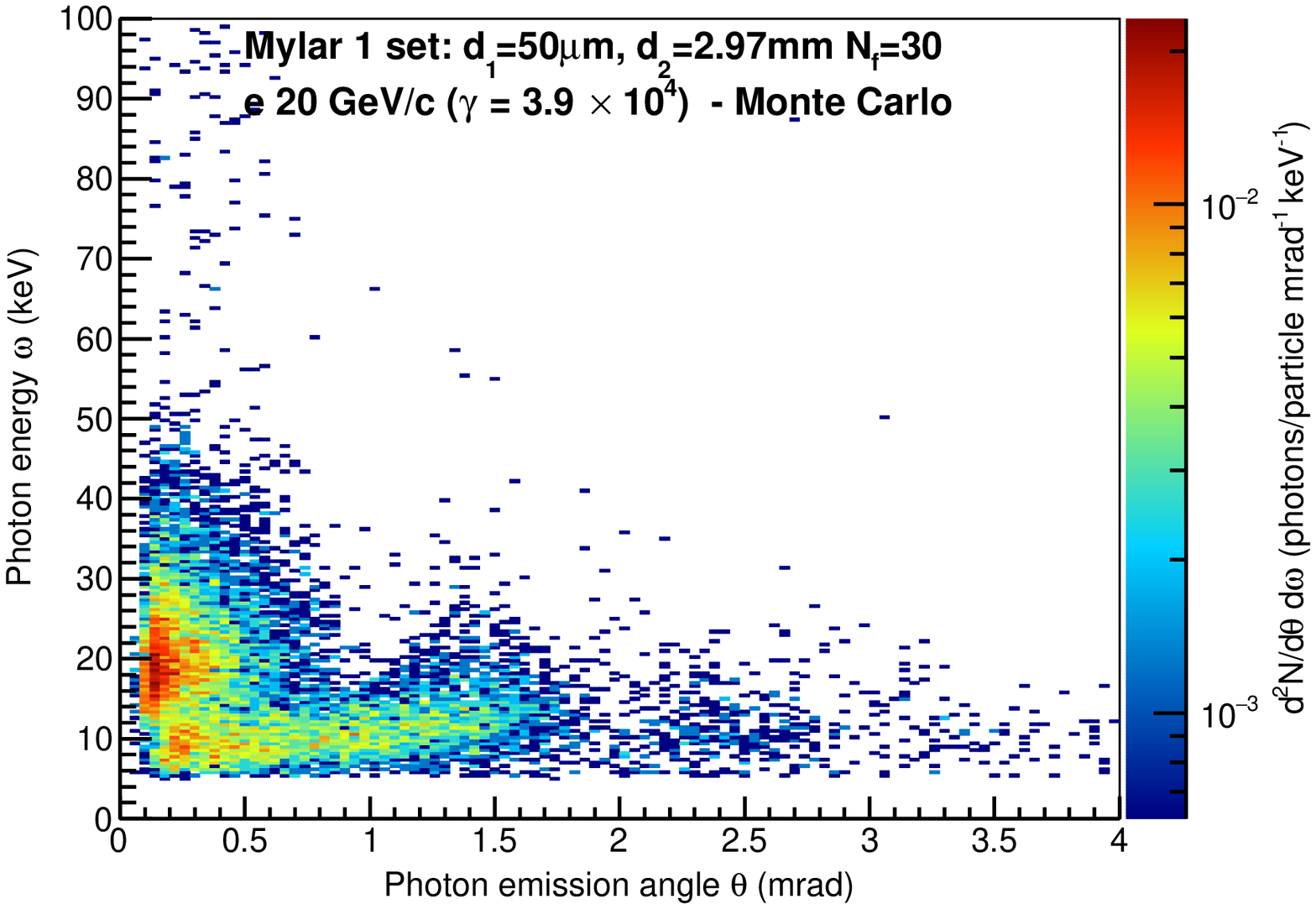}
\end{center}
\caption{Two-dimensional distribution of the TR X-rays produced by $20 \units{GeV/c}$ electrons 
crossing the 1-set mylar radiator. The measured spectrum (left panel) is compared with the 
one predicted  by the simulation (right panel). The measured spectrum has been corrected by subtracting 
the background evaluated in a run with the dummy radiator (see Section~\ref{sec:dummy}).}
\label{fig:mylar1_2d}
\end{figure}

In Figure~\ref{fig:mylar1_2d} a background subtracted two dimensional distribution 
of the TR X-rays obtained in the test with $20 \units{GeV/c}$ electrons crossing
1-set of the mylar radiator is compared with the one obtained from the simulation. 
A qualitative comparison of the two plots shows that the simulation reproduces the experimental data fairly well. 
The Monte Carlo simulations confirm that the highest energy photons ($\omega > 15\units{keV}$) are mostly 
emitted at small angles with respect to the radiating particle, whereas the lower energy photons 
are spread over angles significantly larger than $1/\gamma$.

\begin{figure}[!tbp]
\begin{center}
\includegraphics[width=0.39\columnwidth]{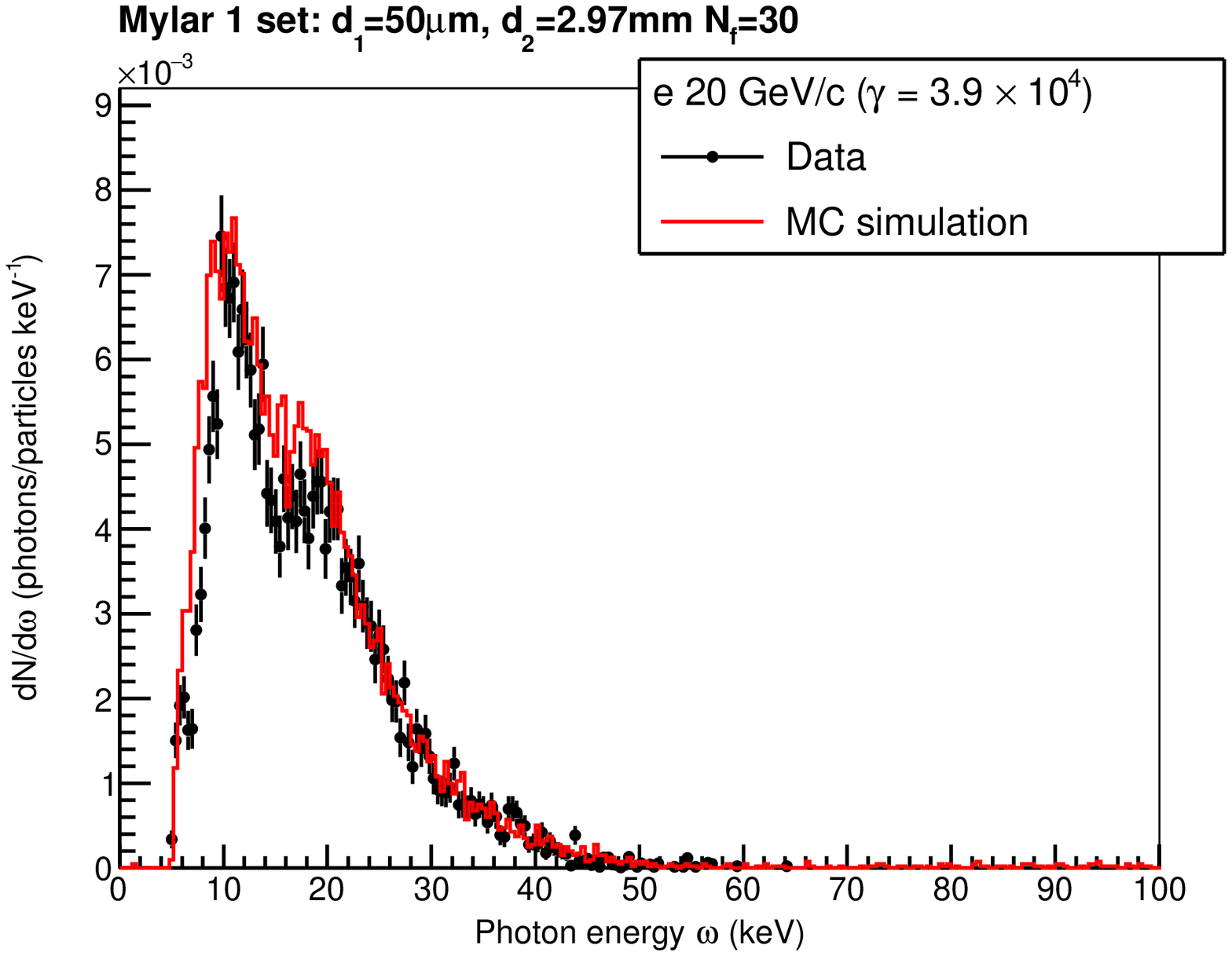}
\includegraphics[width=0.39\columnwidth]{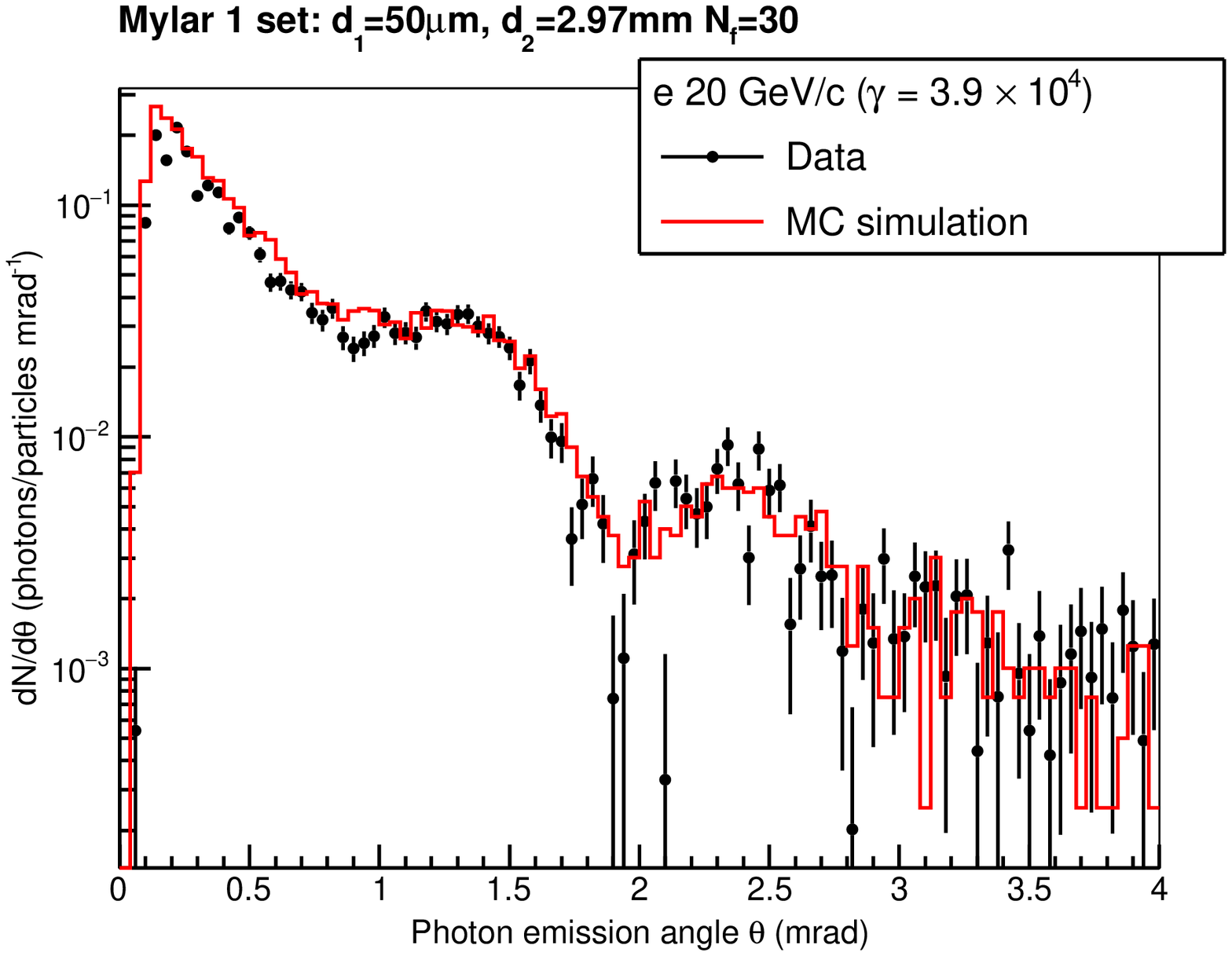}
\includegraphics[width=0.39\columnwidth]{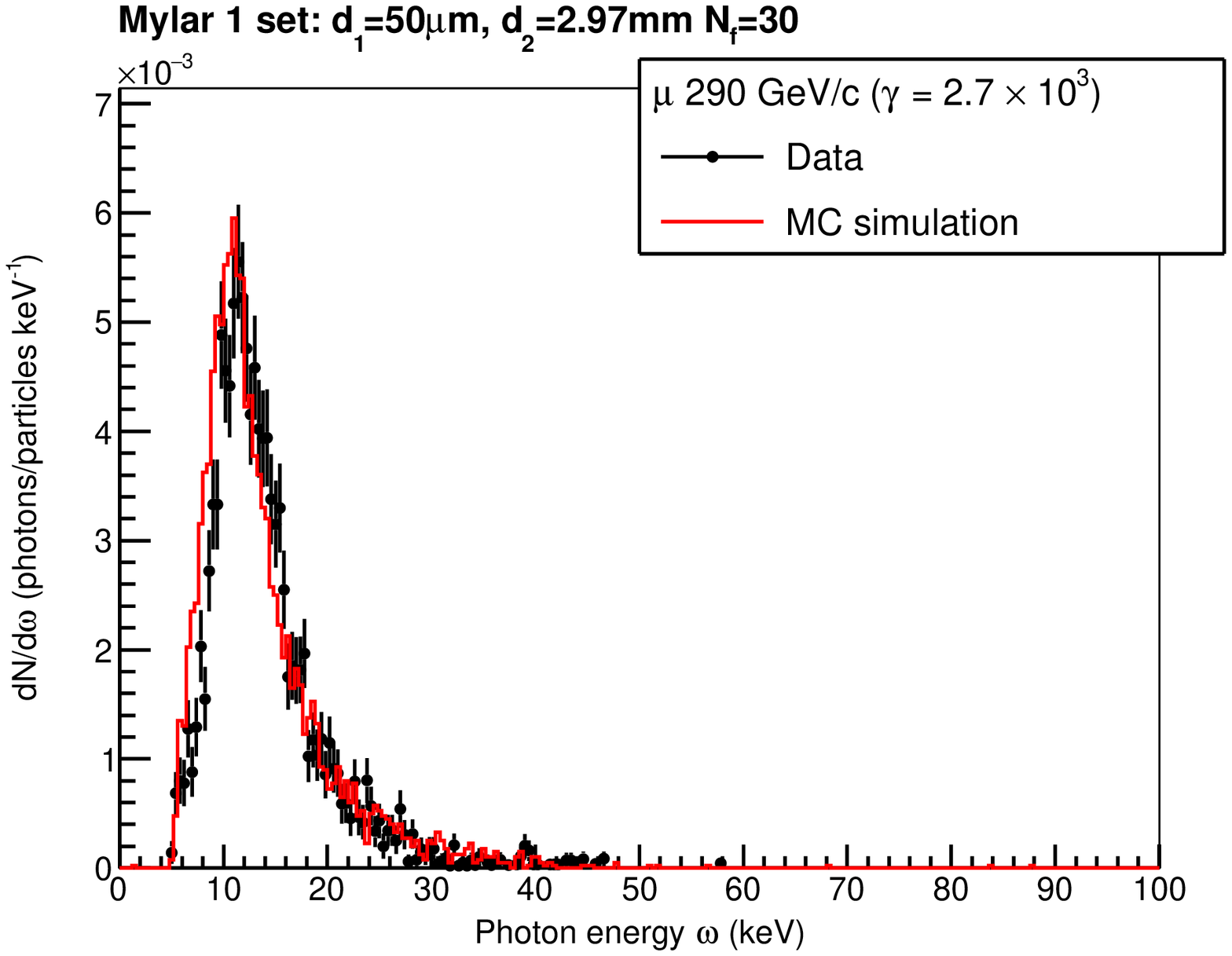}
\includegraphics[width=0.39\columnwidth]{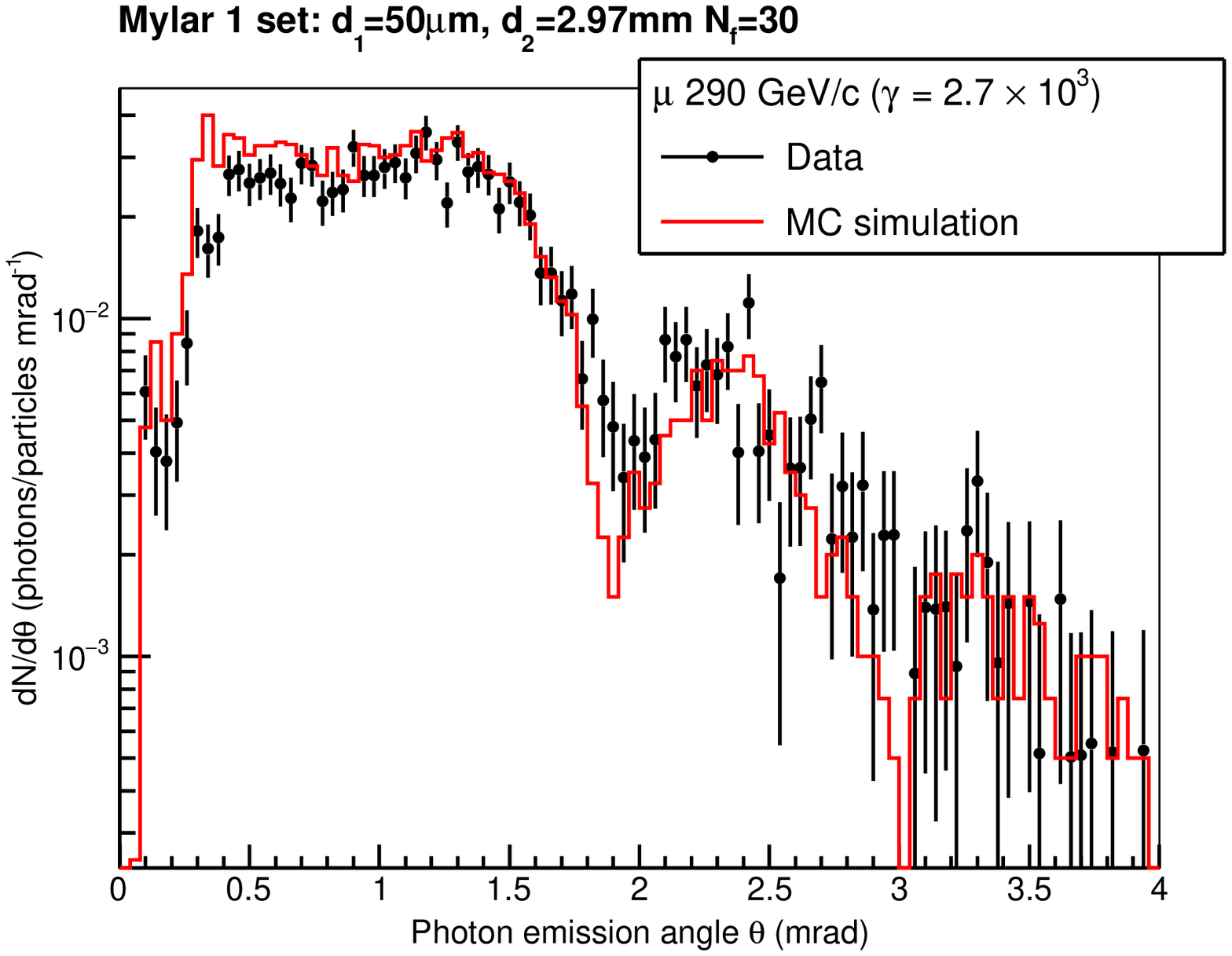}
\includegraphics[width=0.39\columnwidth]{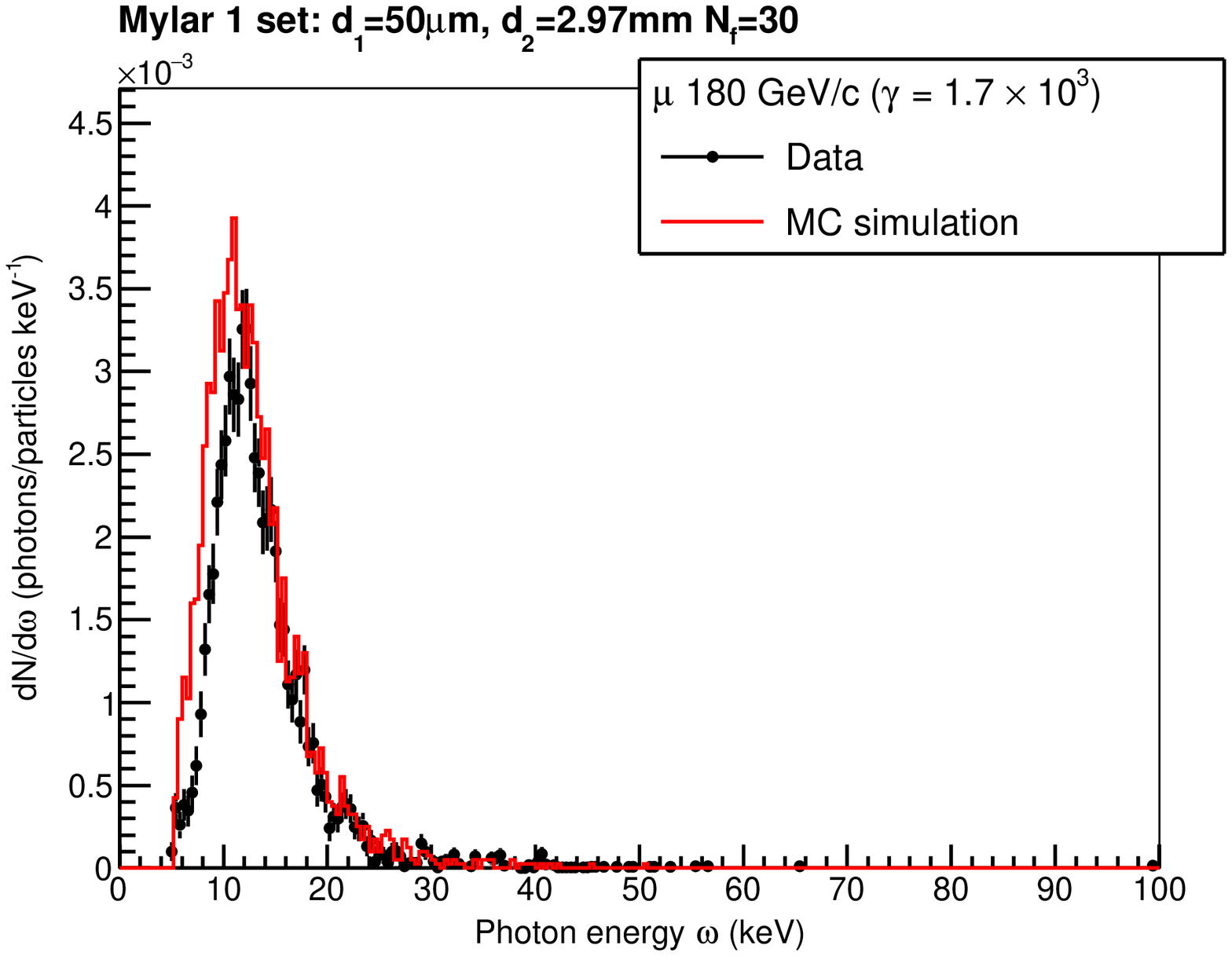}
\includegraphics[width=0.39\columnwidth]{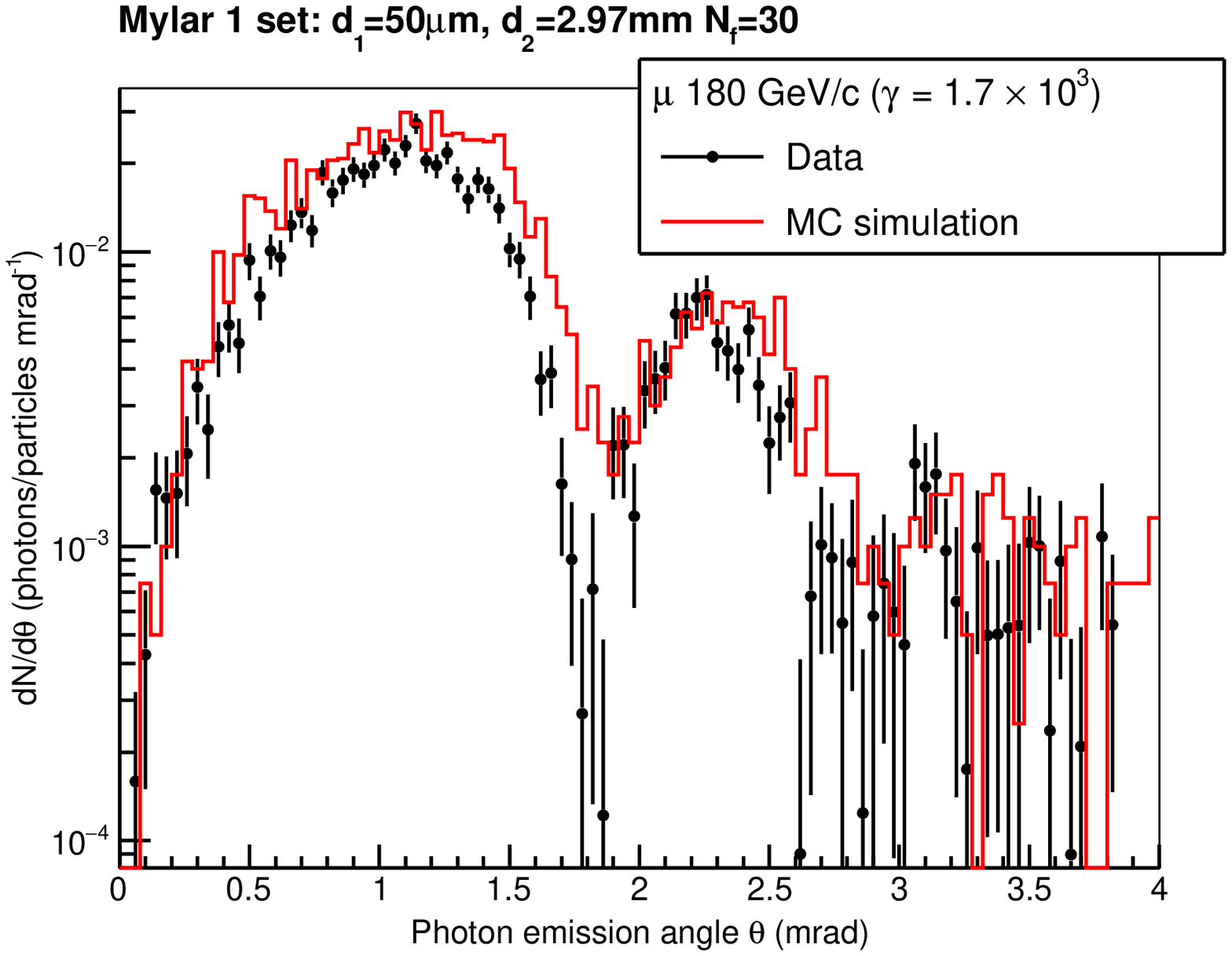}
\includegraphics[width=0.39\columnwidth]{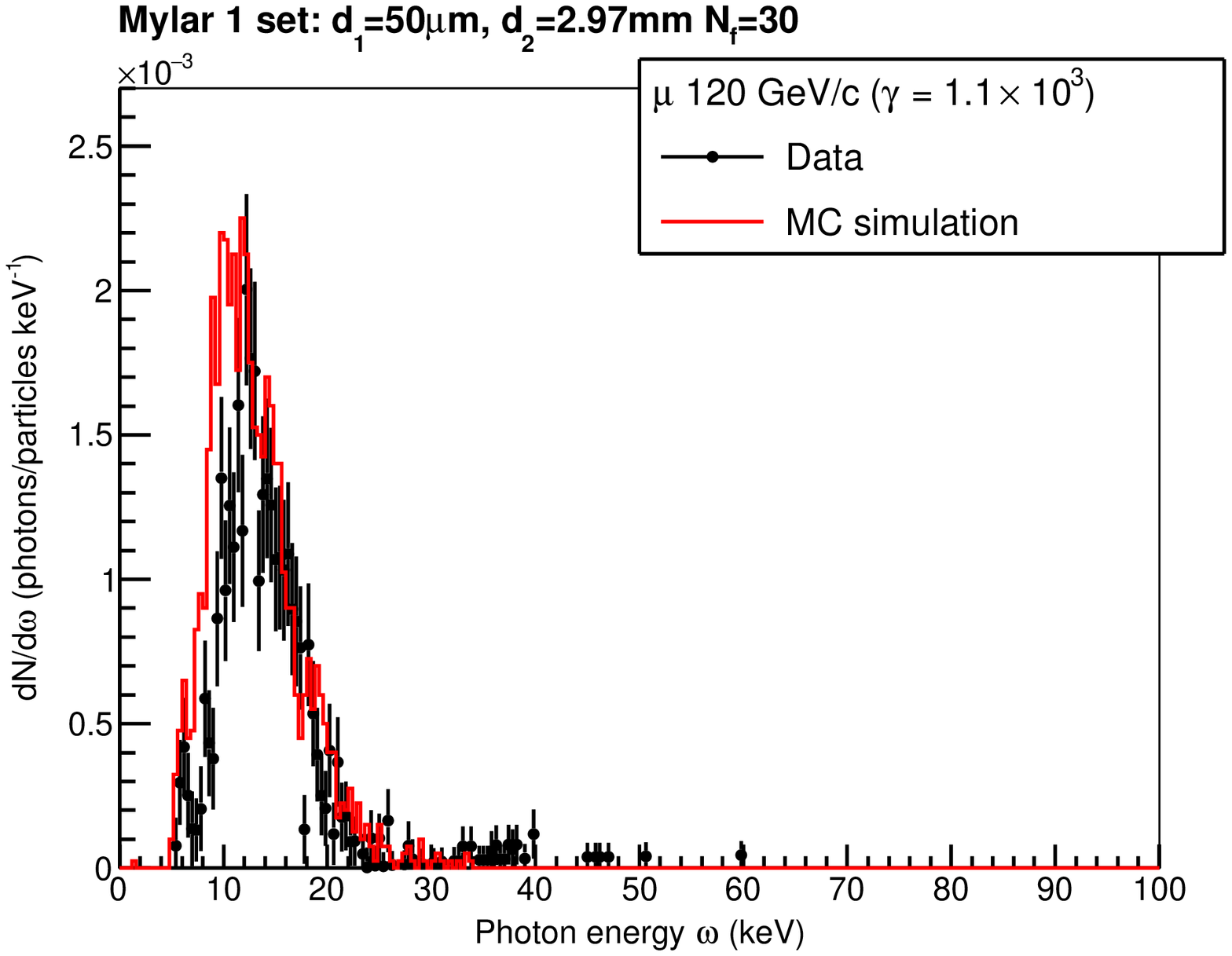}
\includegraphics[width=0.39\columnwidth]{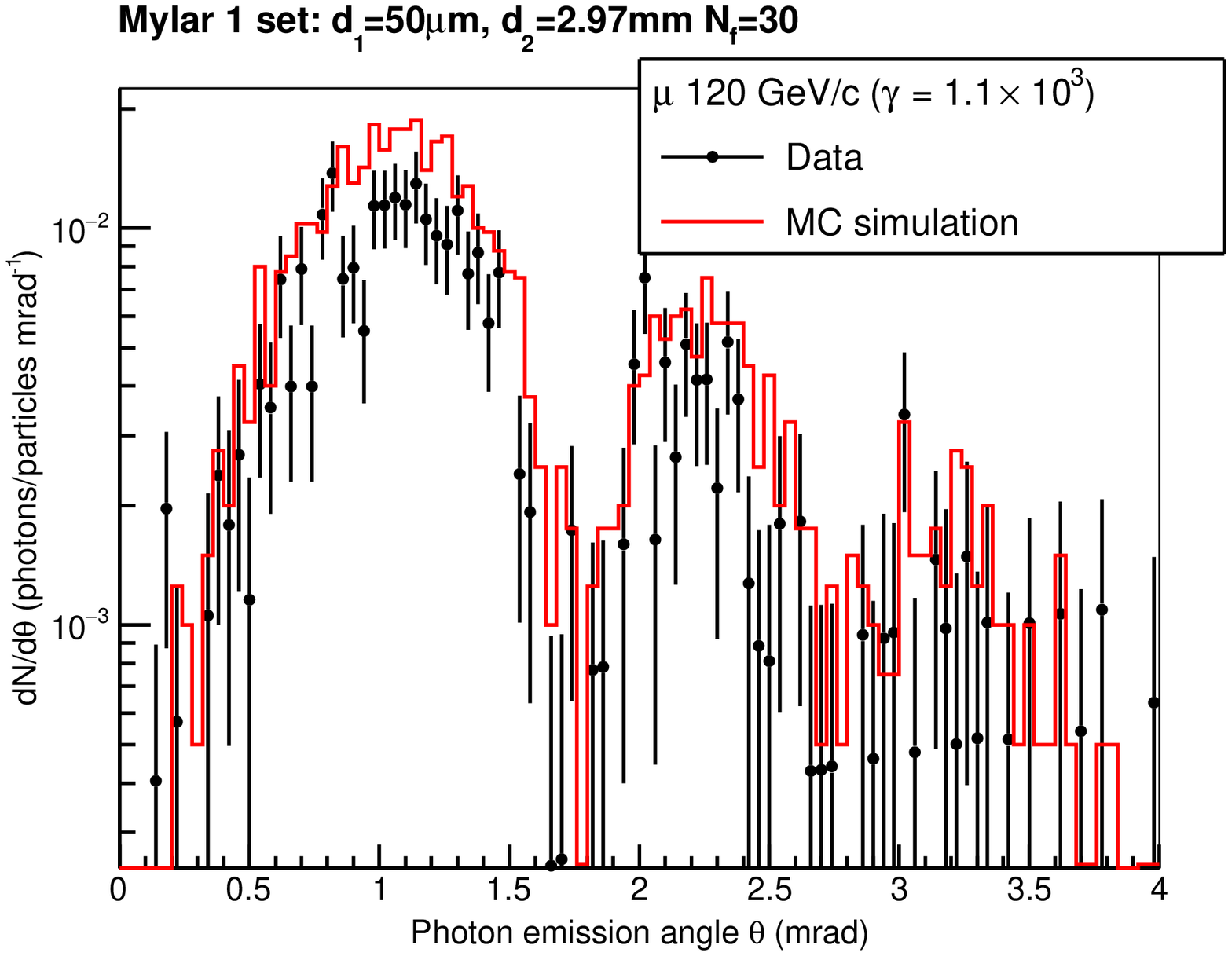}
\end{center}
\caption{Comparison between the measured and simulated differential spectra in energy (left column) 
and in angle (right column) of the TR X-rays produced  by $20 \units{GeV/c}$ electrons and $290$, $180$ and $120 \units{GeV/c}$  
muons crossing the 1-set mylar radiator. The data have been corrected subtracting the background measured in the runs with dummy radiators.}
\label{fig:mylar1}
\end{figure}

\begin{figure}[!tbp]
\begin{center}
\includegraphics[width=0.39\columnwidth]{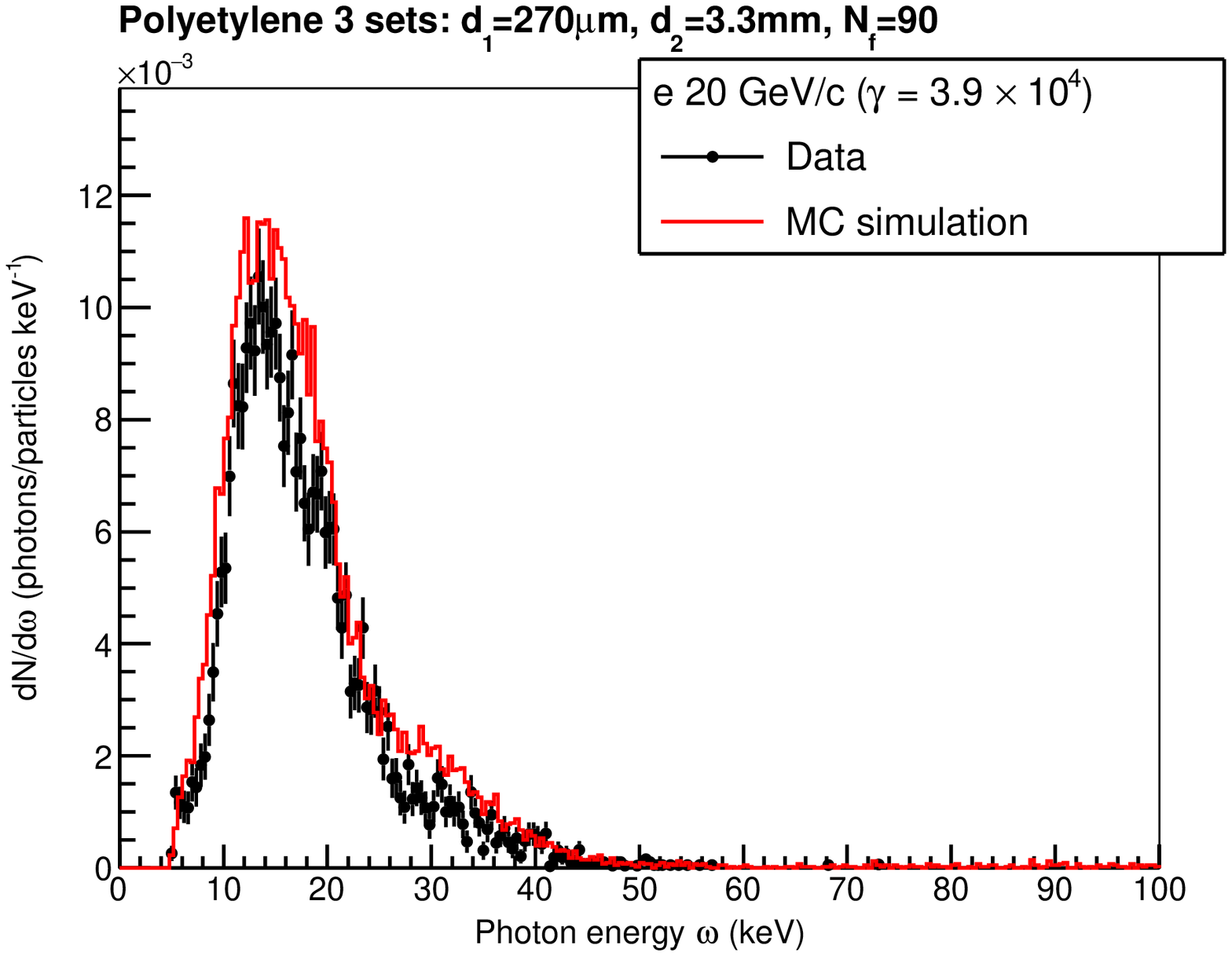}
\includegraphics[width=0.39\columnwidth]{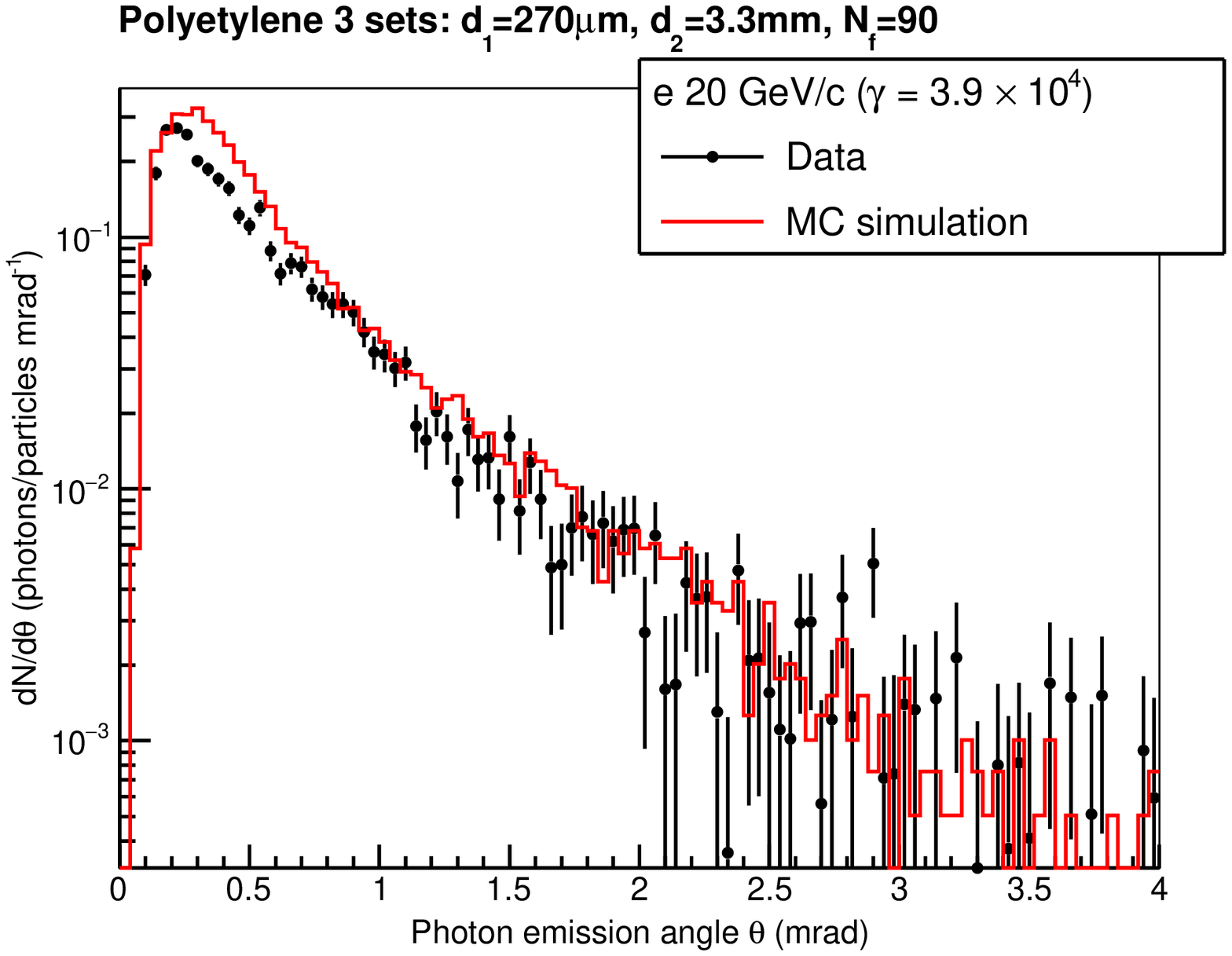}
\includegraphics[width=0.39\columnwidth]{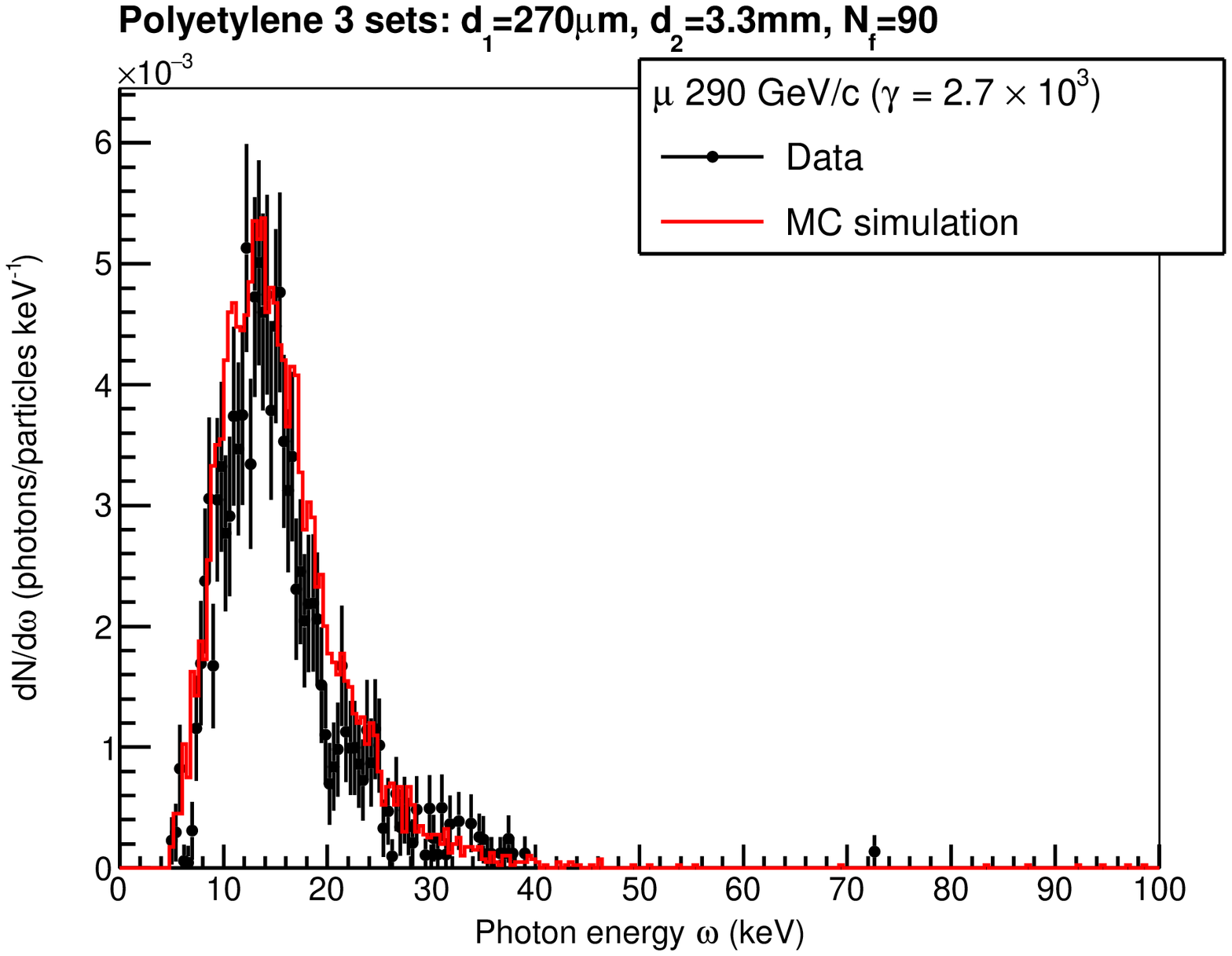}
\includegraphics[width=0.39\columnwidth]{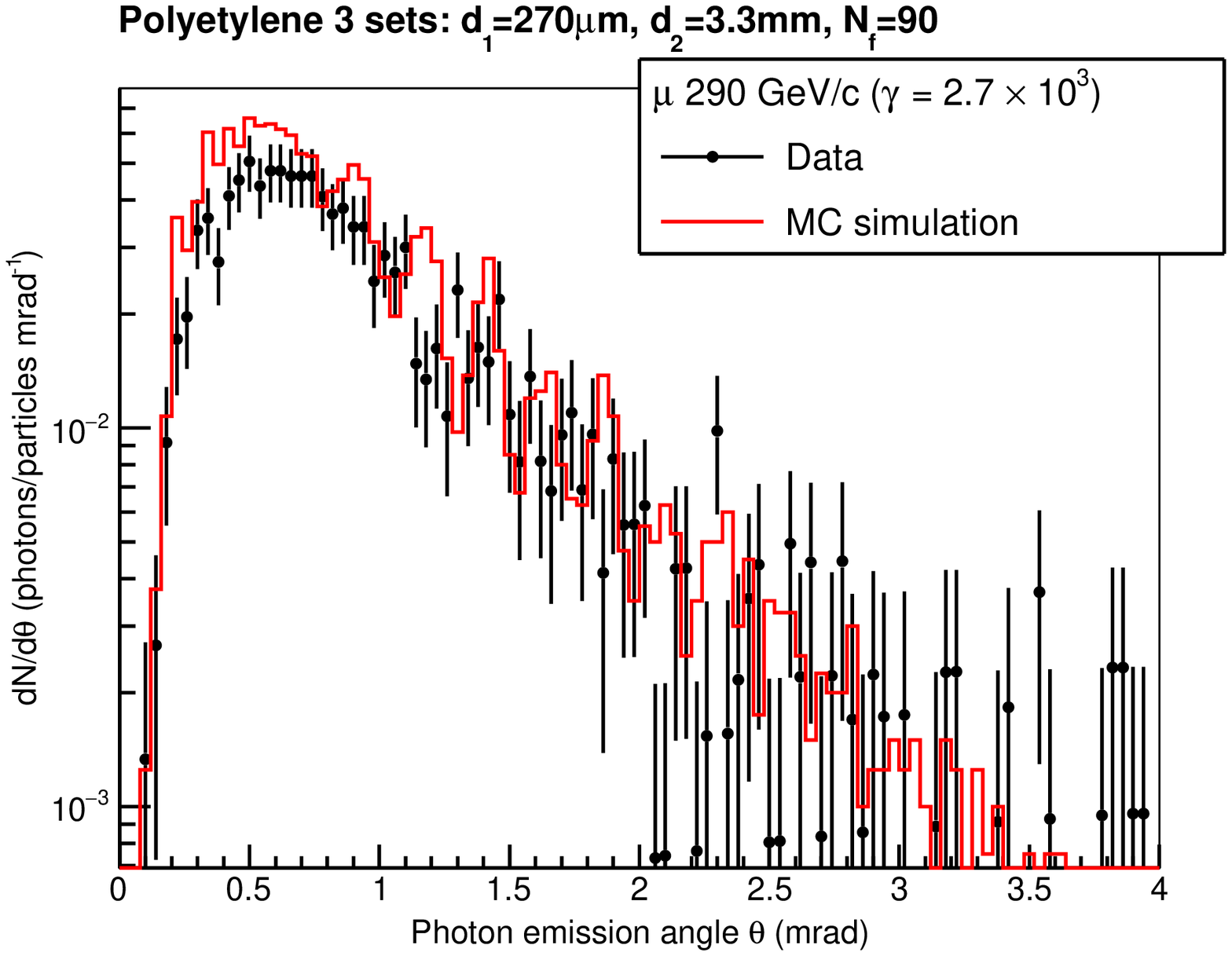}
\includegraphics[width=0.39\columnwidth]{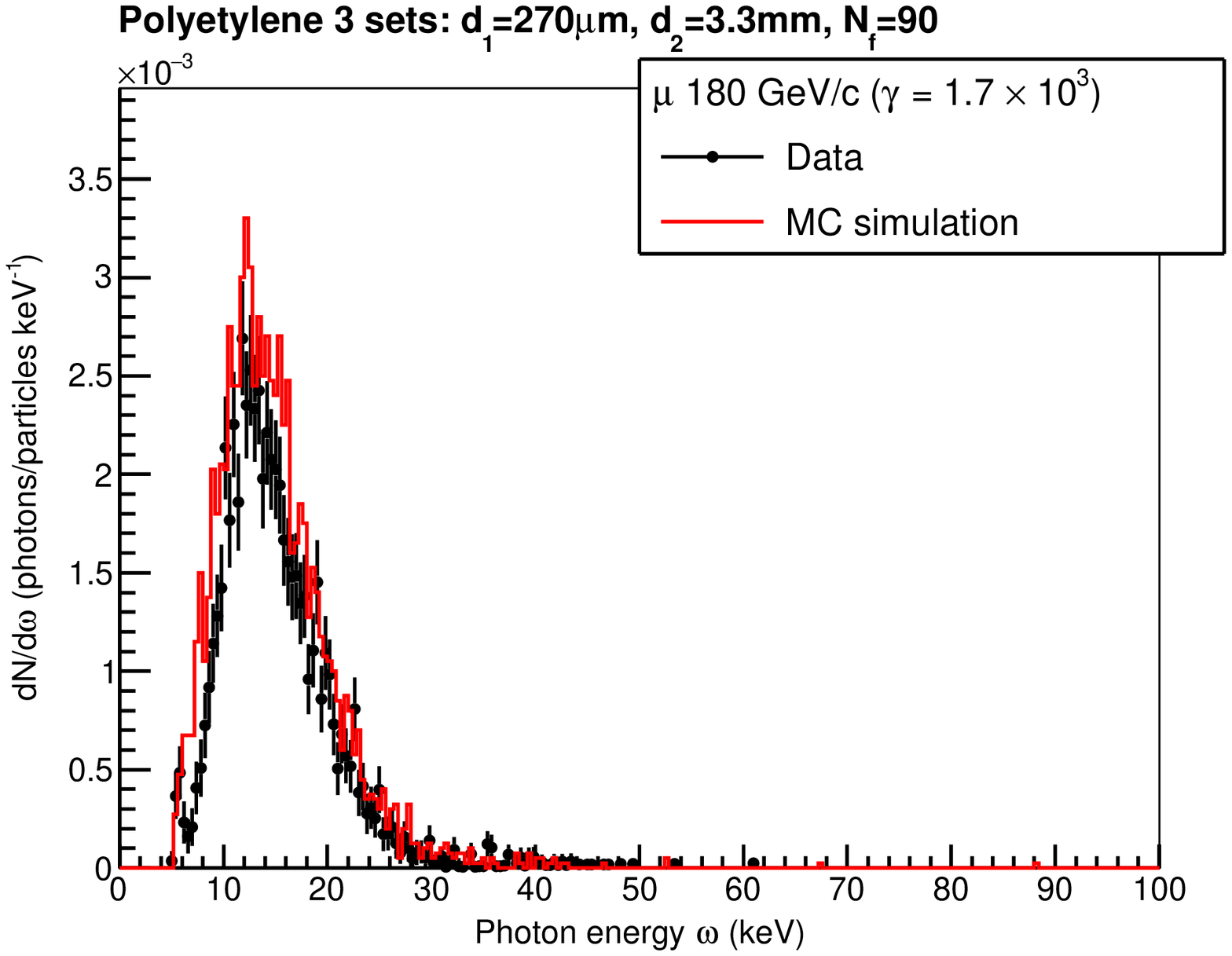}
\includegraphics[width=0.39\columnwidth]{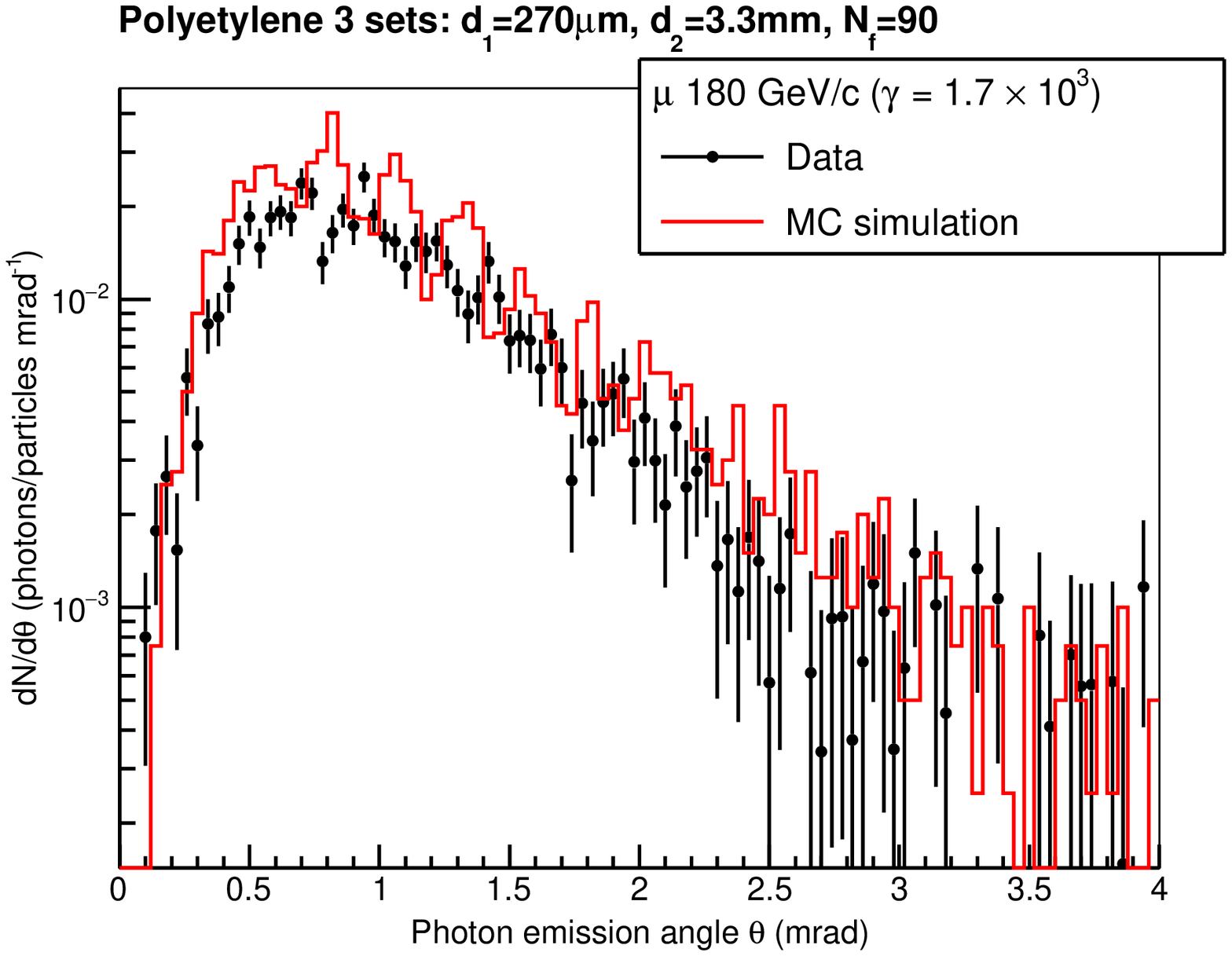}
\includegraphics[width=0.39\columnwidth]{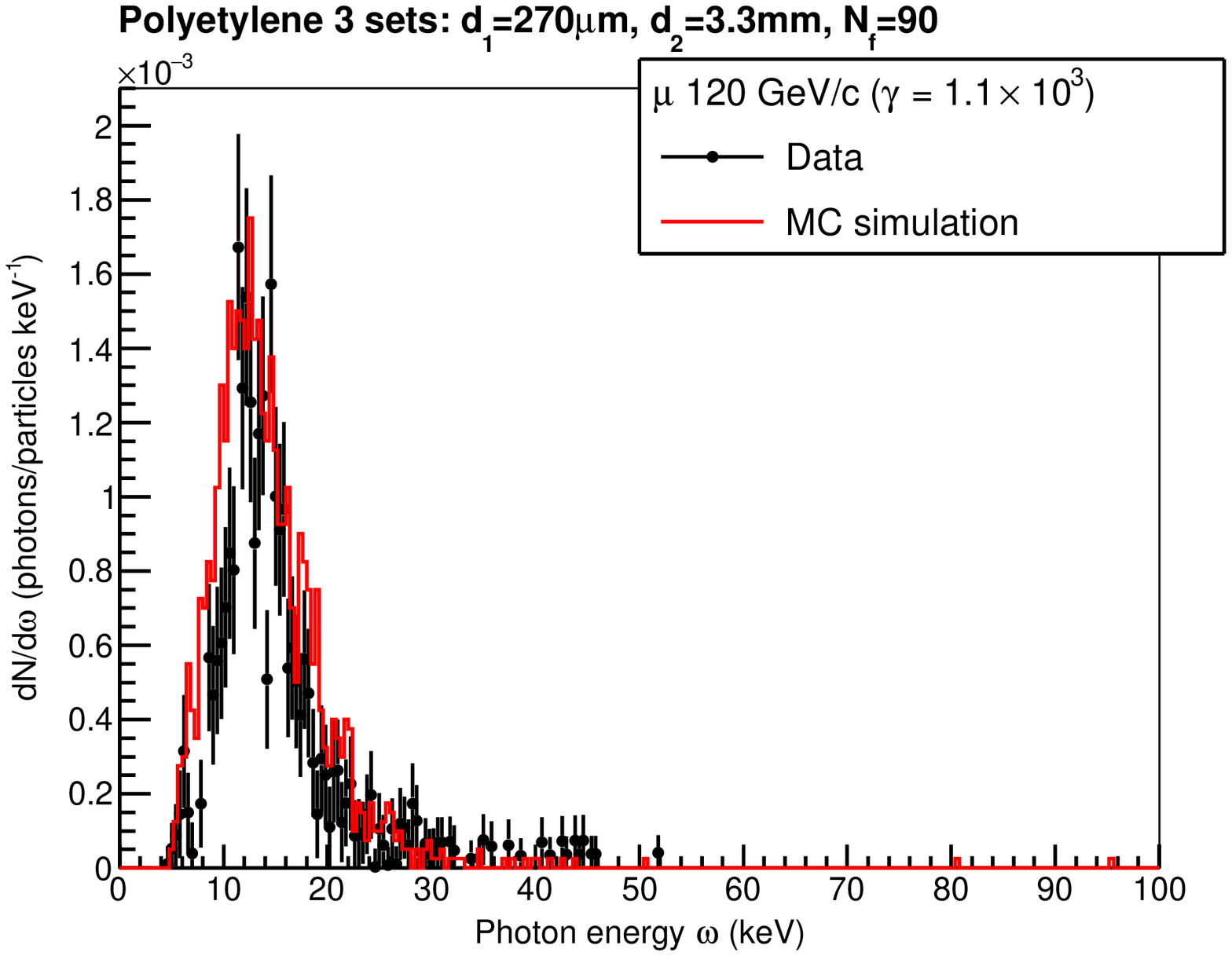}
\includegraphics[width=0.39\columnwidth]{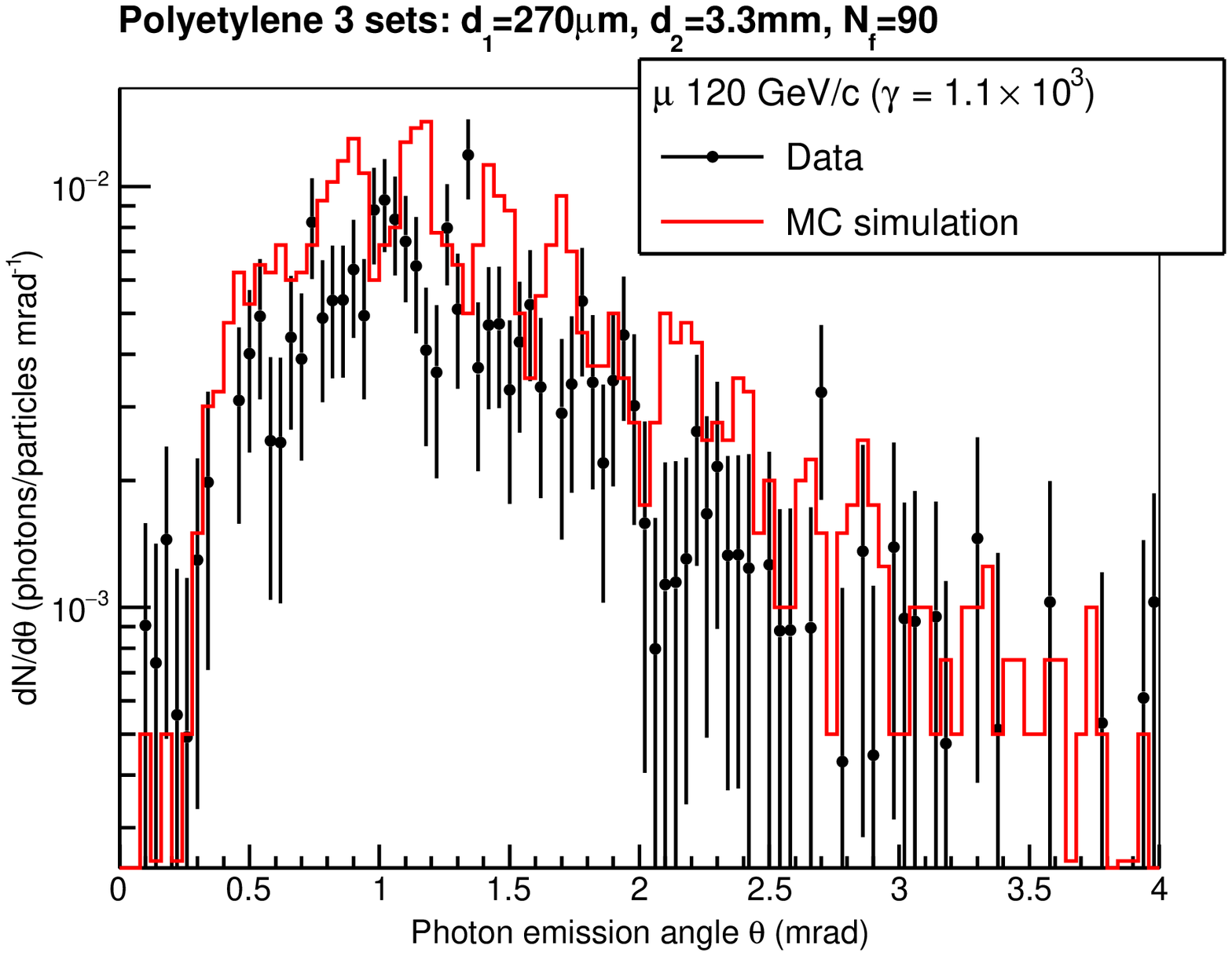}
\end{center}
\caption{Comparison between the measured and simulated differential spectra  in energy (left column) 
and in angle (right column) of the TR X-rays produced  by $20 \units{GeV/c}$ electrons and $290$, $180$ and $120 \units{GeV/c}$  
muons crossing the 3-sets polyethylene radiator. The data have been corrected subtracting the background measured in the runs with dummy radiators.}
\label{fig:polyethylene3}
\end{figure}

Figure~\ref{fig:mylar1} shows the comparison between the measured  energy (left) and angular (right) 
differential spectra of the TR X-rays and the simulation predictions for the 1-set mylar radiator and 
for all types of beam particles. The figure shows that, in general, simulations well reproduce the 
experimental data for  all types of particles, but some discrepancies are observed. 
A comparison of the measured and simulated energy spectra for $20 \units{GeV/c}$ electrons shows
that the simulation predicts a slightly larger number of photons with energies $>15 \units{keV}$. 
This is also seen in the angular spectrum, where the simulation well reproduces the structure 
with multiple peaks, but the first peak below $0.3\units{mrad}$ is more populated than in data. 
These features of the angular and of the energy spectra are clearly correlated, since the higher 
energy photons are emitted at smaller angles with respect to the radiating particles. 
Therefore, an excess of higher energy X-rays corresponds to an excess of photons emitted at small angles. 
The simulation also yields an overestimation of the low-energy photons, 
a feature which is clearly visible for muons.
These discrepancies can  be associated to the inaccurate representation of the material thicknesses 
and values of the absorption lengths of X-rays  which are implemented in the simulation. 
In fact,  from Eq.~\ref{eq:specdet} it is evident that an overestimation, as well as an underestimation 
of the absorption in the materials between the radiator  and the detector, would affect the number of absorbed X-rays. 
This effect is also clearly seen in Figure~\ref{fig:polyethylene3}, where the comparison between the measured spectra and those 
predicted by the Monte Carlo simulation is shown for 3-sets of the polyethylene radiator. 
As in case of the mylar radiator the data are fairly well described by the Monte Carlo simulation. 
The simulation slightly overestimates the number of detected X-rays, particularly 
those with energies above $20\units{keV}$ emitted at small angles.

\subsection{Polypropylene radiator}

\begin{figure}[!tbp]
\begin{center}
\includegraphics[width=0.39\columnwidth]{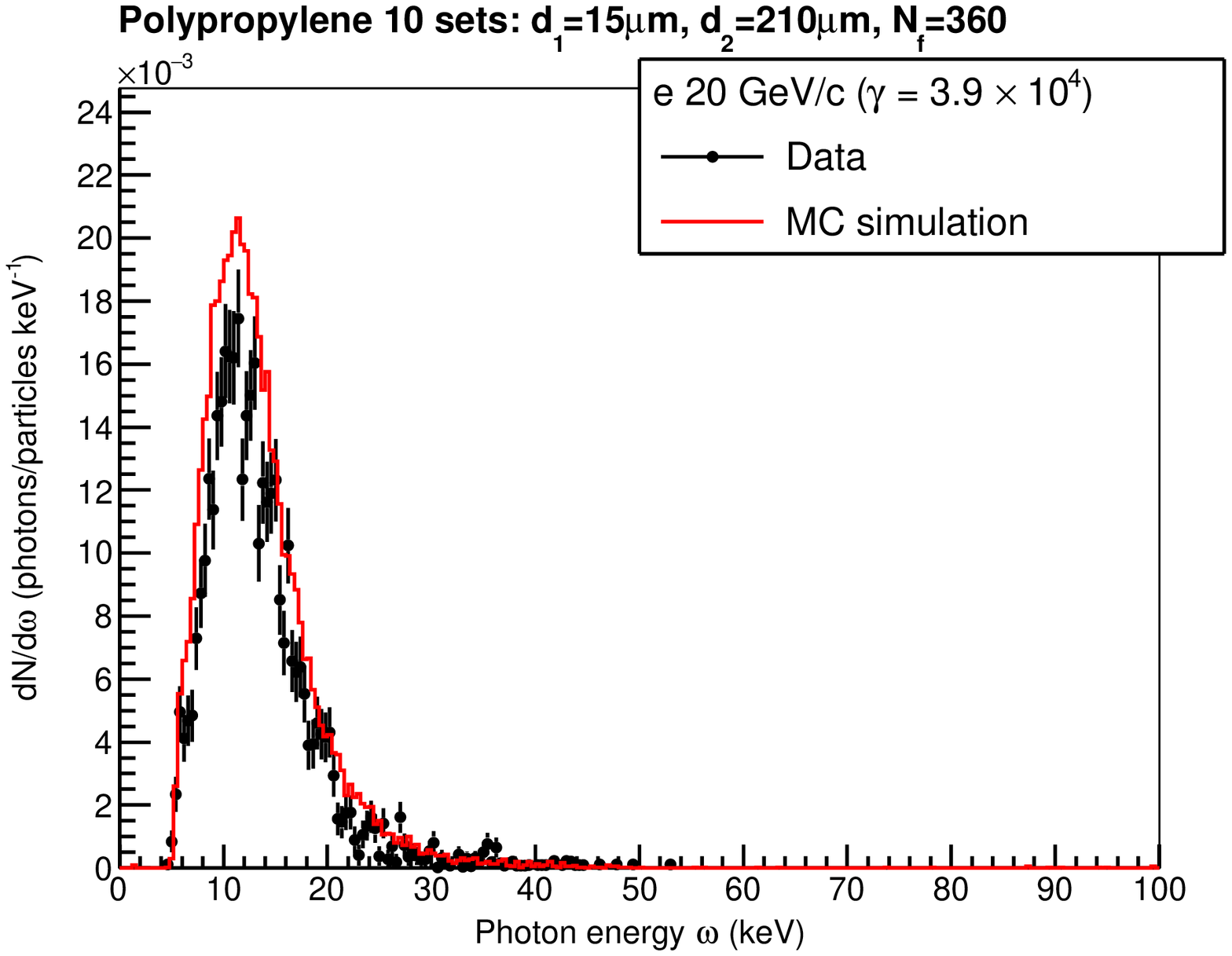}
\includegraphics[width=0.39\columnwidth]{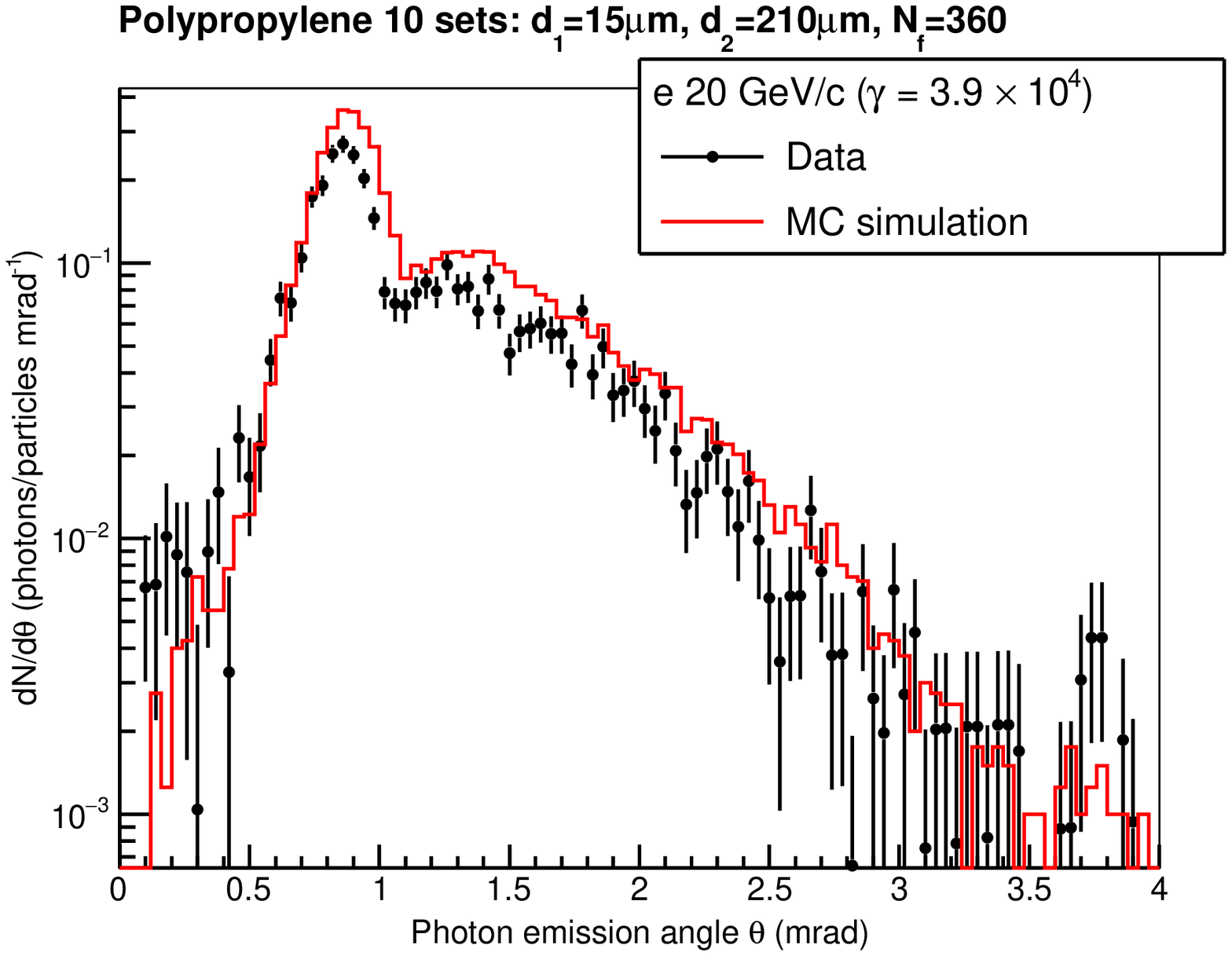}
\includegraphics[width=0.39\columnwidth]{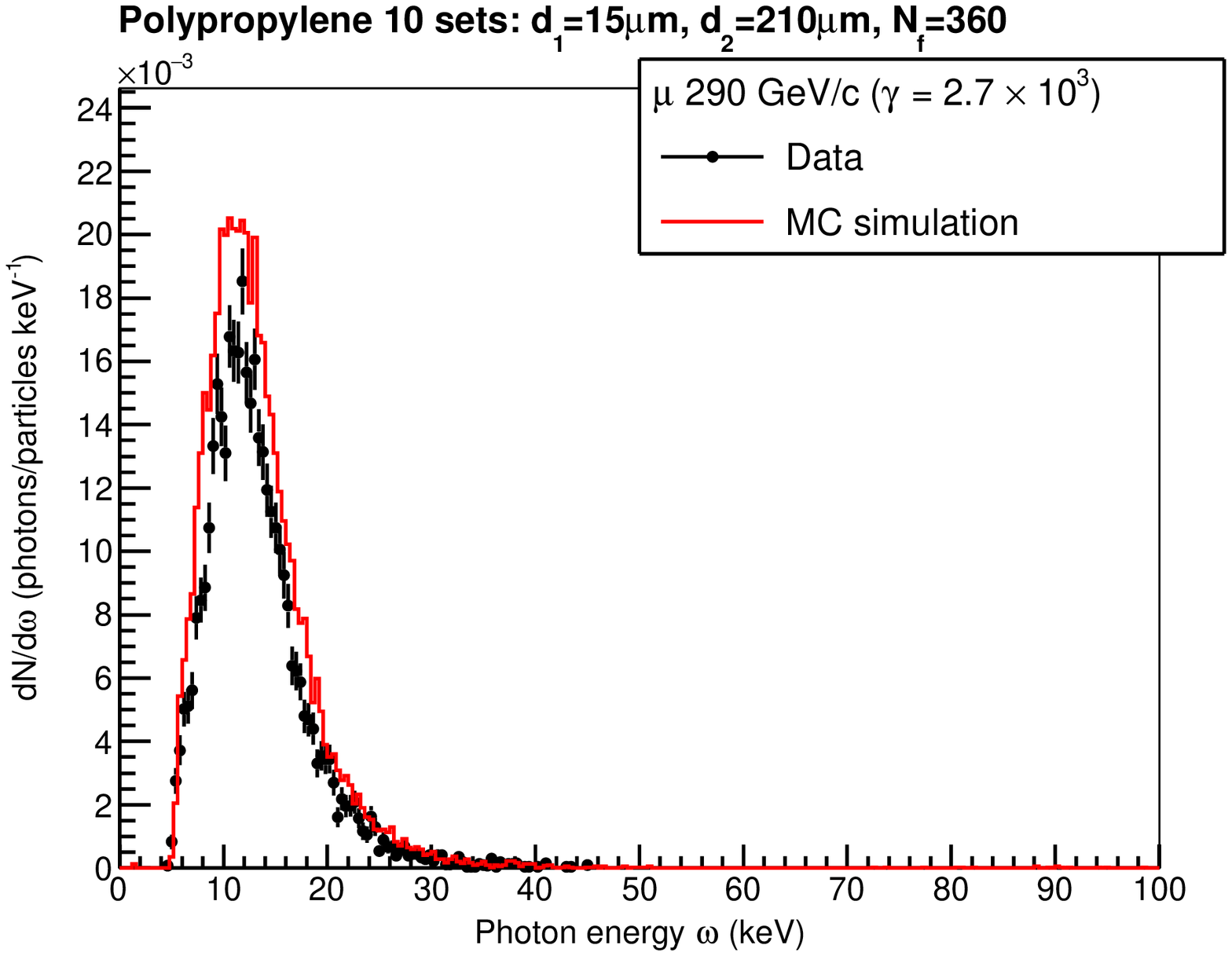}
\includegraphics[width=0.39\columnwidth]{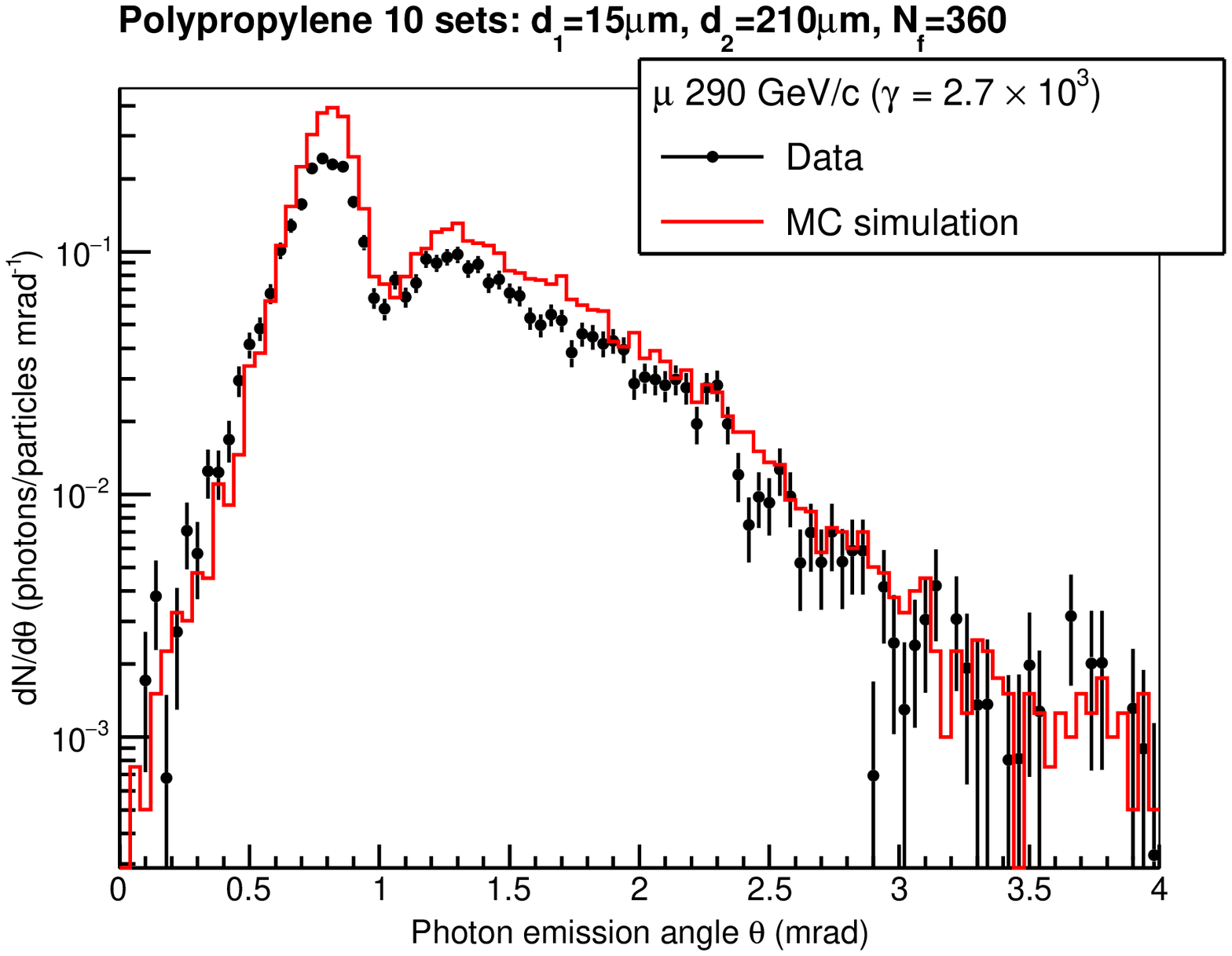}
\includegraphics[width=0.39\columnwidth]{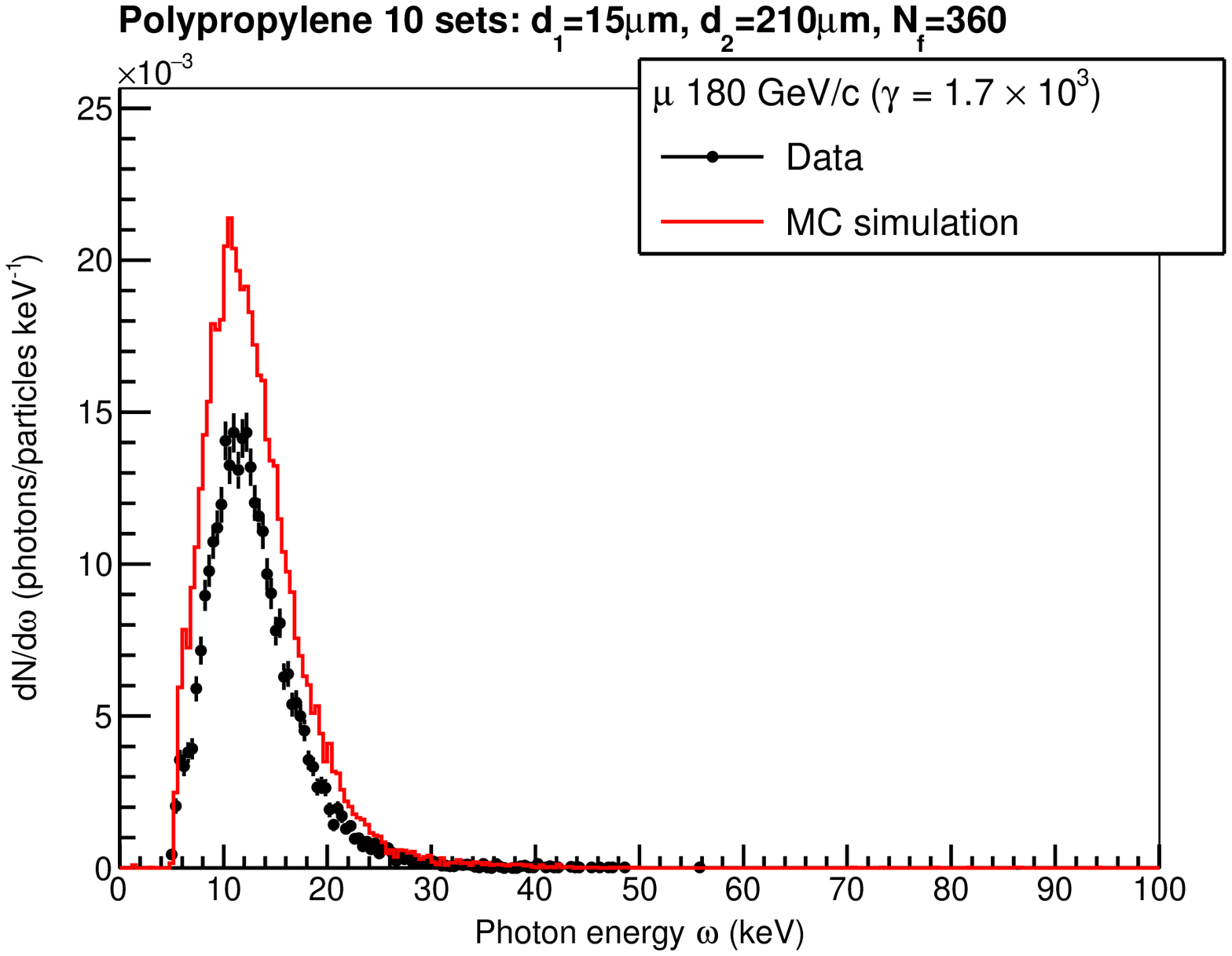}
\includegraphics[width=0.39\columnwidth]{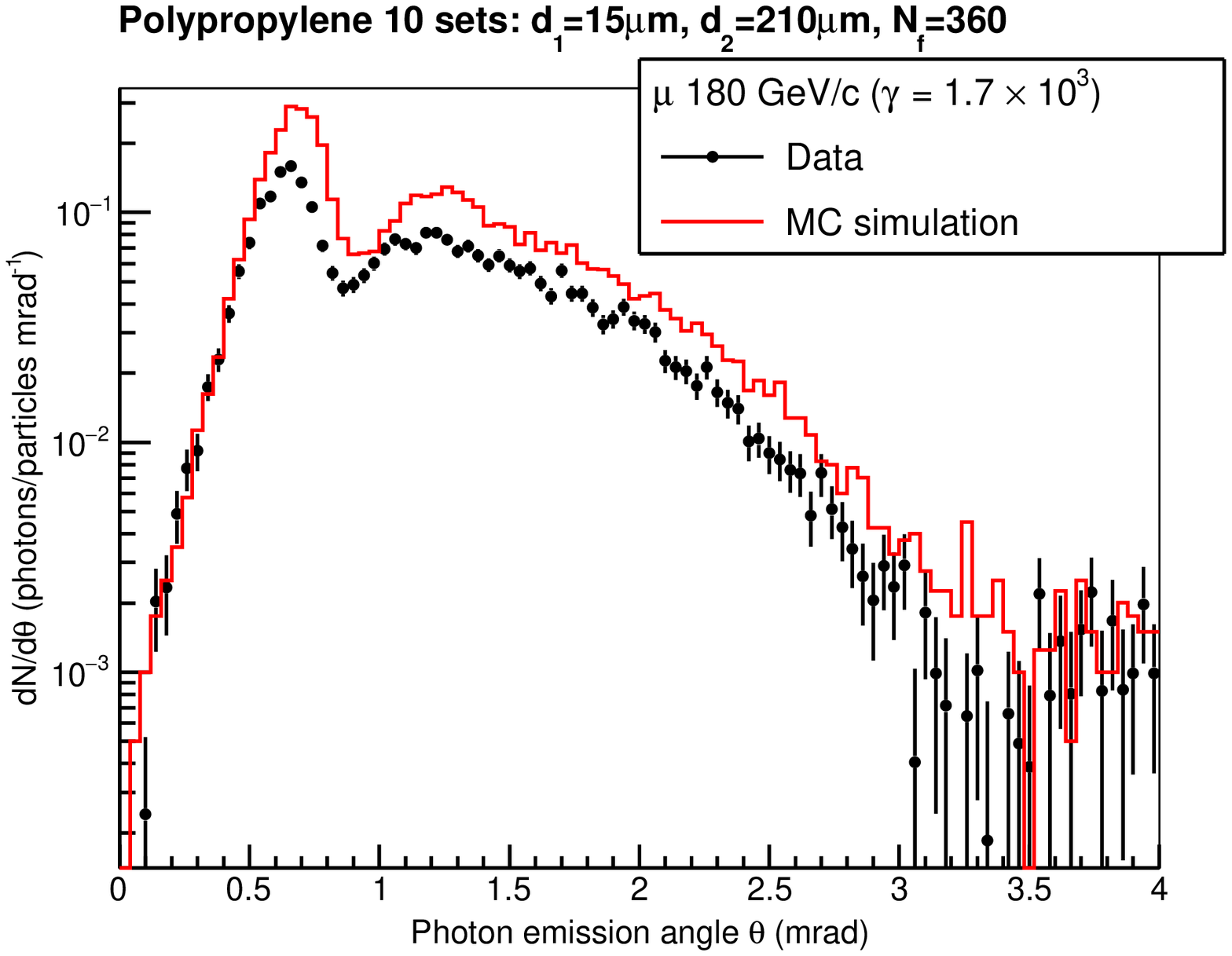}
\includegraphics[width=0.39\columnwidth]{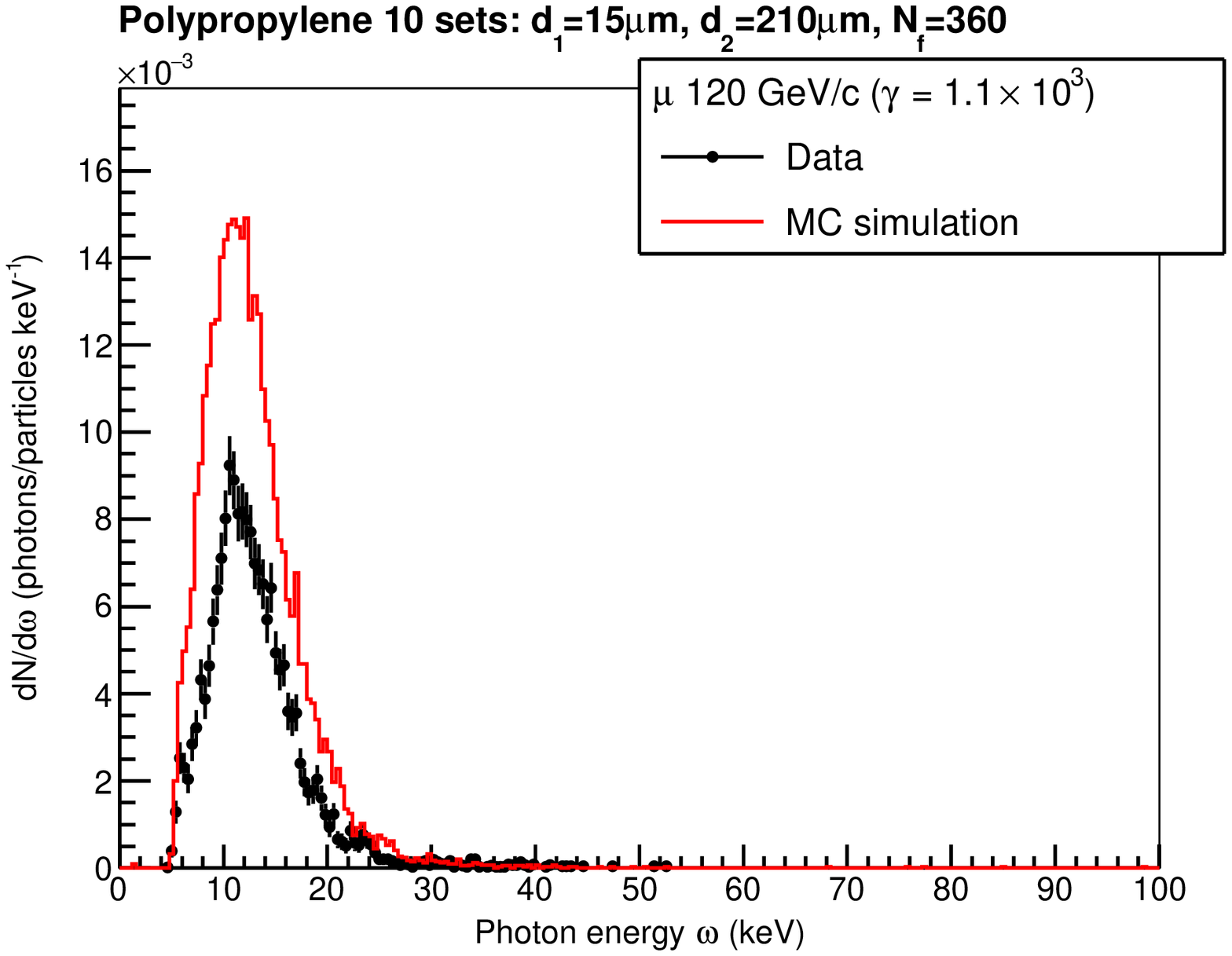}
\includegraphics[width=0.39\columnwidth]{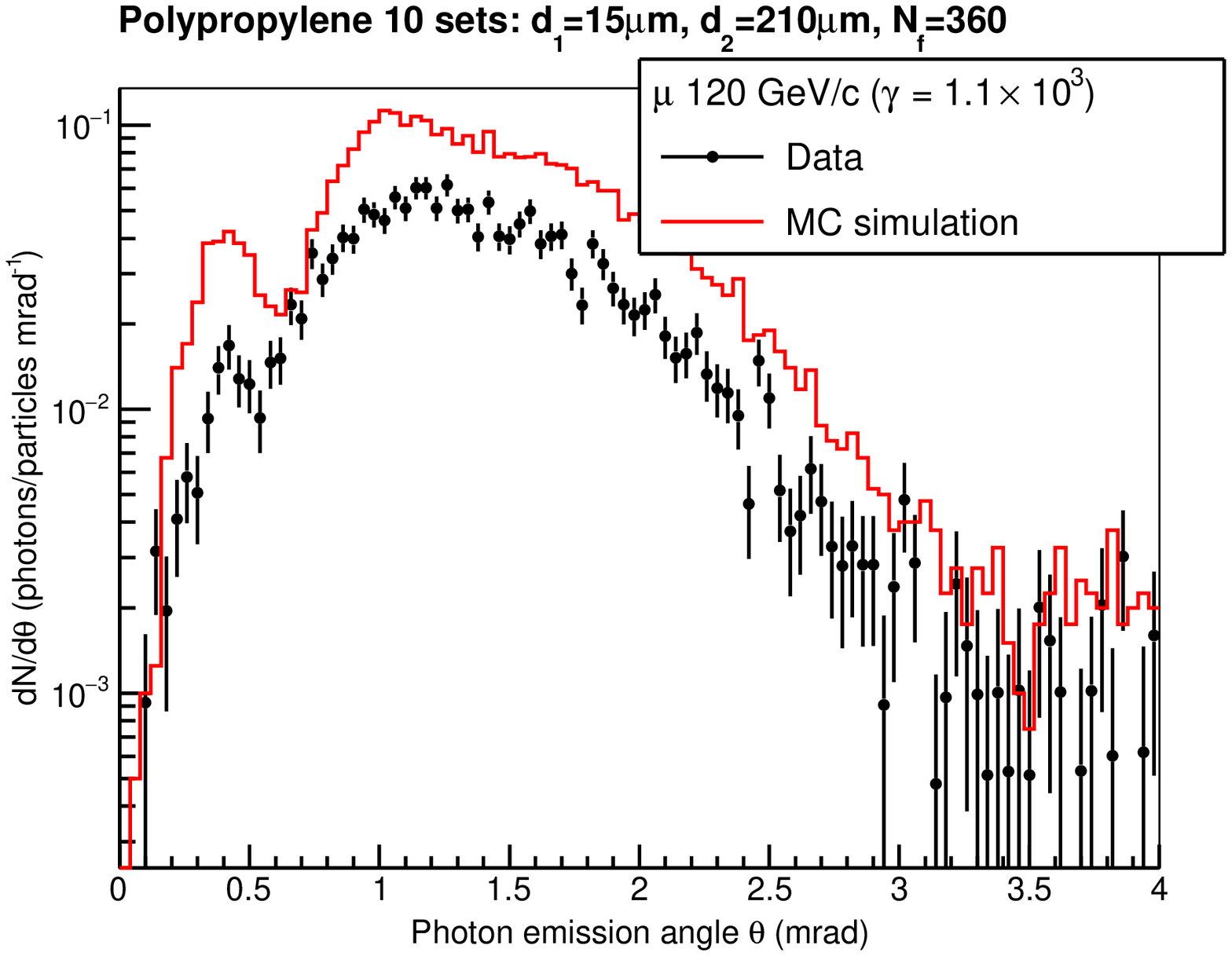}
\end{center}
\caption{Comparison between the measured and simulated differential spectra 
in energy (left column) and in angle (right column) of the TR X-rays produced 
by $20 \units{GeV/c}$ electrons and $290$, $180$ and $120 \units{GeV/c}$ 
muons crossing the 10-sets polypropylene radiator. The measured spectra have been
corrected subtracting the background measured in the runs with dummy radiators.}
\label{fig:polypropylene}
\end{figure}

The polypropylene radiator has a fine structure (see Table~\ref{tab:radiators}) and any 
fluctuations of the foil and gap thicknesses can significantly affect the results of simulation. 
It was found that the best agreement with experimental data is obtained when the polypropylene radiator 
is simulated as a slightly irregular radiator, with $\sigma_{d_{1}}=0.1 d_{1}=1.5\units{\mu m}$ 
and $\sigma_{d_{2}}=0.05 d_{2}=10.5\units{\mu m}$. Figure~\ref{fig:polypropylene} shows a comparison 
of the measured spectra with those obtained from the simulation of the polypropylene radiator for the four 
types of beam particles. As for other radiators the shapes of the spectra are well reproduced, including 
interference peak positions. However, significant discrepancies are observed in the normalizations. 
These discrepancies are at the level of a few percent in the run with $20 \units{GeV/c}$ electrons, 
and increase with decreasing Lorentz factor of the radiating particles. The discrepancies in the normalization 
of the spectra could be due to differences between the polypropylene absorption lengths used in the simulation 
and the real ones. This may also reflect the fact that exact TR turn-on curve strongly depends on the foil 
thickness and its density. Significant fluctuations of these parameters may lead to a change of the TR yield 
below the saturation level.

\subsection{Fiber radiator}

\begin{figure}[!tbp]
\begin{center}
\includegraphics[width=0.39\columnwidth]{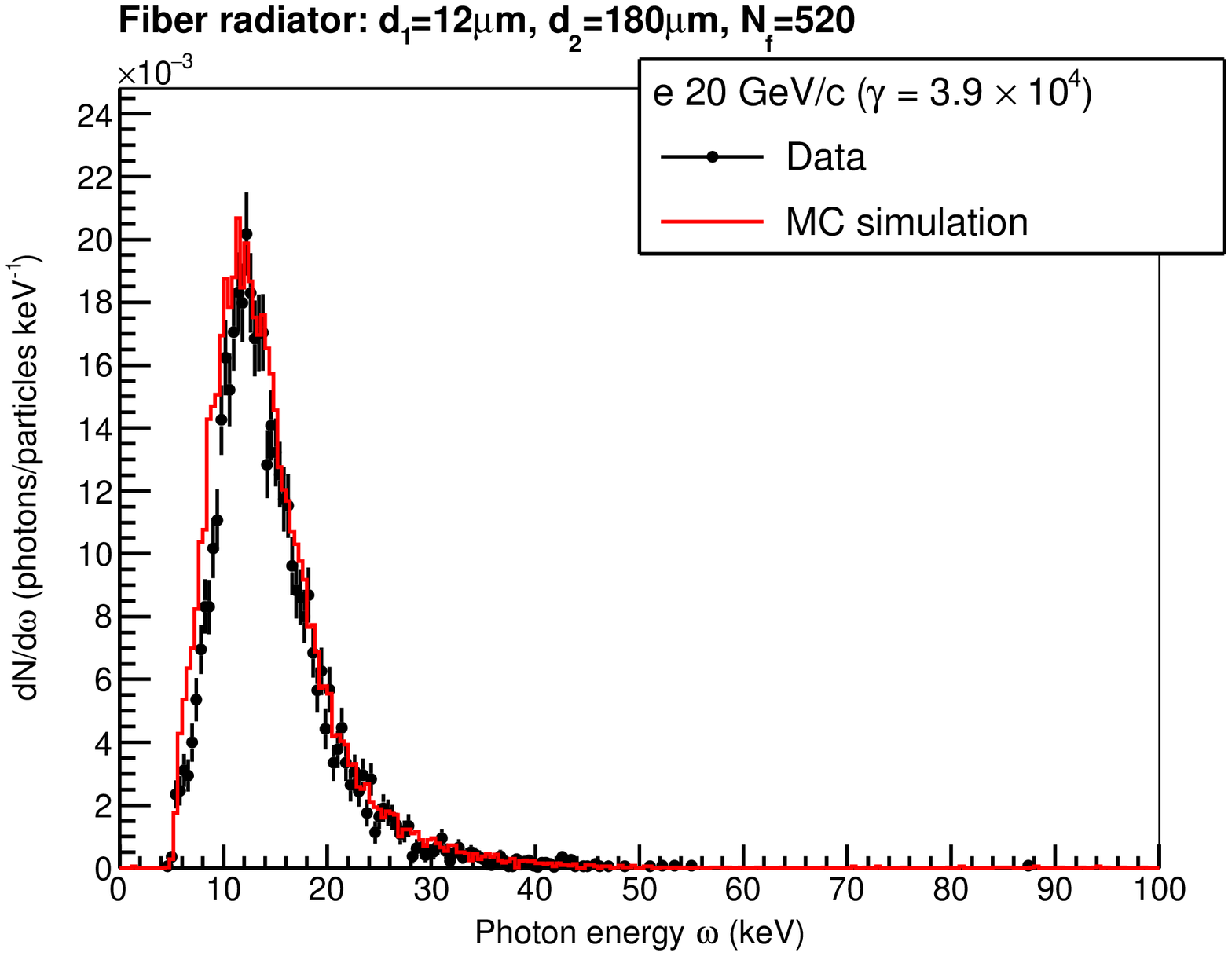}
\includegraphics[width=0.39\columnwidth]{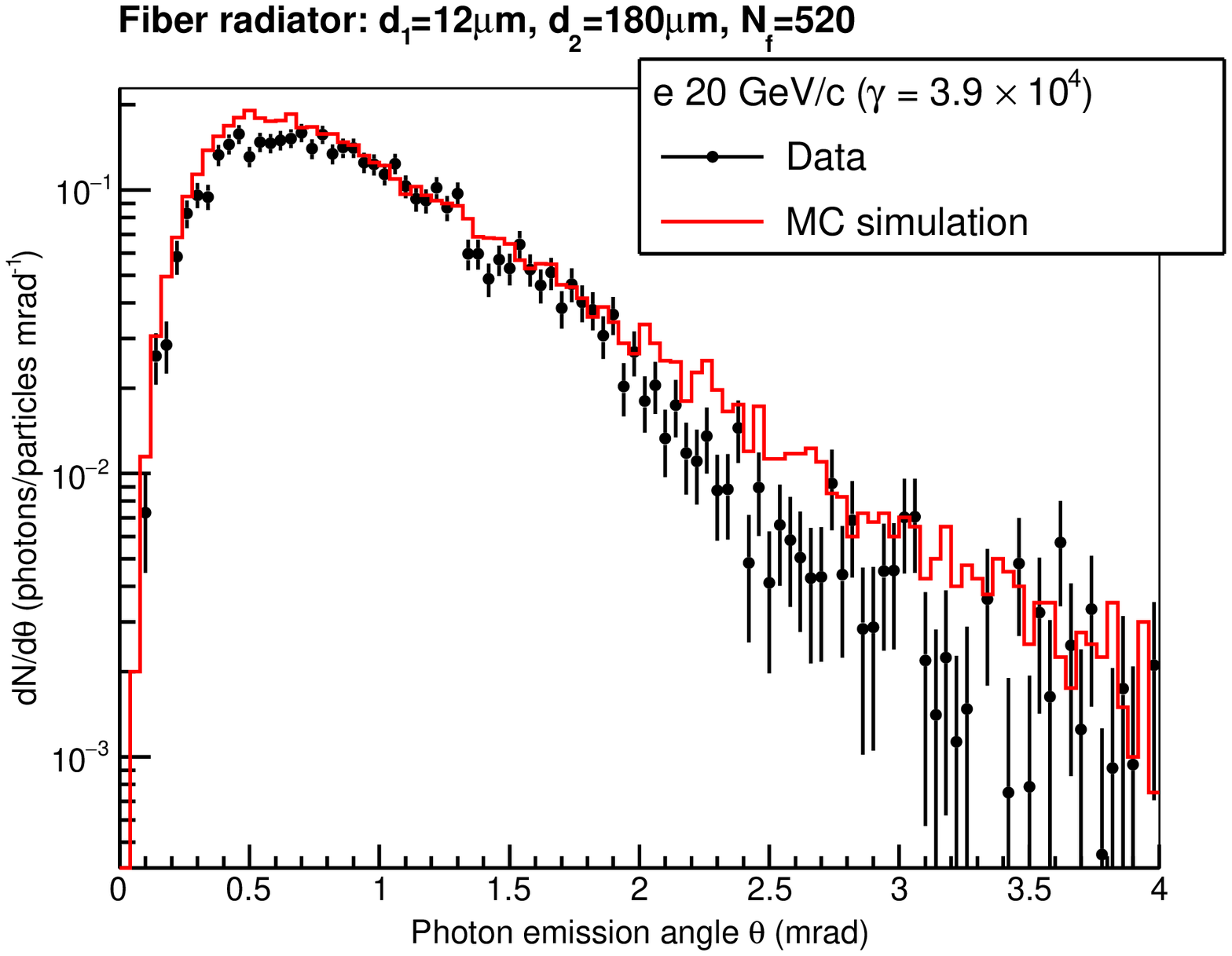}
\includegraphics[width=0.39\columnwidth]{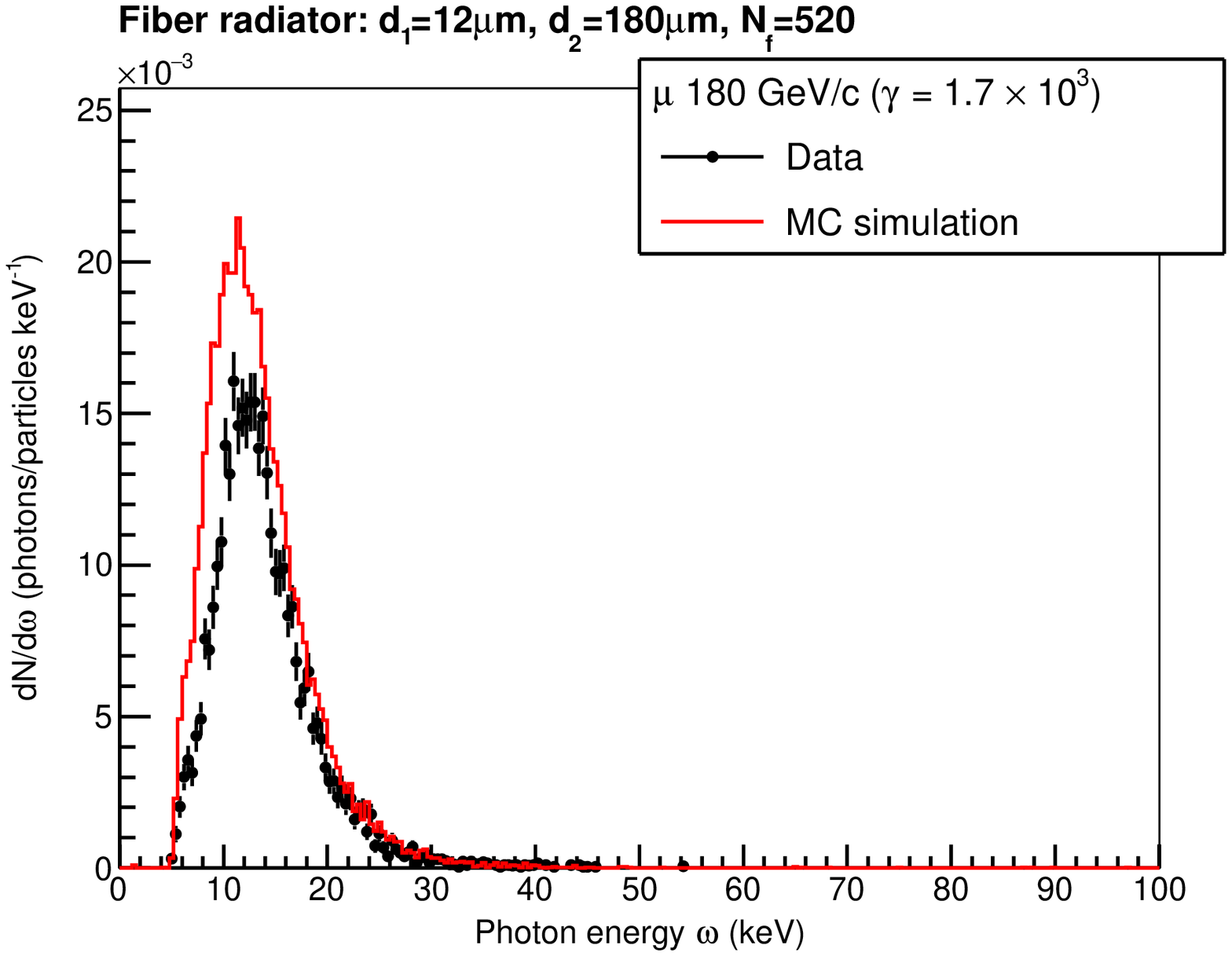}
\includegraphics[width=0.39\columnwidth]{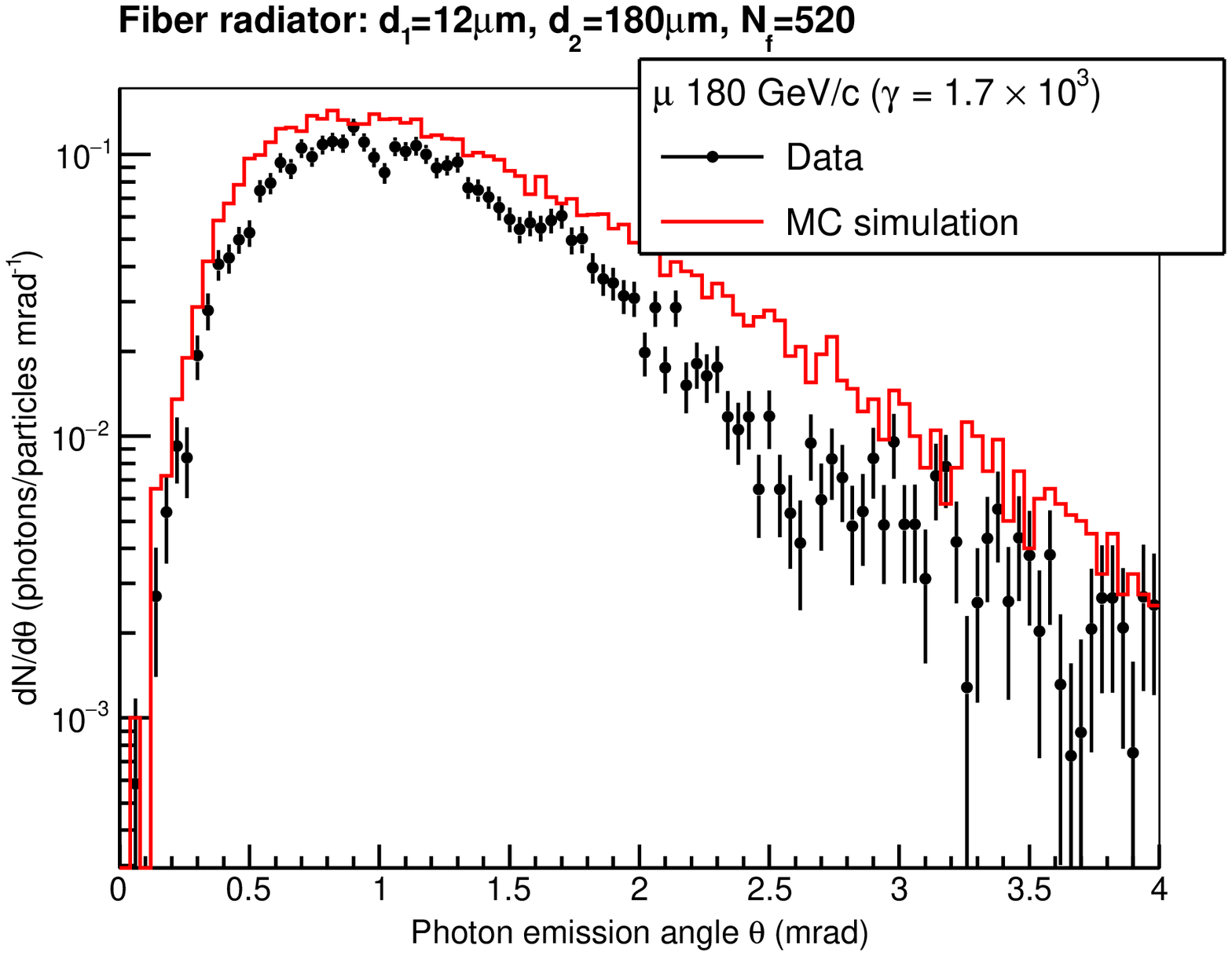}
\includegraphics[width=0.39\columnwidth]{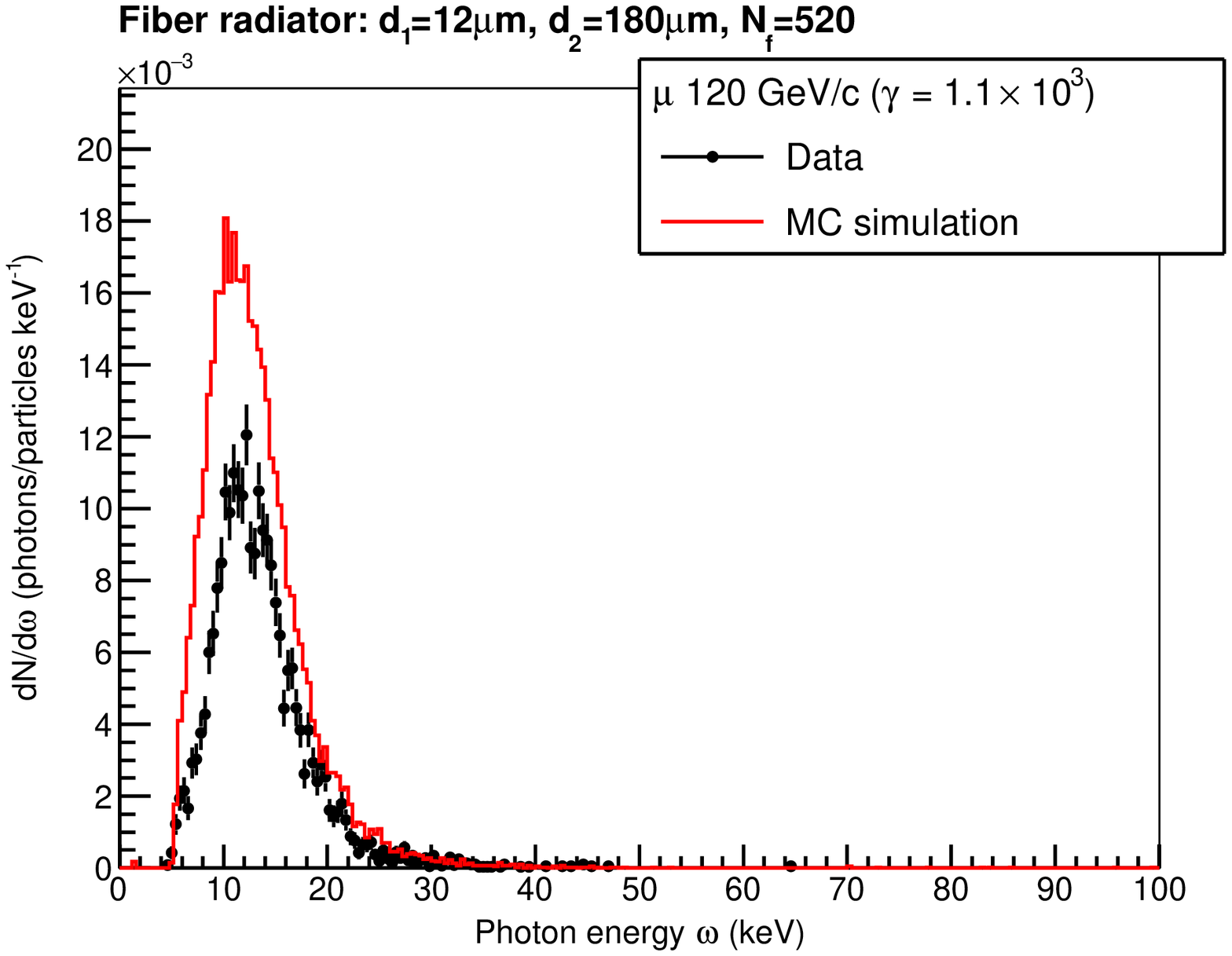}
\includegraphics[width=0.39\columnwidth]{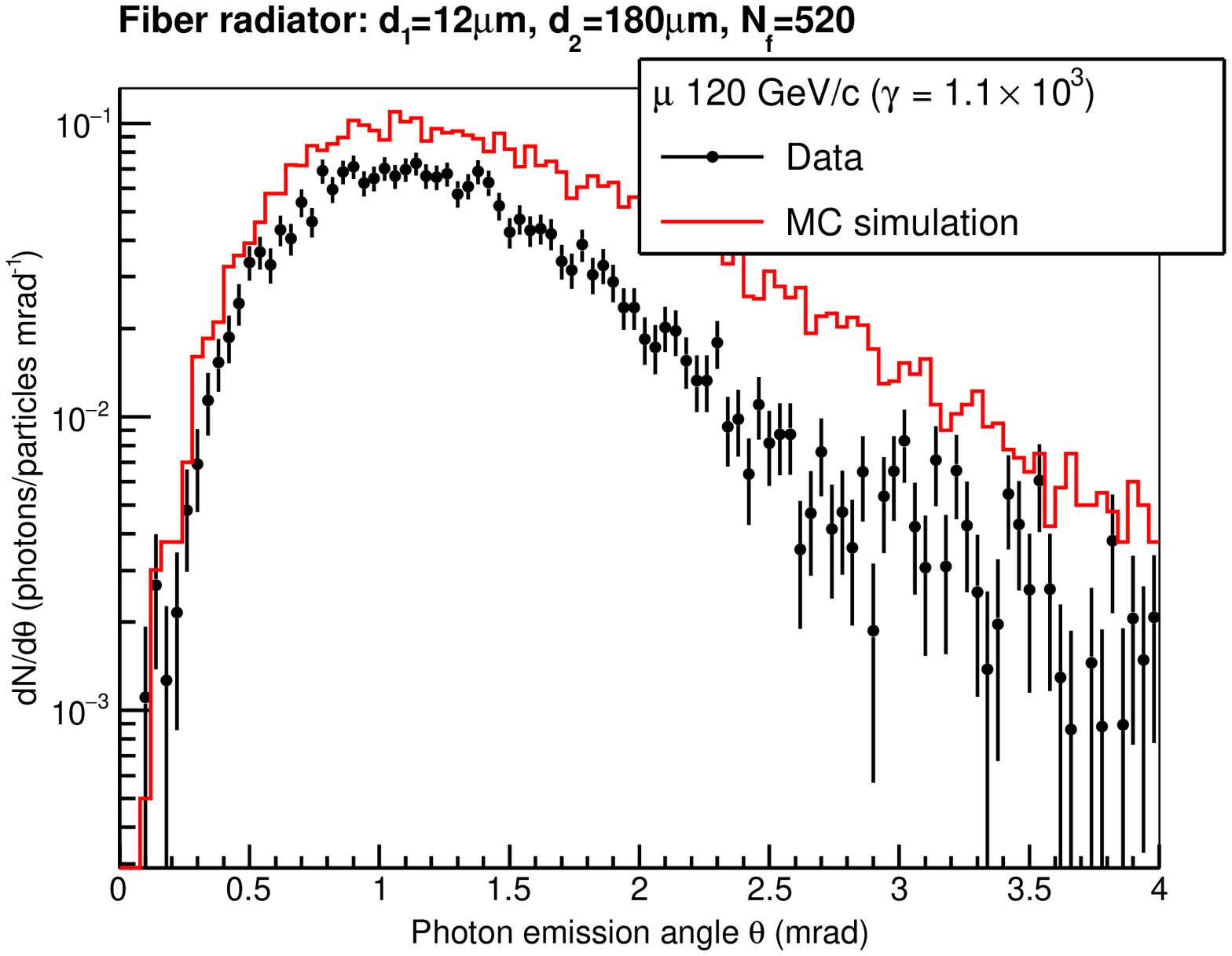}
\end{center}
\caption{Comparison between the measured and simulated differential spectra 
in energy (left column) and in angle (right column) of the TR X-rays produced 
by $20 \units{GeV/c}$ electrons and $180$ and $120 \units{GeV/c}$ 
muons crossing the fiber radiator. The measured spectra have been
corrected subtracting the background measured in the runs with dummy radiators.}
\label{fig:fiber}
\end{figure}

The fiber radiator is made of polypropylene fibers with diameter of about $15\units{\mu m}$. 
The density of this radiator is close to that of the polypropylene one. In general, one should expect similar 
TR production properties of these two radiators. The measured and the simulated TR spectra obtained with this radiator 
are shown in Figure~\ref{fig:fiber}. As the fiber radiator is highly irregular, in the simulation large variations of 
its parameters were assumed: $\sigma_{d_{1}}=d_{1}$ and $\sigma_{{d_2}}=d_{2}$. 
As in the case of the polypropylene regular radiator, we see that the shapes of the spectra are quite well reproduced, 
even though also in this case discrepancies (increasing with decreasing Lorentz factor) are observed  in the normalizations.

\section{Conclusions}

Simultaneous measurements of the energy and emission angle of TR X-rays produced by fast particles crossing different 
types of radiators were performed using a double-sided silicon strip detector. Tests were done with particles covering a 
wide range  of Lorentz factors from about $10^{3}$ to about $4 \times 10^{4}$. The test beam data were compared  with the results 
of dedicated Monte Carlo simulations, which use the  parameterizations of the TR yield derived from theoretical calculations. 
The simulations reproduce very well the shapes of the measured energy and angular distributions of TR X-rays for all radiators 
and types of beam particles. However, the simulations systematically overestimate the number of photons, particularly for particles 
with low Lorentz factors. This phenomenon requires a dedicated study. 

Our data also show that interference effects between foils may significantly change 
the angular distribution of the TR. For instance, for $20\units{GeV/c}$ electrons crossing the polypropylene radiator, 
the most probable TR emission angle is about $0.8 \units{mrad}$ instead of $0.025 \units{mrad}$, 
as expected from the commonly used law $\theta \approx 1/\gamma$. 

These studies show that simultaneous measurements of both the energy and the emission angles of the TR X-rays could be 
used to enhance the  particle identification performances of TRDs. For example, new generation TRDs based on high-granularity 
detectors, might be able to discriminate different species of hadrons in the gamma-factor ranges where conventional identification methods are unreliable. 

\section*{Acknowledgments}
We gratefully acknowledge the~financial support from Russian Science
Foundation grant (project No.\ 16-12-10277). We would also like to thank
the CERN staff for providing logistics and technical support at the SPS
beam line.

\bibliographystyle{elsarticle-num} 
\bibliography{article.bib}

\end{document}